\documentclass[a4paper,12pt]{iopart}
\usepackage{iopams}
\usepackage{setstack}
\usepackage{graphicx}
\usepackage{bm}
\usepackage{epsfig}

\newcommand{\diag}{{\rm diag\,}}
\newcommand{\sign}{{\rm sign\,}}

\newcommand{\Pf}{{\rm Pf\,}}

\newcommand{\U}{{\rm U\,}}

\newcommand{\W}{{\rm W\,}}
\newcommand{\rt}{{\rm r\,}}
\newcommand{\lt}{{\rm l\,}}

\newcommand{\RE}{{\rm Re\,}}

\newcommand{\IM}{{\rm Im\,}}

\newcommand{\eins}{\leavevmode\hbox{\small1\kern-3.8pt\normalsize1}}
\eqnobysec

\begin{document}

\title[Mixing in orthogonal polynomial theory]{Mixing of orthogonal and skew-orthogonal polynomials and its relation to Wilson RMT}
\author{Mario Kieburg}
\address{Department of Physics and Astronomy, State University of New York at Stony Brook, NY 11794-3800, USA}
\eads{\mailto{mario.kieburg@stonybrook.edu}}

\date{\today}

\begin{abstract}
The unitary Wilson random matrix theory is an interpolation between the chiral Gaussian unitary ensemble and the Gaussian unitary ensemble. This new way of interpolation is also reflected in the orthogonal polynomials corresponding to such a random matrix ensemble.  Although the chiral Gaussian unitary ensemble as well as the Gaussian unitary ensemble are associated to the Dyson index $\beta=2$ the intermediate ensembles exhibit a mixing of orthogonal polynomials and skew-orthogonal polynomials. We consider the Hermitian as well as the non-Hermitian Wilson random matrix and derive the corresponding polynomials, their recursion relations, Christoffel-Darboux-like formulas, Rodrigues formulas and representations as random matrix averages in a unifying way. With help of these results we derive the unquenched $k$-point correlation function of the Hermitian and the non-Hermitian Wilson random matrix in terms of two-flavor partition functions only. This representation is due to a Pfaffian factorization. It drastically simplifies the expressions which can be easily numerically evaluated. It also serves as a good starting point for studying the Wilson-Dirac operator in the $\epsilon$-regime of lattice quantum chromodynamics.
\end{abstract}

\submitto{\JPA}

\section{Introduction}\label{sec1}

In the microscopic limit chiral random matrix theory ($\chi$RMT) can be directly mapped to the $\epsilon$-regime of quantum chromodynamics (QCD) and is successfully applied to it since the 90's \cite{Shuryak:1992pi,Verbaarschot:1993ze}. Both theories share the same universality class which is the reason for the existence of this equivalence. $\chi$RMT was also extended to a non-zero chemical potential by adding a scalar proportional to $\gamma_0$ \cite{Stephanov:1996ki,Verbaarschot:1998zn,Halasz:1999gc,Splittorff:2006fu,Han:2008xj}.  In the last decade  a second approach was pursued. A second chiral random matrix was introduced yielding the chiral analogue of the Ginibre ensembles \cite{Akemann:2003by,Osborn:2004rf,Akemann:2004dr,Akemann:2005fd,Osborn:2008jp,Akemann:2009my,Akemann:2009fc,Akemann:2010mt}.
A quantitative analysis of the sign problem in Monte-Carlo simulations was quite elusive until it was solved in $\chi$RMT \cite{Splittorff:2006fu,Han:2008xj,Bloch:2008cf,Bloch:2011ny}. The hope is now to extend these new insights to  QCD at non-zero lattice spacing.

Recently, random matrix theories for lattice QCD became the focus of interest. The idea is to derive analytical results of lattice artefacts in the data. One important realization of lattice QCD is by means of staggered fermions. In Refs.~\cite{Osborn:2003dr,Kieburg:2011yg}, a $\chi$RMT was considered which is equivalent to the $\epsilon$-regime of these fermions. Unfortunately, this model is highly involved due to the high number of low energy constants and, hence, of the coupling constants in the random matrix model.

The Wilson Dirac operator is another realization of lattice QCD. It proved that the corresponding random matrix model \cite{Damgaard:2010cz,Akemann:2010zp,Splittorff:2011bj,Akemann:2011kj,Kieburg:2011yg,Kieburg:2011uf} is much better accessible for analytical calculations than the one of the staggered fermions. The Wilson term which is given by a Laplace operator \cite{Wilson77,GatLan10} explicitly breaks chiral symmetry and is Hermitian. Thus the main idea was to add on the diagonal of $\chi$RMT two Hermitian matrices to simulate the same effect \cite{Damgaard:2010cz} and it proved to be in the same universality class as the Wilson Dirac operator in the $\epsilon$-regime \cite{Sharpe:1998xm,RS02,Bar:2003mh,Sharpe06}. Actually one can consider a Hermitian version \cite{Damgaard:2010cz,Akemann:2010zp,Splittorff:2011bj,Akemann:2011kj,Deuzeman:2011dh,Damgaard:2011eg} of this random matrix ensemble which is numerically cheaper in lattice simulations. However only the non-Hermitian version \cite{Kieburg:2011yg,Kieburg:2011uf,Damgaard:2011eg} is directly related to the chiral symmetry breaking by a finite lattice spacing. The Hermitian version can also be considered as an interpolation between a chiral Gaussian unitary ensemble ($\chi$GUE) and a Gaussian unitary ensemble (GUE). The coupling constant is then the lattice spacing.

Quite recently this new kind of random matrix model has given new insights on the signs of the low energy constants in the chiral Lagrangian of the Wilson Dirac operator \cite{Akemann:2010zp,Splittorff:2012gp,KSV12}. These signs are controversial since they are crucial to decide if an Aoki phase \cite{Aoki:1983qi} exists or not. Such a phase is a pure lattice artefact and has no analogue in continuum QCD. Therefore a large  analytical \cite{Sharpe:1998xm,Hansen:2011kk,Splittorff:2012gp,KSV12} as well as numerical \cite{Aoki:2007zza,Baron:2009wt,Necco:2011vx,Deuzeman:2011dh,Damgaard:2011eg} effort was made to determine the low energy constants.

Orthogonal polynomial theory \cite{Meh60,MehGau60,Dys62,Akemann:2002vy,Meh04,Akemann2004,Bergere:2004cp,Kuijlaars10} was as successfully applied to RMT as the supersymmetry method \cite{Efe83,VWZ85,Efe97,Fyodorov2002,Guh06,Som07,LSZ08,KGG09,KSG09,Guh11}. In particular the combination of both methods with the recently developed method of an algebraical rearrangement of the joint probability density with quotients of characteristic polynomials \cite{KieGuh09a} are quite efficient to find compact and simple analytical results of the spectral correlations of random matrix ensembles. In this work we address the $k$-point correlation functions of the Hermitian as well as the non-Hermitian version and make use of such a combination.

An interesting point of view of Wilson RMT appears when we study it with help of orthogonal polynomial theory. In Ref.~\cite{Akemann:2011kj} the authors considered the Hermitian Wilson RMT and found that the construction of the skew-orthogonal polynomials strongly depend on the index $\nu$ of the random matrix which is the number of zero modes in the continuum limit. They only explicitly constructed these polynomials for $\nu=0,1$. In this article we construct these polynomials for an arbitrary index and for both version of Wilson RMT in a unifying way. We also successfully look for a recursion relation, Christoffel Darboux-like formulas, Rodrigues formulas and their explicit expression as random matrix integrals. By this study we get a complete picture what these orthogonal and skew-orthogonal polynomials are and how they are related to the orthogonal polynomials of some limits, in particular the continuum limit and the limit of a large lattice spacing.

First we specify what are the conditions the orthogonal and skew-orthogonal polynomials have to fulfill. Thereby we recognize that the corresponding weight has to satisfy a particular property, too. Luckily we are able to modify the weight in the joint probability densities without changing the partition functions and the $k$-point correlation functions such that this property can be achieved. In the second step we construct the polynomials with help of Pfaffians whose anti-symmetry under permutations of rows and columns proves quite useful.

After we show some useful properties of the orthogonal and skew-orthogonal polynomials we derive a Pfaffian factorization of the $k$-point correlation functions. This factorization is for numerical evaluations advantageous because it reduces the complexity of the integrand to an average over two characteristic polynomials. In combination with the supersymmetry method \cite{Guh06,LSZ08,KGG09,Guh11} one may simplify the whole problem to two-fold integrals. Factorizations to determinants and Pfaffians were found for many random matrix ensembles of completely different symmetries \cite{Meh60,MehGau60,Dys62,BreHik00,Mehta2001,Fyodorov2003,Baik2003,Meh04,Bergere:2004cp,BorStr05,KieGuh09a}. A Pfaffian for the eigenvalue correlations of the Hermitian Wilson random matrix ensemble was already shown in Ref.~\cite{Akemann:2011kj}. We prove the existence of such a structure for the non-Hermitian version, too. Furthermore we identify the kernels of both Pfaffians with two-flavor partition functions. The identification as well as the structure carry over to the microscopic limit which makes them also applicable to the chiral Lagrangian of the Wilson-Dirac operator.

We consider unquenched Wilson RMT, i.e. a finite number of fermionic flavors. Recently, the partition function with one fermionic flavor and the corresponding microscopic level density was studied  in Ref.~\cite{Larsen:2011sd}. In our calculations the number of fermions may be arbitrary. Nevertheless all eigenvalue correlations, also the one of the unquenched theory, can be expressed by two-flavor partition functions because of the Pfaffian factorization.

Moreover such a Pfaffian determinant of the $k$-point correlation functions comes in handy when calculating the individual eigenvalue distributions. The authors of Ref.~\cite{Akemann:2012pn} were able to express the gap probability of the eigenvalues of the Hermitian Wilson random matrix ensemble as a Fredholm-Pfaffian only due to this structure. Hence a similar simplification is highly desirable for the non-Hermitian Wilson random matrix ensemble.
 
The outline of this article is as follows. In Sec.~\ref{sec2} we briefly introduce the Wilson random matrix model and its two kinds of joint probability densities corresponding to the Hermitian and the non-Hermitian version. With help of the $k$-point correlation function we propose the problem. In particular we will list the conditions the polynomials have to fulfill. In Sec.~\ref{sec3} we construct the orthogonal and skew-orthogonal polynomials. Thereby we derive a recursion relation which is helpful to proof a Christoffel Darboux-like formula. Moreover we show a representation of the polynomials and the Christoffel Darboux-like formula as random matrix averages. Such a representation is useful to study the microscopic limit of the random matrix ensemble by means of the supersymmetry method. Hence we calculate the asymptotics of the polynomials and the Christoffel Darboux-like formula. In Sec.~\ref{sec4} we apply the derived results to the $k$-point correlation functions of the Hermitian and the non-Hermitian version of Wilson RMT and identify the kernels of the Pfaffian with two-flavor partition functions. Readers only interested in the $k$-point correlation functions of the Wilson random matrix ensemble can jump to this section because it contains the main results which can be mostly understood without the technical details in Sec.~\ref{sec3} due to the identification of the kernels with two-flavor partition functions. The conclusions are made in Sec.~\ref{sec5} and the details of the calculations are given in the appendices.

\section{Two joint probability densities for one random matrix theory}\label{sec2}

The models we want to consider are motivated by the Wilson Dirac operator in lattice QCD \cite{Damgaard:2010cz}. The corresponding random matrix theory consists of the matrix
\begin{eqnarray}\label{2.1}
 D_\W=\left(\begin{array}{cc} A & W \\ -W^\dagger & B \end{array}\right)
\end{eqnarray}
distributed by the Gaussian
\begin{eqnarray}\label{2.2}
 \fl P(D_\W)&=&\left(\frac{n}{2\pi a^2}\right)^{[n^2+(n+\nu)^2]/2}\left(-\frac{n}{2\pi}\right)^{n(n+\nu)}\exp\left[-\frac{a^2}{2}\left(\mu_\rt^2+\frac{n+\nu}{n}\mu_\lt^2\right)\right]\\
 \fl&\times&\exp\left[-\frac{n}{2a^2}(\tr A^2+\tr B^2)-n\tr WW^\dagger+\mu_\rt\tr A+\mu_\lt\tr B\right].\nonumber
\end{eqnarray}
The Hermitian matrices $A$ and $B$ have the dimensions $n\times n$ and $(n+\nu)\times(n+\nu)$ and explicitly break chiral symmetry,
\begin{eqnarray}\label{2.3}
 \gamma_5\left.D_{\rm W}\right|_{m=0}\gamma_5\neq-\left.D_{\rm W}\right|_{m=0}\quad\mathrm{with}\quad\gamma_5=\diag(\eins_{n},-\eins_{n+\nu}).
\end{eqnarray}
The matrix $W$ is a $n\times(n+\nu)$ complex matrix with independent entries. The variable $a$ plays the role of the lattice spacing and the Gaussian of $A$ and $B$ yields one low energy constant known as $W_8$ \cite{Damgaard:2010cz,Akemann:2010zp,Splittorff:2011bj,Kieburg:2011yg,Kieburg:2011uf,KSV12}. The variables $\mu_{\rt/\lt}$ might be also considered as Gaussian distributed random variables and generate two additional low energy constants, $W_6$ and $W_7$ \cite{Akemann:2010zp,Splittorff:2011bj,KSV12}, in chiral perturbation theory of the Wilson Dirac operator \cite{Sharpe:1998xm,RS02,Bar:2003mh,Sharpe06}. Here we consider them as fixed constants to keep the calculation as simple as possible but the model is general enough to introduce also the Gaussian integrals for $\mu_{\rt/\lt}$ at the end of the day. They originate from a shift of the matrices $A$ and $B$ by mass terms. The case when we do not integrate over $\mu_{\rt/\lt}$ by Gaussians and keep them as constants corresponds to the low energy constants $W_6=W_7=0$.

The parameter $\nu$ is called the index of the Dirac operator and is the number of the generic real modes of $D_{\W}$. Since $D_\W$ is $\gamma_5$-Hermitian, i.e. $D_\W^\dagger=\gamma_5D_\W\gamma_5$, the matrix
\begin{equation}\label{2.4}
 D_5=\gamma_5D_\W
\end{equation}
is Hermitian. Moreover the complex eigenvalues of $D_\W$ come in complex conjugated pairs only. The number of these pairs, $l$, varies from $0$ to $n$.

The matrix $D_5=VxV^{-1}$ can be diagonalized by a unitary matrix $V\in\U(2n+\nu)$ whereas the matrix $D_\W=UZ_lU^{-1}$ can only be quasi diagonalized by a matrix in the non-compact unitary group $U\in\U(n,n+\nu)$, i.e. $U^{-1}=\gamma_5U^\dagger\gamma_5$. The diagonal matrix $x=\diag(x_1,\ldots,x_{2n+\nu})$ consists of real eigenvalues, only. The quasi-diagonal matrix
\begin{eqnarray}\label{2.5}
 Z_l=\left(\begin{array}{cc|cc} x^{(1)} & 0 & 0 & 0 \\ 0 & x^{(2)} & y^{(2)} & 0 \\ \hline 0 & -y^{(2)} & x^{(2)} & 0 \\ 0 & 0 & 0 & x^{(3)} \end{array}\right),
\end{eqnarray}
depends on the real diagonal matrices $x^{(1)}=\diag(x_1^{(1)},\ldots,x_{n-l}^{(1)})$, $x^{(2)}=\diag(x_1^{(2)},\ldots,$ $ x_{l}^{(2)})$, $y^{(2)}=\diag(y_1^{(2)},\ldots, y_{l}^{(2)})$ and $x^{(3)}=\diag(x_1^{(3)},\ldots,x_{n+\nu-l}^{(3)})$ with the dimensions $n-l$, $l$, $l$ and $n+\nu-l$, respectively. Then the complex conjugated eigenvalue pairs of $D_\W$ are $(z^{(2)},z^{*(2)})=(x^{(2)}+\imath y^{(2)},x^{(2)}-\imath y^{(2)})$. The $n+1$ different sectors of different numbers of the complex conjugated pairs are labelled by  $l$. 

The joint probability density is one of the best quantities for analyzing the eigenvalue correlations of random matrices. It is also the starting point of our discussions in the ensuing sections. The Hermitian, $D_5$, and the non-Hermitian, $D_\W$, Wilson random matrix have different joint probability densities. Though these densities have a completely different form, we will see that their orthogonal and skew-orthogonal polynomials have much in common, see Sec.~\ref{sec3}.

The joint probability density of $D_5$ is \cite{Akemann:2011kj}
\begin{eqnarray}\label{2.6}
 \fl p_5(x)d[x]&=&c_-(1-a^2)^{-n(n+\nu-1/2)}a^{-n-\nu^2}\exp\left[-\frac{a^2}{2}\left(\mu_\rt^2+\left(1+\frac{\nu}{n}\right)\mu_{\lt}^2\right)+\frac{n\widehat{m}_{6-}^2}{8\widehat{a}_-^2}\right]\hspace*{-0.1cm}\Delta_{2n+\nu}(x)\nonumber\\
 \fl&\times&\Pf\left[\begin{array}{cc} \left\{g^{(-)}_2(x_i,x_j)\right\}_{1\leq i,j\leq 2n+\nu} & \displaystyle\{x_i^{j-1}g_1^{(-)}(x_i)\}\underset{1\leq j \leq \nu}{\underset{1\leq i\leq 2n+\nu}{}} \\ \displaystyle\{-x_j^{i-1}g_1^{(-)}(x_j)\}\underset{1\leq j \leq 2n+\nu}{\underset{1\leq i\leq \nu}{}} & 0 \end{array}\right]\prod\limits_{j=1}^{2n+\nu}dx_j,
\end{eqnarray}
where
\begin{eqnarray}
 \fl g^{(-)}_2(x_1,x_2)&=&\exp\left[-\frac{n}{4a^2}(x_1+x_2)^2-\frac{n}{4}(x_1-x_2)^2+\frac{n\widehat{\lambda}_{7-}}{4\widehat{a}_-^2}(x_1+x_2)\right]\label{2.7}\\
 \fl&\times&{\rm erf}\left[\frac{1}{\sqrt{8}\widehat{a}_-}\left(n(x_2-x_1)-\widehat{m}_{6-}\right),\frac{1}{\sqrt{8}\widehat{a}_-}\left(n(x_1-x_2)-\widehat{m}_{6-}\right)\right],\nonumber\\
 \fl g_1^{(\pm)}(x)&=&\exp\left[-\frac{n}{2a^2}x^2\pm\mu_\lt x\right].\label{2.8}
\end{eqnarray}
We define the constants
\begin{eqnarray}\label{2.9}
 \fl\widehat{a}_\pm&=&\sqrt{\frac{na^2}{2(1\pm a^2)}},\\
 \fl\widehat{m}_{6\pm}&=&\frac{2\widehat{a}_\pm^2}{n}(\mu_\rt+\mu_\lt),\label{2.9a}\\
 \fl\widehat{\lambda}_{7\pm}&=&\frac{2\widehat{a}_\pm^2}{n}(\mu_\rt-\mu_\lt),\label{2.9b}\\
 \fl\frac{1}{c_-}&=&\left(\frac{16\pi}{n}\right)^{n/2}(2\pi)^{\nu/2}n^{-\nu^2/2-n(n+\nu)}(2n+\nu)!\prod\limits_{j=0}^{n-1}j!\prod\limits_{j=0}^{n+\nu-1}j!\,.\label{2.9d}
\end{eqnarray}
The notation of $\widehat{m}_{6\pm}$ and $\widehat{\lambda}_{7\pm}$ reflects the nature of their symmetries. The constant $\widehat{m}_{6\pm}$ acts as an effective mass and $\widehat{\lambda}_{7\pm}$ as an effective axial mass, i.e. a source term proportional to $\gamma_5$.  They refer to the low-energy constants $W_6$ and $W_7$ which are found in the microscopic limit \cite{Damgaard:2010cz,Akemann:2010zp,Splittorff:2011bj,Kieburg:2011yg,Kieburg:2011uf,KSV12}, i.e. $n\to\infty$, $\widehat{a}=\sqrt{n}a/\sqrt{2}=\widehat{a}_\pm={\rm const.}$, $\widehat{m}_{6}=\widehat{m}_{6\pm}={\rm const.}$, $\widehat{\lambda}_{7}=\widehat{\lambda}_{7\pm}={\rm const.}$ and $\widehat{z}=2n z={\rm const.}$  We emphasize that we have to integrate over $\widehat{m}_{6}$ and $\widehat{\lambda}_{7}$ to obtain the low energy constants $W_6$ and $W_7$, respectively. Please notice that our notation differs from the one in Refs.~\cite{Damgaard:2010cz,Akemann:2010zp} where the source terms proportional to $\gamma_5$ are denoted by $z$. To avoid confusion with the complex eigenvalues of $D_\W$ we denote these variables by $\lambda$ in the present article. Furthermore we renamed the variables $y_6$ and $y_7$ to $\widehat{m}_{6}$ and $\widehat{\lambda}_{7}$, respectively, since the former notation can create a confusion with the imaginary parts of the complex eigenvalues of $D_\W$.

 The Vandermonde determinant is given by
\begin{equation}\label{2.9c}
 \fl\Delta_{2n+\nu}(x)=\prod\limits_{1\leq i<j\leq 2n+\nu}(x_i-x_j)=(-1)^{(2n+\nu)(2n+\nu-1)/2}\det\left[x_i^{j-1}\right]_{1\leq i,j\leq 2n+\nu}.
\end{equation}
The function ${\rm erf}(x_1,x_2)={\rm erf}(x_2)-{\rm erf}(x_1)$ is the generalized error function.

The Pfaffian in $p_5$, see Eq.~\eref{2.6}, is due to the symmetrization of the eigenvalues. The two-point weight $g_2^{(-)}$ is anti-symmetric and is a strong interaction of two different eigenvalues. In the continuum limit, $a\to0$, $g_2^{(-)}$ generates a Dirac delta function enforcing that we have always an eigenvalue pair $(\lambda,-\lambda)$ of the Dirac operator if $\lambda\neq0$. The two off-diagonal blocks  are reminiscent of Vandermonde determinants and are artefacts of the zero modes at $a=0$.

The joint probability density of $D_\W$ is
\begin{eqnarray}
 \fl p_\W(Z)d[Z]&=&\hspace*{-0.1cm}c_+(1+a^2)^{-n(n+\nu-1/2)}a^{-n-\nu^2}\hspace*{-0.2cm}\exp\left[-\frac{a^2}{2}\left(\mu_\rt^2+\left(1+\frac{\nu}{n}\right)\mu_{\lt}^2\right)+\frac{n\widehat{\lambda}_{7+}^2}{8\widehat{a}_+^2}\right]\hspace*{-0.15cm}\Delta_{2n+\nu}(Z)\nonumber\\
 \fl&\times&\det\left[\begin{array}{c} \displaystyle\{g^{(+)}_2(z_i^{(\rt)},z_j^{(\lt)})\}\underset{1\leq j \leq n+\nu}{\underset{1\leq i\leq n}{}} \\ \displaystyle\{(x_j^{(\lt)})^{i-1}g_1^{(+)}(x_j^{(\lt)})\delta(y_j^{(\lt)})\}\underset{1\leq j \leq n+\nu}{\underset{1\leq i\leq \nu}{}} \end{array}\right]\prod\limits_{j=1}^nd x_j^{(\rt)}d y_j^{(\rt)}\prod\limits_{j=1}^{n+\nu}d x_j^{(\lt)}d y_j^{(\lt)},\nonumber\\
 \fl&&\label{2.10}
\end{eqnarray}
where
\begin{eqnarray}
 \fl g^{(+)}_2(z_1,z_2)&=&g_{\rm r}(x_1,x_2)\delta(y_1)\delta(y_2)+g_{\rm c}(z_1)\delta(x_1-x_2)\delta(y_1+y_2),\label{2.11}\\
 \fl g_{\rm r}(x_1,x_2)&=&\exp\left[-\frac{n}{4a^2}(x_1+x_2)^2+\frac{n}{4}(x_1-x_1)^2+\frac{n\widehat{m}_{6+}}{4\widehat{a}_+^2}(x_1+x_2)\right]\nonumber\\
 \fl&\times&\left(\frac{x_1-x_2}{|x_1-x_2|}-{\rm erf}\left[\frac{1}{\sqrt{8}\widehat{a}_+}\left(n(x_1-x_2)-\widehat{\lambda}_{7+}\right)\right]\right),\label{2.12}\\
 \fl g_{\rm c}(z)&=&-2\imath\frac{y}{|y|}\exp\left[-\frac{n}{a^2}x^2-ny^2+\frac{n\widehat{m}_{6+}}{2\widehat{a}_+^2}x\right],\label{2.13}\\
 \fl\frac{1}{c_+}&=&(-1)^{\nu(\nu-1)/2+n(n-1)/2}\left(\frac{16\pi}{n}\right)^{n/2}(2\pi)^{\nu/2}n^{-\nu^2/2-n(n+\nu)}\prod\limits_{j=0}^{n}j!\prod\limits_{j=0}^{n+\nu}j!.\label{2.14}
\end{eqnarray}
Note that the one point weight $g_1^{(\pm)}$ of $D_5$ and of $D_\W$ is apart from the sign of the linear shift in the exponent the same. Also the other distributions show similarities with each other.

Comparing $p_\W$ with $p_5$ we recognize the major difference is the determinant which replaces the Pfaffian. The reason is a broken permutation symmetry in the eigenvalues of $D_\W$. We have to symmetrize over the eigenvalues $z^{(\rt)}$ and  $z^{(\lt)}$ separately. Since the two-point weight $g_2^{(+)}$ only couples $z^{(\rt)}$ with  $z^{(\lt)}$ but not two eigenvalues of one and the same set the symmetrization yields a determinant. Another crucial difference of $p_\W$ to $p_5$ is the distinction of real and complex eigenvalues reflecting the non-Hermiticity of $D_\W$. Interestingly the complex conjugated pairs only enter the two-point weight $g_2^{(+)}$. In the continuum limit the interaction of a pair of real eigenvalues, $g_{\rm r}$, is suppressed and the term for  the complex eigenvalues, $g_{\rm c}$, enforces the pairing of non-zero eigenvalues, $(\imath\lambda,-\imath\lambda)$, along the imaginary axis. Again a block resembling the Vandermonde determinant appears and is again a relict  of the former zero modes.

In the next two subsection we motivate the polynomials constructed in Sec.~\ref{sec3}. For this we consider the $k$-point correlation functions of $D_\W$ and $D_5$.

\subsection{The $k$-point correlation function of $D_5$}\label{sec2.1}

First, we consider the fermionic partition function of $D_5$ with $N_{\rm f}$ axial masses (characteristic polynomials of $D_5$), $\lambda=\diag(\lambda_1,\ldots,\lambda_{N_{\rm f}})$,
\begin{eqnarray}\label{2.1.1}
 Z_{N_{\rm f}}^{(n,\nu,-)}(\lambda)&\propto&\int d[D_\W]P(D_\W)\prod\limits_{j=1}^{N_{\rm f}}\det(D_5+\lambda_j\eins_{2n+\nu}).
\end{eqnarray}
The unit matrix of dimension $2n+\nu$ is denoted by $\eins_{2n+\nu}$. In the microscopic limit this partition function corresponds to the integral \cite{Damgaard:2010cz,Akemann:2010zp}
\begin{eqnarray}\label{2.1.2}
 \fl &&Z_{N_{\rm f}}^{(n,\nu,-)}\left(\frac{\lambda}{2n}\right)\overset{n\gg1}{\propto}\int\limits_{\U(N_{\rm f})} d\mu(U){\det}^{\nu}U\,\\
 \fl&\times&\exp\left[\frac{\widehat{m}_6}{2}\tr(U+U^{-1})+\frac{1}{2}\tr(\widehat{\lambda}_7\eins_{N_{\rm f}}+\widehat{\lambda})(U-U^{-1})-\widehat{a}^2\tr(U^2+U^{-2})\right].\nonumber
\end{eqnarray}
This is the effective Lagrangian of the Wilson-Dirac operator of the partition function with $N_{\rm f}$ fermionic quarks with a degenerate quark mass $\widehat{m}_6$ and $N_{\rm f}$ source terms $\widehat{\lambda}_7\eins_{N_{\rm f}}+\widehat{\lambda}$ proportional to $\gamma_5$, cf. Refs.~\cite{Sharpe:1998xm,RS02,Bar:2003mh,Sharpe06}. An integration over the variables $\widehat{m}_6$ and $\widehat{\lambda}_7$ weighted by two additional Gaussian will yield the two low energy constants $W_6$ and $W_7$ proportional to two squared trace terms \cite{Damgaard:2010cz,Akemann:2010zp,KSV12}. Here we will not consider these integrals.

Employing the joint probability density $p_5$, see Eq.~\eref{2.6}, we combine the Vandermonde determinant and the characteristic polynomials to a quotient of two  Vandermonde determinants. Then we rewrite the finite $n$ partition function~\eref{2.1.1} as
\begin{eqnarray}\label{2.1.3}
 \fl Z_{N_{\rm f}}^{(n,\nu,-)}(\lambda)&\propto&\int\limits_{\mathbb{R}^{2n+\nu}} d[x]\frac{\Delta_{2n+\nu+N_{\rm f}}(x,-\lambda)}{\Delta_{N_{\rm f}}(\lambda)}\\
 &\times&\Pf\left[\begin{array}{cc} \left\{g^{(-)}_2(x_i,x_j)\right\}_{1\leq i,j\leq 2n+\nu} & \displaystyle\{x_i^{j-1}g_1^{(-)}(x_i)\}\underset{1\leq j \leq \nu}{\underset{1\leq i\leq 2n+\nu}{}} \\ \displaystyle\{-x_j^{i-1}g_1^{(-)}(x_j)\}\underset{1\leq j \leq 2n+\nu}{\underset{1\leq i\leq \nu}{}} & 0 \end{array}\right].\nonumber
\end{eqnarray}
We want to consider a little bit more than the partition function namely the $k$-point correlation function. For this purpose we  only integrate over $2n+\nu-k$ variables, $\widetilde{x}=\diag(x_{k+1},\ldots,x_{2n+\nu})$. The remaining variables $x^\prime=\diag(x_1,\ldots,x_k)$ are the $k$ levels  we look at, i.e.
\begin{eqnarray}\label{2.1.3a}
 \fl R_{N_{\rm f}, k}^{(n,\nu,-)}(x^\prime,\lambda)&\propto&\int\limits_{\mathbb{R}^{2n+\nu-k}} \prod\limits_{j=k+1}^{2n+\nu}dx_j\frac{\Delta_{2n+\nu+N_{\rm f}}(x,-\lambda)}{\Delta_{N_{\rm f}}(\lambda)}\\
 &\times&\Pf\left[\begin{array}{cc} \left\{g^{(-)}_2(x_i,x_j)\right\}_{1\leq i,j\leq 2n+\nu} & \displaystyle\{x_i^{j-1}g_1^{(-)}(x_i)\}\underset{1\leq j \leq \nu}{\underset{1\leq i\leq 2n+\nu}{}} \\ \displaystyle\{-x_j^{i-1}g_1^{(-)}(x_j)\}\underset{1\leq j \leq 2n+\nu}{\underset{1\leq i\leq \nu}{}} & 0 \end{array}\right].\nonumber
\end{eqnarray}

The idea is the following. In the Vandermonde determinant of the numerator we can build an arbitrary basis of polynomials from order $0$ to order $2n+\nu+N_{\rm f}-1$,
\begin{eqnarray}
 \fl\Delta_{2n+\nu+N_{\rm f}}(x,-\lambda)&=&(-1)^{(2n+\nu+N_{\rm f})(2n+\nu+N_{\rm f}-1)/2} \nonumber\\
 &\times&\det\left[\begin{array}{cc} \left\{p_j^{(-)}(x_i)\right\}\underset{0\leq j \leq \nu-1}{\underset{1\leq i\leq 2n+\nu}{}} & \left\{q^{(-)}_{\nu+j}(x_i)\right\}\underset{0\leq j \leq 2n+N_{\rm f}-1}{\underset{1\leq i\leq 2n+\nu}{}}\\ \left\{p_j^{(-)}(-\lambda_i)\right\}\underset{0\leq j \leq \nu-1}{\underset{1\leq i\leq N_{\rm f}}{}} & \left\{q^{(-)}_{\nu+j}(-\lambda_i)\right\}\underset{0\leq j \leq 2n+N_{\rm f}-1}{\underset{1\leq i\leq N_{\rm f}}{}} \end{array}\right].\label{2.1.4}
\end{eqnarray}
Also the entries of the Pfaffian can be transformed by adding rows and columns with each other,
\begin{eqnarray}
 \fl&&\Pf\left[\begin{array}{cc} \left\{g^{(-)}_2(x_i,x_j)\right\}_{1\leq i,j\leq 2n+\nu} & \displaystyle\{x_i^{j-1}g_1^{(-)}(x_i)\}\underset{1\leq j \leq \nu}{\underset{1\leq i\leq 2n+\nu}{}} \\ \displaystyle\{-x_j^{i-1}g_1^{(-)}(x_j)\}\underset{1\leq j \leq 2n+\nu}{\underset{1\leq i\leq \nu}{}} & 0 \end{array}\right]\nonumber\\
 \fl&=&\Pf\left[\begin{array}{cc} \left\{G^{(-)}_2(x_i,x_j)\right\}_{1\leq i,j\leq 2n+\nu} & \displaystyle\{p_j^{(-)}(x_i)g_1^{(-)}(x_i)\}\underset{0\leq j \leq \nu-1}{\underset{1\leq i\leq 2n+\nu}{}} \\ \displaystyle\{-p_i^{(-)}(x_j)g_1^{(-)}(x_j)\}\underset{1\leq j \leq 2n+\nu}{\underset{0\leq i\leq \nu-1}{}} & 0 \end{array}\right],\label{2.1.5}
\end{eqnarray}
where we change the basis of the monomials to the polynomials $p_j^{(-)}$ and the two-point weight $g^{(-)}_2$ to $G^{(-)}_2$.

To shorten the notation we define the scalar product of two integrable functions $f_1$ and $f_2$ with the one-point weight $g_1^{(\pm)}$
\begin{eqnarray}\label{2.1.6}
 \langle f_1|f_2\rangle_{g_1^{(\pm)}}=\int\limits_{\mathbb{C}}d[z] f_1(z)f_2(z) g_1^{(\pm)}(x)\delta(y)
\end{eqnarray}
with $d[z]=dxdy$. The same can be done for the two-point weight $G^{(-)}_2$. We define the anti-symmetric product
\begin{eqnarray}\label{2.1.7}
 \fl\left(f_1|f_2\right)_{G^{(-)}_2}=\frac{1}{2}\int\limits_{\mathbb{C}^2}d[z_1]d[z_2]\det\left[\begin{array}{cc} f_1(z_1) & f_2(z_1) \\ f_1(z_2) & f_2(z_2) \end{array}\right] G^{(-)}_2(z_1,z_2)\delta(y_1)\delta(y_2).
\end{eqnarray}
Both definitions are extended to the complex plane by Dirac delta functions because we want to discuss the situation for both random matrices $D_5$ and $D_\W$ in a unifying way.

In the next step we employ the de Bruijn-like integration theorem derived in \ref{app1.1} which yields
\begin{eqnarray}\label{2.1.8}
 R_{N_{\rm f}, k}^{(n,\nu,-)}(x^\prime,\lambda)\propto\frac{1}{\Delta_{N_{\rm f}}(\lambda)}\Pf\left[\begin{array}{cc} M_-^{(11)} & M_-^{(12)} \\ -(M_-^{(12)})^T & M_-^{(22)} \end{array}\right]
\end{eqnarray}
for the $k$-point correlation function. The matrices in the Pfaffian determinant are
\begin{eqnarray}
 M_-^{(11)}&=&\left[\begin{array}{c|c|c} \left(p_i^{(-)}|p_j^{(-)}\right)_{G^{(-)}_2} & \left(p_i^{(-)}|q^{(-)}_{\nu+j}\right)_{G^{(-)}_2} & \langle p_i^{(-)}|p_j^{(-)}\rangle_{g_1^{(-)}}  \\ \hline \left(q^{(-)}_{\nu+i}|p_j^{(-)}\right)_{G^{(-)}_2} & \left(q^{(-)}_{\nu+i}|q^{(-)}_{\nu+j}\right)_{G^{(-)}_2} & \langle q^{(-)}_{\nu+i}|p_j^{(-)}\rangle_{g_1^{(-)}} \\ \hline -\langle p_i^{(-)}|p_j^{(-)}\rangle_{g_1^{(-)}} & -\langle p_i^{(-)}|q^{(-)}_{\nu+j}\rangle_{g_1^{(-)}} & 0\end{array}\right],\label{2.1.9a}\\
 M_-^{(12)}&=&\left[\begin{array}{c|c|c} \int\limits_{\mathbb{R}}d\widetilde{x}p_i^{(-)}(\widetilde{x})G^{(-)}_2(\widetilde{x},x_j) & p_i^{(-)}(x_j) & p_i^{(-)}(-\lambda_j)  \\ \hline  \int\limits_{\mathbb{R}}d\widetilde{x}q^{(-)}_{\nu+i}(\widetilde{x})G^{(-)}_2(\widetilde{x},x_j) & q^{(-)}_{\nu+i}(x_j) & q^{(-)}_{\nu+i}(-\lambda_j) \\ \hline -p_i^{(-)}(x_j)g_1^{(-)}(x_j) & 0 & 0\end{array}\right],\label{2.1.9b}\\
 M_-^{(22)}&=&\left[\begin{array}{c|c|c} G^{(-)}_2(x_i,x_j) & 0 &  0  \\ \hline 0 & 0 & 0 \\ \hline 0 & 0 & 0 \end{array}\right].\label{2.1.9c}
\end{eqnarray}
In the Pfaffian~\eref{2.1.8} the indices $i$ and $j$ of the rows and columns are $(0\ldots \nu-1,0\ldots2n+N_{\rm f}-1,0\ldots\nu-1,1\ldots k,1\ldots k,1\ldots N_{\rm f})$ from top to bottom and left to right. Please notice that regardless what the polynomials $p_l^{(-)}$ and $q_{\nu+l}^{(-)}$ and the modified two-point weight $G^{(-)}_2$ are Eq.~\eref{2.1.8} tells us that the joint probability density $p_5$ can also be written as a single Pfaffian. We have only to choose $k=2n+\nu$ to see that this statement is true. However the representation~\eref{2.1.8} is quite cumbersome. A more compact one is given in subsection~\ref{sec4.1}.

The aim is now to choose $q^{(-)}_{\nu+i}$, $p_i^{(-)}$ and $G^{(-)}_2$ such that the matrix $M_-^{(11)}$ becomes quasi-diagonal since we want to invert this matrix. A quasi-diagonal structure is equivalent to the conditions
\begin{eqnarray}
 \fl\langle p_i^{(-)}|p_j^{(-)}\rangle_{g_1^{(-)}}&=&h_j^{(-)}\delta_{ij},\quad{\rm for}\quad 0\leq i,j\leq \nu-1,\label{2.1.10}\\
 \fl\langle p_i^{(-)}|q^{(-)}_{\nu+j}\rangle_{g_1^{(-)}}&=&0,\quad{\rm for}\quad 0\leq i\leq \nu-1,0\leq j\leq2n+N_{\rm f}-1,\label{2.1.11}\\
 \fl\left(p_i^{(-)}|p_j^{(-)}\right)_{G^{(-)}_2}&=&0,\quad{\rm for}\quad 0\leq i,j\leq \nu-1,\label{2.1.12}\\
 \fl\left(p_i^{(-)}|q^{(-)}_{\nu+j}\right)_{G^{(-)}_2}&=&0,\quad{\rm for}\quad 0\leq i\leq \nu-1,0\leq j\leq2n+N_{\rm f}-1,\label{2.1.13}\\
 \fl\left(q^{(-)}_{\nu+2i+1}|q^{(-)}_{\nu+2j+1}\right)_{G^{(-)}_2}&=&0,\quad{\rm for}\quad0\leq i,j\leq2n+N_{\rm f}-1,\label{2.1.14}\\
 \fl\left(q^{(-)}_{\nu+2i}|q^{(-)}_{\nu+2j}\right)_{G^{(-)}_2}&=&0,\quad{\rm for}\quad0\leq i,j\leq2n+N_{\rm f}-1,\label{2.1.15}\\
 \fl\left(q^{(-)}_{\nu+2i}|q^{(-)}_{\nu+2j+1}\right)_{G^{(-)}_2}&=&o^{(-)}_j\delta_{ij},\quad{\rm for}\quad0\leq i,j\leq2n+N_{\rm f}-1.\label{2.1.16}
\end{eqnarray}
The constants $h_j^{(-)}$ and $o^{(-)}_j$ are the normalization constants of the polynomials. In Sec.~\ref{sec3} we will see that this system of equations have indeed a solution. We will give an explicit construction of them.

Please note that the solution of the odd skew-orthogonal polynomials, $q^{(-)}_{\nu+2j+1}$, exhibits an ambiguity. The polynomials $q^{(-)}_{\nu+2j+1}+c_j q^{(-)}_{\nu+2j}$ are also a solution of the Eqs.~(\ref{2.1.10}-\ref{2.1.16}) with arbitrary constants $c_j\in\mathbb{C}$ as it was already found in Ref.~\cite{Eynard2001} for pure skew-orthogonal polynomials.

\subsection{The $(k_\rt,k_\lt)$-point correlation function of $D_\W$}\label{sec2.2}

The next case we want to consider is the fermionic partition function of $D_\W$ with $N_{\rm f}$ quark masses, $m=\diag(m_1,\ldots,m_{N_{\rm f}})$,
\begin{eqnarray}\label{2.2.1}
 \fl Z_{N_{\rm f}}^{(n,\nu,+)}(m)&\propto&\int d[D_\W]P(D_\W)\prod\limits_{j=1}^{N_{\rm f}}\det(D_\W+m_j\eins_{2n+\nu}).
\end{eqnarray}
In the microscopic limit it corresponds to \cite{Damgaard:2010cz,Akemann:2010zp}
\begin{eqnarray}\label{2.2.2}
 \fl &&Z_{N_{\rm f}}^{(n,\nu,+)}\left(\frac{m}{2n}\right)\overset{n\gg1}{\propto}\int\limits_{\U(N_{\rm f})} d\mu(U){\det}^{\nu}U\,\\
 \fl&\times&\exp\left[\frac{1}{2}\tr(\widehat{m}_6\eins_{N_{\rm f}}+\widehat{m})(U+U^{-1})+\frac{\widehat{\lambda}_7}{2}\tr(U-U^{-1})-\widehat{a}^2\tr(U^2+U^{-2})\right].\nonumber
\end{eqnarray}
This is the effective Lagrangian of the Wilson-Dirac operator of the partition function with $N_{\rm f}$ fermionic quarks with non-degenerate quark masses $\widehat{m}_6\eins_{N_{\rm f}}+\widehat{m}$ and one source term $\widehat{\lambda}_7$ proportional to $\gamma_5$ \cite{Sharpe:1998xm,RS02,Bar:2003mh,Sharpe06}. Again one can integrate over the two variables $\widehat{m}_6$ and $\widehat{\lambda}_7$ weighted by Gaussians to obtain the two low energy constants $W_6$ and $W_7$ but we will consider $Z_{N_{\rm f}}^{(n,\nu,+)}$ without these integrals.

The partition function with the joint probability density $p_\W$ reads
\begin{eqnarray}\label{2.2.3}
 \fl Z_{N_{\rm f}}^{(n,\nu,+)}(m)&\propto&\int\limits_{\mathbb{C}^{2n+\nu}} d[Z]\frac{\Delta_{2n+\nu+N_{\rm f}}(Z,-m)}{\Delta_{N_{\rm f}}(m)}\\
 &\times&\det\left[\begin{array}{c} \displaystyle\{g^{(+)}_2(z_i^{(\rt)},z_j^{(\lt)})\}\underset{1\leq j \leq n+\nu}{\underset{1\leq i\leq n}{}} \\ \displaystyle\{(x_j^{(\lt)})^{i-1}g_1^{(+)}(x_j^{(\lt)})\delta(y_j^{(\lt)})\}\underset{1\leq j \leq n+\nu}{\underset{1\leq i\leq \nu}{}} \end{array}\right].\nonumber
\end{eqnarray}
Since the permutation symmetry in the eigenvalues of $D_\W$ is broken we have to consider a two parameter set of eigenvalue correlation functions. The number of eigenvalues $z^{(\rt)}$ is independent of the number for the eigenvalues $z^{(\lt)}$. Let $Z^\prime=\diag(z_1^{(\rt)},\ldots,z_{k_\rt}^{(\rt)},z_1^{(\lt)},\ldots,z_{k_\lt}^{(\lt)})$. Hence we define the $(k_\rt,k_\lt)$-point correlation function,
\begin{eqnarray}\label{2.2.3a}
 \fl R_{N_{\rm f}, k_\rt,k_\lt}^{(n,\nu,+)}(Z^\prime,m)&\propto&\int\limits_{\mathbb{C}^{2n+\nu-k_\rt-k_\lt}} \prod\limits_{j=k_\rt+1}^{n}d[z_j^{(\rt)}]\prod\limits_{j=k_\lt+1}^{n+\nu}d[z_j^{(\lt)}]\frac{\Delta_{2n+\nu+N_{\rm f}}(Z,-m)}{\Delta_{N_{\rm f}}(m)}\\
 &\times&\det\left[\begin{array}{c} \displaystyle\{g^{(+)}_2(z_i^{(\rt)},z_j^{(\lt)})\}\underset{1\leq j \leq n+\nu}{\underset{1\leq i\leq n}{}} \\ \displaystyle\{(x_j^{(\lt)})^{i-1}g_1^{(+)}(x_j^{(\lt)})\delta(y_j^{(\lt)})\}\underset{1\leq j \leq n+\nu}{\underset{1\leq i\leq \nu}{}} \end{array}\right].\nonumber
\end{eqnarray}

As in subsection~\ref{sec2.1} we construct an arbitrary basis of polynomials in the Vandermonde determinant to inflict some conditions on them later on,
\begin{eqnarray}
 \fl\Delta_{2n+\nu+N_{\rm f}}(Z,-m)&=&(-1)^{(2n+\nu+N_{\rm f})(2n+\nu+N_{\rm f}-1)/2}\nonumber\\
 &\times&\det\left[\begin{array}{cc} \left\{p_j^{(+)}(z_i^{(\rt)})\right\}\underset{0\leq j \leq \nu-1}{\underset{1\leq i\leq n}{}} & \left\{q^{(+)}_{\nu+j}(z_i^{(\rt)})\right\}\underset{0\leq j \leq 2n+N_{\rm f}-1}{\underset{1\leq i\leq n}{}} \\ \left\{p_j^{(+)}(z_i^{(\lt)})\right\}\underset{0\leq j \leq \nu-1}{\underset{1\leq i\leq n+\nu}{}} & \left\{q^{(+)}_{\nu+j}(z_i^{(\lt)})\right\}\underset{0\leq j \leq 2n+N_{\rm f}-1}{\underset{1\leq i\leq n+\nu}{}} \\ \left\{p_j^{(+)}(-m_i)\right\}\underset{0\leq j \leq \nu-1}{\underset{1\leq i\leq N_{\rm f}}{}} & \left\{q^{(+)}_{\nu+j}(-m_i)\right\}\underset{0\leq j \leq 2n+N_{\rm f}-1}{\underset{1\leq i\leq N_{\rm f}}{}} \end{array}\right].\label{2.2.4}
\end{eqnarray}
Also the other determinant in the numerator can be transformed,
\begin{eqnarray}
 \fl&&\det\left[\begin{array}{c} \displaystyle\{g^{(+)}_2(z_i^{(\rt)},z_j^{(\lt)})\}\underset{1\leq j \leq n+\nu}{\underset{1\leq i\leq n}{}} \\ \hspace*{-0.1cm}\displaystyle\{(x_j^{(\lt)})^{i-1}g_1^{(+)}(x_j^{(\lt)})\delta(y_j^{(\lt)})\}\underset{1\leq j \leq n+\nu}{\underset{1\leq i\leq \nu}{}}\hspace*{-0.1cm} \end{array}\right]=\det\left[\begin{array}{c} \displaystyle\{G^{(+)}_2(z_i^{(\rt)},z_j^{(\lt)})\}\underset{1\leq j \leq n+\nu}{\underset{1\leq i\leq n}{}} \\ \hspace*{-0.1cm}\displaystyle\{p_i^{(+)}(x_j^{(\lt)})g_1^{(+)}(x_j^{(\lt)})\delta(y_j^{(\lt)})\}\underset{1\leq j \leq n+\nu}{\underset{0\leq i\leq \nu-1}{}}\hspace*{-0.1cm} \end{array}\right].\nonumber\\
 \fl&&\label{2.2.5}
\end{eqnarray}
The whole procedure works analogous to the one for $D_5$, cf. subsection~\ref{sec2.1}.

Let
\begin{eqnarray}
 \left(f_1|f_2\right)_{G^{(+)}_2}&=&\frac{1}{2}\int\limits_{\mathbb{C}^2}d[z^{(\rt)}]d[z^{(\lt)}]\det\left[\begin{array}{cc} f_1(z^{(\rt)}) & f_2(z^{(\rt)}) \\ f_1(z^{(\lt)}) & f_2(z^{(\lt)}) \end{array}\right]\nonumber\\
 &\times&\left(G^{(+)}_2(z^{(\rt)},z^{(\lt)})-G^{(+)}_2(z^{(\lt)},z^{(\rt)}) \right)\label{2.2.6}
\end{eqnarray}
be the anti-symmetric scalar product of two integrable functions $f_1$ and $f_2$ with respect to the two-point weight $G^{(+)}_2$. Notice that $G^{(+)}_2$ as well as $g^{(+)}_2$ are not anti-symmetric under a permutation of their entries whereas the two-point weight $G^{(-)}_2$ is anti-symmetric. The reason for this is again the breaking of the permutation symmetry in the eigenvalues of $D_\W$.

Considering the $(k_\rt,k_\lt)$-point correlation function we apply the de Bruijn-like integration theorem derived in \ref{app1.2} to the partition function~\eref{2.2.3} and find
\begin{eqnarray}\label{2.2.7}
 R_{N_{\rm f}, k_\rt,k_\lt}^{(n,\nu,+)}(Z^\prime,m)&\propto&\frac{1}{\Delta_{N_{\rm f}}(m)}\Pf\left[\begin{array}{cc} M_+^{(11)} & M_+^{(12)} \\ -(M_+^{(12)})^T & M_+^{(22)} \end{array}\right],
\end{eqnarray}
where the three matrices are
\begin{eqnarray}
 \fl M_+^{(11)}&=&\left[\begin{array}{c|c|c} \left(p_i^{(+)}|p_j^{(+)}\right)_{G^{(+)}_2} & \left(p_i^{(+)}|q^{(+)}_{\nu+j}\right)_{G^{(+)}_2} & \langle p_i^{(+)}|p_j^{(+)}\rangle_{g_1^{(+)}}  \\ \hline \left(q^{(+)}_{\nu+i}|p_j^{(+)}\right)_{G^{(+)}_2} & \left(q^{(+)}_{\nu+i}|q^{(+)}_{\nu+j}\right)_{G^{(+)}_2} & \langle q^{(+)}_{\nu+i}|p_j^{(+)}\rangle_{g_1^{(+)}} \\ \hline -\langle p_i^{(+)}|p_j^{(+)}\rangle_{g_1^{(+)}} & -\langle p_i^{(+)}|q^{(+)}_{\nu+j}\rangle_{g_1^{(+)}} & 0\end{array}\right],\label{2.2.7a}\\
 \fl (M_+^{(12)})^T\hspace*{-0.2cm}&=&\hspace*{-0.3cm}\left[\begin{array}{c|c|c} \hspace*{-0.3cm}\int\limits_{\mathbb{C}}d[\widetilde{z}]p_i^{(+)}(\widetilde{z})G^{(+)}_2(z_j^{(\rt)},\widetilde{z})\hspace*{-0.2cm} & \hspace*{-0.1cm}\int\limits_{\mathbb{C}}d[\widetilde{z}]q^{(+)}_{\nu+i}(\widetilde{z})G^{(+)}_2(z_j^{(\rt)},\widetilde{z})\hspace*{-0.2cm} & 0 \\ \hline \hspace*{-0.2cm}\int\limits_{\mathbb{C}}d[\widetilde{z}]p_i^{(+)}(\widetilde{z})G^{(+)}_2(\widetilde{z},z_j^{(\lt)})\hspace*{-0.2cm} & \hspace*{-0.2cm}\int\limits_{\mathbb{C}}d[\widetilde{z}]q^{(+)}_{\nu+i}(\widetilde{z})G^{(+)}_2(\widetilde{z},z_j^{(\lt)})\hspace*{-0.2cm} & \hspace*{-0.2cm}-p_i^{(+)}(x_j^{(\lt)})g_1^{(+)}(x_j^{(\lt)})\delta(y_j^{(\lt)})\hspace*{-0.2cm} \\ \hline p_i^{(+)}(z_j^{(\rt)}) & q^{(+)}_{\nu+i}(z_j^{(\rt)}) & 0 \\ \hline  p_i^{(+)}(z_j^{(\lt)}) & q^{(+)}_{\nu+i}(z_j^{(\lt)}) & 0 \\ \hline p_i^{(+)}(-\lambda_j) & q^{(+)}_{\nu+i}(-\lambda_j) & 0 \end{array}\right]\hspace*{-0.2cm},\nonumber\\
 \fl&&\label{2.2.7b}\\
 \fl M_+^{(22)}&=&\left[\begin{array}{c|c|c|c|c} 0 & -G^{(+)}_2(z_i^{(\rt)},z_j^{(\lt)}) & 0 & 0 & 0 \\ \hline G^{(+)}_2(z_j^{(\rt)},z_i^{(\lt)}) & 0 & 0 & 0 & 0 \\ \hline 0 & 0 & 0 & 0 & 0 \\ \hline 0 & 0 & 0 & 0 & 0 \\ \hline 0 & 0 & 0 & 0 & 0 \end{array}\right].\label{2.2.7c}
\end{eqnarray}
In the Pfaffian~\eref{2.2.7} the indices $i$ and $j$ are in the range $(0\ldots\nu-1,0\ldots2n+N_{\rm f}-1,0\ldots\nu-1,1\ldots k_\rt,1\ldots k_\lt,1\ldots k_\rt,1\ldots k_\lt,1\ldots N_{\rm f})$ from top to bottom and from left to right.

Please notice the similarity of Eq.~\eref{2.2.7}  with Eq.~\eref{2.1.8}. If $k_\rt=n$ and $k_\lt=n+\nu$ the correlation function is equal to the joint probability density $p_\W$. Hence $p_\W$ can also be written as a single Pfaffian which can be cast into a more compact form, see subsection~\ref{sec4.2}. 

As in subsection~\ref{sec2.1} we want to invert and, thus, quasi-diagonalize the matrix $M_+^{(11)}$.
This yields the following system of equations
\begin{eqnarray}
 \fl\langle p_i^{(+)}|p_j^{(+)}\rangle_{g_1^{(+)}}&=&h_i^{(+)}\delta_{ij},\quad{\rm for}\quad 0\leq i,j\leq \nu-1,\label{2.2.9}\\
 \fl\langle p_i^{(+)}|q^{(+)}_{\nu+j}\rangle_{g_1^{(+)}}&=&0,\quad{\rm for}\quad 0\leq i\leq \nu-1,0\leq j\leq2n+N_{\rm f}-1,\label{2.2.10}\\
 \fl\left(p_i^{(+)}|p_j^{(+)}\right)_{G^{(+)}_2}&=&0,\quad{\rm for}\quad 0\leq i,j\leq \nu-1,\label{2.2.11}\\
 \fl\left(p_i^{(+)}|q^{(+)}_{\nu+j}\right)_{G^{(+)}_2}&=&0,\quad{\rm for}\quad 0\leq i\leq \nu-1,0\leq j\leq2n+N_{\rm f}-1,\label{2.2.12}\\
 \fl\left(q^{(+)}_{\nu+2i+1}|q^{(+)}_{\nu+2j+1}\right)_{G^{(+)}_2}&=&0,\quad{\rm for}\quad0\leq i,j\leq2n+N_{\rm f}-1,\label{2.2.13}\\
 \fl\left(q^{(+)}_{\nu+2i}|q^{(+)}_{\nu+2j}\right)_{G^{(+)}_2}&=&0,\quad{\rm for}\quad0\leq i,j\leq2n+N_{\rm f}-1,\label{2.2.14}\\
 \fl\left(q^{(+)}_{\nu+2i}|q^{(+)}_{\nu+2j+1}\right)_{G^{(+)}_2}&=&o^{(+)}_i\delta_{ij},\quad{\rm for}\quad0\leq i,j\leq2n+N_{\rm f}-1\label{2.2.15}
\end{eqnarray}
with the normalization constants $h_j^{(+)}$ and $o^{(+)}_j$. Comparing this system of equations with the one of $D_5$ we recognize that they are of the same form. Hence, if we solve them in a general setting we solve them for both random matrices, $D_5$ and $D_\W$.

As for $D_5$ the odd skew-orthogonal polynomials, $q^{(-)}_{\nu+2j+1}$, can be added by the polynomials $c_j q^{(-)}_{\nu+2j}$ with arbitrary constants $c_j\in\mathbb{C}$. They solve the same set of equations.

\section{Construction of the polynomials and some of their properties}\label{sec3}

In subsection~\ref{sec3.1} we construct the orthogonal polynomials as well as the skew-orthogonal ones starting from the conditions~(\ref{2.1.10}-\ref{2.1.16}) and (\ref{2.2.9}-\ref{2.2.15}). Furthermore we give explicit expressions of the modified two-point weights, $G^{(\pm)}_2$, and specify the constants $h_l$ and $o_{\nu+2l}^{(\pm)}$. Recursion relations of the polynomials are shown in subsection~\ref{sec3.2}. With aid of these relations we derive the Christoffel Darboux-like formula, in subsection~\ref{sec3.3}. In subsection~\ref{sec3.4}, we rewrite the polynomials as well as the Christoffel Darboux-like formula to random matrix averages and take the microscopic limit of them in subsection~\ref{sec3.5}.

\subsection{The polynomials}\label{sec3.1}

The starting point of the construction are the monomials
\begin{eqnarray}\label{3.1.1}
 \mathfrak{m}_j(z)=z^j\quad{\rm with}\quad j\in\mathbb{N}_0.
\end{eqnarray}
With help of the general formula for the orthogonal polynomials of the one-point weight $g_1^{(\pm)}$~\cite{Meh04} as a quotient of determinants we find
\begin{eqnarray}\label{3.1.2}
 p_l^{(\pm)}(z)={\det}^{-1}\left[\langle \mathfrak{m}_i|\mathfrak{m}_j\rangle_{g_1^{(\pm)}}\right]_{1\leq i,j\leq l-1}\det\left[\begin{array}{c} \left\{\langle \mathfrak{m}_i|\mathfrak{m}_j\rangle_{g_1^{(\pm)}}\right\}\underset{0\leq j\leq l}{\underset{0\leq i\leq l-1}{}} \\ \left\{\mathfrak{m}_j(z)\right\}\underset{0\leq j\leq l}{} \end{array}\right]
\end{eqnarray}
in monic normalization, i.e. $p_l^{(\pm)}(z)=z^l+\ldots$ Since $g_1$ is a shifted Gaussian the polynomials $p_l^{(\pm)}$ are shifted Hermite polynomials, $H_l$, in monic normalization,
\begin{eqnarray}\label{3.1.3}
 p_l^{(\pm)}(z)=\left(\frac{a^2}{n}\right)^{l/2}H_l\left(\sqrt{\frac{n}{a^2}}z\mp\sqrt{\frac{a^2}{n}}\mu_l\right).
\end{eqnarray}
This agrees with Refs.~\cite{Damgaard:2010cz,Akemann:2010zp,Splittorff:2011bj,Akemann:2011kj} where a mixing of the eigenvalue statistics with a finite dimensional GUE was found. The normalization constant is
\begin{eqnarray}\label{3.1.3a}
 h_l^{(\pm)}=h_l=\sqrt{2\pi}\left(\frac{a^2}{n}\right)^{l+1/2}l!\exp\left[\frac{a^2\mu_\lt^2}{2n}\right].
\end{eqnarray}
Thus the normalization constants of the orthogonal polynomials are the same for $D_\W$ and for $D_5$.

Starting from the orthogonal polynomials $p_l^{(\pm)}$ we want to construct the polynomials $q^{(-)}_{\nu+l}$ and $q^{(+)}_{\nu+l}$ fulfilling  the orthogonality conditions~\eref{2.1.11} and \eref{2.2.10}, respectively. Let $N_{\rm f}$ be even for simplicity. If it is odd the anti-symmetric matrices $M_{\pm}^{(11)}$ are never invertible because their dimensions are odd. In such a case we extend the partition function by one  fermionic flavor and remove it at the end of the day by sending its mass to infinity.

As for the orthogonal polynomials we begin with an intuitive definition,
\begin{eqnarray}
 \fl &&q^{(\pm)}_{\nu+2l}(z)=\Pf^{-1}\left[\left(p_i^{(\pm)}|p_j^{(\pm)}\right)_{g^{(\pm)}_2}\right]_{\nu\leq i,j\leq\nu+2l-1}\label{3.1.4}\\
 \fl&\times&\Pf\left[\begin{array}{cc} \left\{\left(p_i^{(\pm)}|p_j^{(\pm)}\right)_{g^{(\pm)}_2}\right\}\underset{\nu\leq i,j\leq \nu+2l}{} & \left\{p_i^{(\pm)}(z)\right\}\underset{\nu\leq i\leq \nu+2l}{} \\ \left\{-p_j^{(\pm)}(z)\right\}\underset{\nu\leq j\leq \nu+2l}{} & 0 \end{array}\right],\nonumber\\
 \fl &&q^{(\pm)}_{\nu+2l+1}(z)=\Pf^{-1}\left[\left(p_i^{(\pm)}|p_j^{(\pm)}\right)_{g^{(\pm)}_2}\right]_{\nu\leq i,j\leq\nu+2l-1}\label{3.1.5}\\
 \fl&\times&\hspace*{-0.1cm}\Pf\hspace*{-0.2cm}\left[\begin{array}{ccc} \hspace*{-0.6cm}\left\{\left(p_i^{(\pm)}|p_j^{(\pm)}\right)_{g^{(\pm)}_2}\right\}\underset{\nu\leq i,j\leq \nu+2l-1}{} & \hspace*{-0.75cm}\left\{\left(p_i^{(\pm)}|p_{\nu+2l+1}^{(\pm)}\right)_{g^{(\pm)}_2}\right\}\underset{\nu\leq i\leq \nu+2l-1}{} & \hspace*{-0.4cm}\left\{p_i^{(\pm)}(z)\right\}\underset{\nu\leq i\leq \nu+2l-1}{}\hspace*{-0.2cm} \\ \hspace*{-0.3cm}\left\{\left(p_{\nu+2l+1}^{(\pm)}|p_j^{(\pm)}\right)_{g^{(\pm)}_2}\right\}\underset{\nu\leq j\leq \nu+2l-1}{} & 0 & p_{\nu+2l+1}^{(\pm)}(z) \\ \left\{-p_j^{(\pm)}(z)\right\}\underset{\nu\leq j\leq \nu+2l-1}{} & -p_{\nu+2l+1}^{(\pm)}(z) & 0 \end{array}\right]\hspace*{-0.15cm},\nonumber
\end{eqnarray}
which is similar to the ansatz of the skew-orthogonal polynomials in $\chi$RMT with non-zero chemical potential and Dyson index $\beta=1,4$, see Ref.~\cite{Akemann:2010mt}. The anti-symmetric products of $g^{(-)}_2$ and $g^{(+)}_2$ are defined similar to Eqs.~\eref{2.1.7} and \eref{2.2.6}, respectively. One can readily prove that the orthogonality conditions~\eref{2.1.11} and \eref{2.2.10} are fulfilled. The multi-linearity of the Pfaffian allows us to pull the scalar product into the Pfaffian. Then one row and one column is zero. For example, let $l\in\{0,1,\ldots,\nu-1\}$. Then the orthogonality of the polynomials $p_l^{(\pm)}$ yields
\begin{eqnarray}
 \fl\langle p_l^{(\pm)}|q^{(\pm)}_{\nu+2b}\rangle_{g_1^{(\pm)}}&=&\frac{\Pf\left[\begin{array}{cc} \left\{\left(p_i^{(\pm)}|p_j^{(\pm)}\right)_{g^{(\pm)}_2}\right\}\underset{\nu\leq i,j\leq \nu+2b}{} & \left\{\langle p_l^{(\pm)}|p_i^{(\pm)}\rangle_{g_1^{(\pm)}}\right\}\underset{\nu\leq i\leq \nu+2b}{} \\ \left\{-\langle p_l^{(\pm)}|p_j^{(\pm)}\rangle_{g_1^{(\pm)}}\right\}\underset{\nu\leq j\leq \nu+2b}{} & 0 \end{array}\right]}{\Pf\left[\left(p_i^{(\pm)}|p_j^{(\pm)}\right)_{g^{(\pm)}_2}\right]_{\nu\leq i,j\leq\nu+2b-1}}\nonumber\\
 \fl&=&\frac{\Pf\left[\begin{array}{cc} \left\{\left(p_i^{(\pm)}|p_j^{(\pm)}\right)_{g^{(\pm)}_2}\right\}\underset{\nu\leq i,j\leq \nu+2b}{} & \left\{0\right\}\underset{\nu\leq i\leq \nu+2b}{} \\ \left\{0\right\}\underset{\nu\leq j\leq \nu+2b}{} & 0 \end{array}\right]}{\Pf\left[\left(p_i^{(\pm)}|p_j^{(\pm)}\right)_{g^{(\pm)}_2}\right]_{\nu\leq i,j\leq\nu+2b-1}}\nonumber\\
 \fl&=&0.\label{3.1.8}
\end{eqnarray}
In a similar way one can prove the other relations.

We underline that the odd skew-orthogonal polynomials~\eref{3.1.5} can be gauged by the even ones~\eref{3.1.4}, i.e. $q^{(\pm)}_{\nu+2l+1}(z)\rightarrow q^{(\pm)}_{\nu+2l+1}(z)+ c_l q^{(\pm)}_{\nu+2l}(z)$ with $c_l\in\mathbb{C}$ arbitrary. This gauge symmetry is similar to the one found for pure skew-orthogonal polynomials \cite{Eynard2001}. The fundamental reason is the anti-symmetry of the two-point weight $g_2^{(\pm)}$ which has always a non-trivial kernel.

The normalization constants in Eqs.~\eref{3.1.4} and \eref{3.1.5} are finite since they are proportional to the constants in Eqs.~\eref{2.9d} and \eref{2.14}, i.e.
\begin{eqnarray}\label{3.1.9}
 \Pf\left[\left(p_i^{(\pm)}|p_j^{(\pm)}\right)_{g^{(\pm)}_2}\right]_{\nu\leq i,j\leq\nu+2l-1}\propto \frac{1}{c_{\pm}}.
\end{eqnarray}
It can be easily shown that the polynomials $q^{(\pm)}_{\nu+l}$ are in monic normalization, too. The constants~\eref{3.1.9} are related to the ones in Eqs.~\eref{2.1.16} and \eref{2.2.15} by
\begin{eqnarray}
 o^{(\pm)}_l&=&\frac{\Pf\left[\left(p_i^{(\pm)}|p_j^{(\pm)}\right)_{g^{(\pm)}_2}\right]_{\nu\leq i,j\leq\nu+2l+1}}{\Pf\left[\left(p_i^{(\pm)}|p_j^{(\pm)}\right)_{g^{(\pm)}_2}\right]_{\nu\leq i,j\leq\nu+2l-1}}.\label{3.1.9a}
\end{eqnarray}
Combining this identity with relation~\eref{3.1.9}  the constants $o^{(\pm)}_l$ are mostly the quotient of two normalization constants of the joint probability density functions. Hence, the constants $o^{(\pm)}_l$ can be directly calculated by the two identities
\begin{eqnarray}
 &&(-1)^{n}(2n+\nu)!\prod\limits_{j=0}^{\nu-1}h_j\prod\limits_{j=0}^{n-1}o_j^{(-)}\label{3.1.9b}\\
 &=&\frac{1}{c_-}(1-a^2)^{n(n+\nu-1/2)}a^{n+\nu^2}\exp\left[\frac{a^2}{2}\left(\mu_\rt^2+\frac{n+\nu}{n}\mu_\lt^2\right)-\frac{n\widehat{m}_{6-}^2}{8\widehat{a}_-^2}\right]\nonumber
\end{eqnarray}
and
\begin{eqnarray}
 &&(-1)^{\nu(\nu-1)/2+n(n+1)/2}n!(n+\nu)!\prod\limits_{j=0}^{\nu-1}h_j\prod\limits_{j=0}^{n-1}o_j^{(+)}\label{3.1.9c}\\
 &=&\frac{1}{c_+}(1+a^2)^{n(n+\nu-1/2)}a^{n+\nu^2}\exp\left[\frac{a^2}{2}\left(\mu_\rt^2+\frac{n+\nu}{n}\mu_\lt^2\right)-\frac{n\widehat{\lambda}_{7+}^2}{8\widehat{a}_+^2}\right].\nonumber
\end{eqnarray}
These identities can be derived with aid of the de Bruijn-like integration theorems in \ref{app1}. With help of  Eqs.~\eref{3.1.9b} and \eref{3.1.9c} we conclude
\begin{eqnarray}\label{3.1.9d}
 \fl o_l^{(\pm)}=-4\,l!(l+\nu)!\sqrt{\frac{\pi}{n(1\pm a^2)}}\left(\frac{1\pm a^2}{n}\right)^{2l+\nu+1}\hspace*{-0.6cm}a\,\exp\left[\frac{a^2}{4n}(\mu_\rt\pm\mu_\lt)^2\pm\frac{a^4}{4n(1\pm a^2)}(\mu_\rt\mp\mu_\lt)^2\right].
\end{eqnarray}
Hence the normalization constant is linear in $a$ for small lattice spacing and is proportional to $n^{-2l-\nu-2}l!(l+\nu)!$ in the microscopic limit.

The polynomials $q^{(\pm)}_{\nu+2b}$ and $q^{(\pm)}_{\nu+2b+1}$ are also orthogonal to $p_{\nu+l}^{(\pm)}$ , $ l\in\{0,\dots,2b\}$ and $ l\in\{0,\dots,2b-1,2b+1\}$, respectively, corresponding to the two-point weight $g^{(\pm)}_2$ since the rows and columns are not linearly independent anymore. For example
\begin{eqnarray}
 \fl\left( p_{\nu+l}^{(\pm)}|q^{(\pm)}_{\nu+2b}\right)_{g^{(\pm)}_2}&=&\frac{\Pf\left[\begin{array}{cc} \left\{\left(p_{\nu+i}^{(\pm)}|p_{\nu+j}^{(\pm)}\right)_{g^{(\pm)}_2}\right\}\underset{0\leq i,j\leq 2b}{} & \left\{\left( p_{\nu+l}^{(\pm)}|p_{\nu+i}^{(\pm)}\right)_{g^{(\pm)}_2}\right\}\underset{0\leq i\leq 2b}{} \\ \left\{-\left( p_{\nu+l}^{(\pm)}|p_{\nu+j}^{(\pm)}\right)_{g^{(\pm)}_2}\right\}\underset{0\leq j\leq 2b}{} & 0 \end{array}\right]}{\Pf\left[\left(p_{\nu+i}^{(\pm)}|p_{\nu+j}^{(\pm)}\right)_{g^{(\pm)}_2}\right]_{0\leq i,j\leq2b-1}}\nonumber\\
 \fl&=&0,\label{3.1.10}
\end{eqnarray}
the $l$th row and column and the last ones are the same. In the same way one can prove the skew-orthogonality of $p_{\nu+l}^{(\pm)}$ , $ l\in\{0,\dots,2b-1,2b+1\}$, with $q^{(\pm)}_{\nu+2b+1}$. Due to the definitions~\eref{3.1.4} and \eref{3.1.5} the polynomials $q^{(\pm)}_{\nu+l}$ are a linear combination of $p_{\nu+l}^{(\pm)}$ with $0\leq l\leq2n+N_{\rm f}-1$. Therefore the polynomials are indeed skew-orthogonal with respect to $g^{(\pm)}_2$. In particular they fulfill the conditions similar to the relations~(\ref{2.1.14}-\ref{2.1.16}) and (\ref{2.2.13}-\ref{2.2.15}) by exchanging $G^{(\pm)}_2\to g^{(\pm)}_2$. However the remaining conditions \eref{2.1.12}, \eref{2.1.13}, \eref{2.2.11} and \eref{2.2.12} are not fulfilled. This is the reason for modifying the two-point weights.

The simplest way to enforce the remaining conditions is the projection of the measures $g^{(\pm)}_2$ onto the polynomials $q^{(\pm)}_{\nu+l}$ only. This means the polynomials $p_l^{(\pm)}$, $0\leq l\leq\nu-1$, have to be in the kernel of $G^{(\pm)}_2$. We make the ansatz
\begin{eqnarray}
 \fl(f_1,f_2)_{G^{(\pm)}_2}&=&(f_1,f_2)_{g^{(\pm)}_2}-\sum\limits_{j=0}^{\nu-1}\frac{1}{h_j}\left(\langle f_1|p_j^{(\pm)}\rangle_{g_1^{(\pm)}}(p_j^{(\pm)},f_2)_{g^{(\pm)}_2}+(f_1,p_j^{(\pm)})_{g^{(\pm)}_2}\langle p_j^{(\pm)}|f_2\rangle_{g_1^{(\pm)}}\right)\nonumber\\
 \fl&+&\sum\limits_{i,j=0}^{\nu-1}\frac{1}{h_jh_i}\langle f_1|p_i^{(\pm)}\rangle_{g_1^{(\pm)}}(p_i^{(\pm)},p_j^{(\pm)})_{g^{(\pm)}_2}\langle p_j^{(\pm)}|f_2\rangle_{g_1^{(\pm)}}\label{3.1.11}.
\end{eqnarray}
Indeed we have
\begin{eqnarray}
 \fl\left(p_l^{(\pm)},f\right)_{G^{(-)}_2}&=&(p_l^{(\pm)},f)_{G^{(+)}_2}=0,\quad{\rm for\  all\ functions}\ f\ {\rm and}\ 0\leq l\leq\nu-1,\label{3.1.13}\\
 \fl\left(q^{(\pm)}_i,q^{(\pm)}_j\right)_{G^{(\pm)}_2}&=&\left(q^{(\pm)}_i,q^{(\pm)}_j\right)_{g^{(\pm)}_2}.\label{3.1.14}
\end{eqnarray}
Thus all orthogonality conditions are fulfilled.

The explicit expressions of $G^{(-)}_2$ and $G^{(+)}_2$ are
\begin{eqnarray}
 \fl G^{(-)}_2(x_1,x_2)&=&g^{(-)}_2(x_1,x_2)\label{3.1.16}\\
 \fl&-&\sum\limits_{j=0}^{\nu-1}\frac{1}{h_j}\left(\int\limits_{\mathbb{R}}dx^\prime p_j^{(-)}(x^\prime)g^{(-)}_2(x^\prime,x_2) p_j^{(-)}(x_1)g_1^{(-)}(x_1)\right.\nonumber\\
 \fl&+&\left.\int\limits_{\mathbb{R}}dx^\prime p_j^{(-)}(x^\prime)g^{(-)}_2(x_1,x^\prime) p_j^{(-)}(x_2)g_1^{(-)}(x_2)\right)+\sum\limits_{i,j=0}^{\nu-1}\nonumber\\
 \fl&\times&\frac{1}{h_jh_i}\int\limits_{\mathbb{R}^2}d[x^\prime]p_i^{(-)}(x_1^\prime)p_j^{(-)}(x_2^\prime)g^{(-)}_2(x_1^\prime,x_2^\prime)p_i^{(-)}(x_1)p_j^{(-)}(x_2)g_1^{(-)}(x_1)g_1^{(-)}(x_2),\nonumber\\
 \fl G^{(+)}_2(z_1,z_2)&=&g^{(+)}_2(z_1,z_2)\label{3.1.17}\\
 \fl&-&\sum\limits_{j=0}^{\nu-1}\frac{1}{h_j}\int\limits_{\mathbb{C}}d[z^\prime]p_j^{(+)}(z^\prime)\left(g^{(+)}_2(z_1,z^\prime)-g^{(+)}_2(z^\prime,z_1)\right)p_j^{(+)}(z_2)g_1^{(+)}(z_2)\nonumber\\
 \fl&+&\sum\limits_{i,j=0}^{\nu-1}\frac{1}{2h_jh_i}\int\limits_{\mathbb{C}^2}d[z_1^\prime]d[z_2^\prime]p_i^{(+)}(z_1^\prime)p_j^{(+)}(z_2^\prime)\left(g^{(+)}_2(z_1^\prime,z_2^\prime)-g^{(+)}_2(z_2^\prime,z_1^\prime)\right)\nonumber\\
 \fl&\times&p_i^{(+)}(z_1)p_j^{(+)}(z_2)g_1^{(+)}(z_1)g_1^{(+)}(z_2).\nonumber
\end{eqnarray}
The change of the two-point measures is restricted by linear combinations with other rows and columns in the Pfaffian~\eref{2.1.5} and the determinant~\eref{2.2.4}. Essentially we add the orthogonal polynomials $p_l^{(\pm)}$ to the weight. Thereby we have to recall that everything which is done with the rows has to be done with the columns in the Pfaffian. This is the reason why $G^{(-)}_2$ stays anti-symmetric whereas $G^{(+)}_2$ is asymmetric in the entries.

\subsection{Recursion relations}\label{sec3.2}

The recursion relations of the orthogonal polynomials are
\begin{eqnarray}
 \frac{\partial p_l^{(\pm)}}{\partial x}(x)&=&lp_{l-1}^{(\pm)}(x),\label{3.2.1}\\
 xp_l^{(\pm)}(x)&=&p_{l+1}^{(\pm)}(x)\pm\frac{a^2\mu_\lt}{n}p_{l}^{(\pm)}(x)+\frac{la^2}{n}p_{l-1}^{(\pm)}(x).\label{3.2.2}
\end{eqnarray}
They result from the orthogonality relation~\eref{2.1.10} and the two identities
\begin{eqnarray}
 \langle D^{(\pm)}f_1|f_2\rangle_{g_1^{(\pm)}}&=&-\langle f_1|D^{(\pm)}f_2\rangle_{g_1^{(\pm)}},\label{3.2.3}\\
 \langle \mathfrak{m}_1f_1|f_2\rangle_{g_1^{(\pm)}}&=&\langle f_1|\mathfrak{m}_1f_2\rangle_{g_1^{(\pm)}}\label{3.2.4}
\end{eqnarray}
for two arbitrary integrable functions $f_1$ and $f_2$. The function $\mathfrak{m}_1$ is the monomial of order one and the differential operator $D^{(\pm)}$ is the creation operator of the harmonic oscillator corresponding to the measure $g_1^{(\pm)}$,
\begin{eqnarray}\label{3.2.5}
 D^{(\pm)}=\frac{\partial}{\partial x}-\frac{n}{2a^2}x\pm\frac{\mu_\lt}{2}.
\end{eqnarray}
Identity~\eref{3.2.4} cannot be extended to the measures $G^{(-)}_2$ and $G^{(+)}_2$ or equivalently $g^{(-)}_2$ and $g^{(+)}_2$, i.e.
\begin{eqnarray}\label{3.2.6}
 \left(\mathfrak{m}_1f_1|f_2\right)_{g^{(\pm)}_2}\neq\left(f_1|\mathfrak{m}_1f_2\right)_{g^{(\pm)}_2}.
\end{eqnarray}
However Eq.~\eref{3.2.3} has an analogue. Defining the differential operator
\begin{eqnarray}\label{3.2.7}
 \widetilde{D}^{(\pm)}=\frac{\partial}{\partial z}-\frac{n}{a^2}z+\frac{\mu_\rt\pm\mu_\lt}{2}
\end{eqnarray}
one can readily verify
\begin{eqnarray}
 \left( \widetilde{D}^{(\pm)} f_1|f_2\right)_{g^{(\pm)}_2}&=&-\left( f_1|\widetilde{D}^{(\pm)} f_2\right)_{g^{(\pm)}_2}.\label{3.2.8}
\end{eqnarray}
The starting point of such a proof is the differential equation
\begin{eqnarray}\label{3.2.9}
 \left[\frac{\partial}{\partial z_1}+\frac{\partial}{\partial z_2}+\frac{n}{a^2}(z_1+z_2)-(\mu_\rt\pm\mu_\lt)\right]g_2^{(\pm)}(z_1,z_2)=0.
\end{eqnarray}
Notice that the differential operator is restricted to the real eigenvalues of $D_5$ and $D_\W$ and to the real part of the complex conjugated pair of $z=z_1=z_2^*$ of $D_\W$ due to the Dirac delta-functions.

Both operators $\widetilde{D}^{(\pm)}$ and $D^{(\pm)}$ are closely related with each other which is quite advantageous. For example the action of $\widetilde{D}^{(\pm)}$ in the scalar product~\eref{2.1.6} is
\begin{eqnarray}\label{3.2.10}
 \langle \widetilde{D}^{(\pm)} f_1|f_2\rangle_{g_1^{(\pm)}}&=&-\langle f_1| f_2^\prime\rangle_{g_1^{(\pm)}}+\frac{\mu_\rt\mp\mu_\lt}{2}\langle f_1| f_2\rangle_{g_1^{(\pm)}},
\end{eqnarray}
where $f_2^\prime$ is the first derivative of $f_2$.

We consider the action of $\widetilde{D}^{(\pm)}$ on the polynomials $p_l^{(\pm)}$ and $q_{\nu+l}^{(\pm)}$. The recursion relations~\eref{3.2.1} and \eref{3.2.2} yield
\begin{eqnarray}
 \widetilde{D}^{(\pm)}p_l^{(\pm)}(z)&=&-\frac{n}{a^2}p_{l+1}^{(\pm)}(z)+\frac{\mu_\rt\mp\mu_\lt}{2} p_l^{(\pm)}(z).\label{3.2.11}
\end{eqnarray}
The polynomial $\widetilde{D}^{(\pm)}q^{(\pm)}_{\nu+l}$ can be expanded in the polynomials $\{p_j^{(\pm)},q^{(\pm)}_{\nu+j}\}$, i.e.
\begin{eqnarray}
 \widetilde{D}^{(\pm)}q^{(\pm)}_{\nu+l}(z)&=&\sum\limits_{j=0}^{l+1}\alpha^{(\pm)}_{lj}q^{(\pm)}_{\nu+j}(z)+\sum\limits_{j=0}^{\nu-1}\beta_{lj}^{(\pm)}p_{j}^{(\pm)}(z),\label{3.2.12}
\end{eqnarray}
where $\alpha^{(\pm)}_{lj}$ and $\beta_{lj}^{(\pm)}$ are the coefficients which have to be found.

In \ref{app2} we derive the recursions
\begin{eqnarray}
 \fl\widetilde{D}q^{(\pm)}_{\nu+2l}(z)&=&-\frac{n}{a^2}q^{(\pm)}_{\nu+2l+1}(z)+\widetilde{\epsilon}^{(\pm)}_lq^{(\pm)}_{\nu+2l}(z),\label{3.2.0.1}\\
 \fl\widetilde{D}q^{(\pm)}_{\nu+2l+1}(z)&=&-\frac{n}{a^2}q^{(\pm)}_{\nu+2l+2}(z)-\widetilde{\epsilon}^{(\pm)}_lq^{(\pm)}_{\nu+2l+1}(z)+\epsilon^{(\pm)}_lq^{(\pm)}_{\nu+2l}(z)-\frac{n}{a^2}\frac{o^{(\pm)}_{l}}{o^{(\pm)}_{l-1}}q^{(\pm)}_{\nu+2l-2}(z),\label{3.2.0.2}
\end{eqnarray}
with the coefficients
\begin{eqnarray}
 \fl\widetilde{\epsilon}^{(\pm)}_l&=&(2l+1)\frac{\mu_\rt\mp\mu_\lt}{2},\label{3.2.0.3}\\
 \fl\epsilon^{(\pm)}_l&=&\frac{n}{a^2}\left(\frac{\langle q_{\nu+2l+3}^{(\pm)}|p_{\nu+2l+1}\rangle_{g_1}}{h_{\nu+2l+1}}-\frac{\langle q_{\nu+2l+1}^{(\pm)}|p_{\nu+2l-1}\rangle_{g_1}}{h_{\nu+2l-1}}\right)-\frac{(l+1)^2(\mu_\rt\mp\mu_\lt)^2a^2}{n}.\label{3.2.0.4}
\end{eqnarray}
The recursion formula for $q^{(\pm)}_{\nu+2l+1}$, see Eq.~\eref{3.2.0.2}, is restricted to $l\geq1$. For $l=0$ we have to omit the last term, i.e. the constant $1/o^{(\pm)}_{-1}$ is zero, see Eq.~\eref{3.1.9d} when replacing the factorial by Euler's Gamma-function. This formula is quite useful to find Christoffel Darboux-like formulas, see subsection~\ref{sec3.3}.

\subsection{A Christoffel Darboux-like formula}\label{sec3.3}

For the calculation of spectral correlations the Christoffel Darboux formula is quite useful. However searching for such a formula of skew-orthogonal polynomials proved as a difficult task~\cite{Gho09}. The same is true for the polynomials $q_{\nu+l}^{(\pm)}$ for which we want to simplify the sum
\begin{eqnarray}\label{3.3.1}
 \Sigma_{n-1}^{(\pm)}(z_1,z_2)=\sum\limits_{l=0}^{n-1}\frac{1}{o_l^{(\pm)}}\det\left[\begin{array}{cc} q_{\nu+2l}^{(\pm)}(z_1) & q_{\nu+2l+1}^{(\pm)}(z_1) \\ q_{\nu+2l}^{(\pm)}(z_2) & q_{\nu+2l+1}^{(\pm)}(z_2) \end{array}\right].
\end{eqnarray}
For the orthogonal polynomials $p_l^{(\pm)}$ we already know such a result,
\begin{eqnarray}\label{3.3.2}
 \sum\limits_{l=0}^{\nu-1}\frac{1}{h_l}p_l^{(\pm)}(z_1)p_l^{(\pm)}(z_2)=\frac{1}{h_{\nu-1}}\frac{p_\nu^{(\pm)}(z_1)p_{\nu-1}^{(\pm)}(z_2)-p_\nu^{(\pm)}(z_2)p_{\nu-1}^{(\pm)}(z_1)}{z_1-z_2}.
\end{eqnarray}
Identity~\eref{3.3.2} is a direct consequence of the three term recursion relation~\eref{3.2.2}. Hence we pursue the same idea for Eq.~\eref{3.3.1} which is done in \ref{app5}. The Christoffel Darboux-like formula for the skew-orthogonal polynomials is
\begin{eqnarray}\label{3.3.5}
 \fl\Sigma_{n-1}^{(\pm)}(z_1,z_2)&=&\frac{n}{a^2o_{n-1}^{(\pm)}}\int\limits_{0}^\infty d\widetilde{x}\exp\left[-\frac{n}{a^2}\widetilde{x}^2+\left(\mu_\rt\pm\mu_\lt-\frac{n}{a^2}(z_1+z_2)\right)\widetilde{x}\right]\\
 \fl&\times&\det\left[\begin{array}{cc} q_{\nu+2n-2}^{(\pm)}(z_1+\widetilde{x}) & q_{\nu+2n}^{(\pm)}(z_1+\widetilde{x}) \\ q_{\nu+2n-2}^{(\pm)}(z_2+\widetilde{x}) & q_{\nu+2n}^{(\pm)}(z_2+\widetilde{x}) \end{array}\right].\nonumber
\end{eqnarray}
This result only depends on a few polynomials as it is already well known for the original Christoffel Darboux formula, cf. Eq.~\eref{3.3.2}.

\subsection{Representation as random matrix averages}\label{sec3.4}

As we have already seen in subsection~\ref{sec3.2} all skew-orthogonal polynomials $q_{\nu+l}^{(\pm)}$ are easy to derive if we know a compact expression for $l$ even. For this purpose we want to derive a representation as an integral over a random matrix. For the orthogonal polynomials the well known expression of this kind is
\begin{eqnarray}\label{3.4.1}
 \fl p_l^{(\pm)}(z)=\left(\frac{2\pi a^2}{n}\right)^{l^2}\int d[H]\det(z\eins_l-H)\exp\left[-\frac{n}{2a^2}\tr \left(H\mp\frac{a^2}{n}\mu_\lt\right)^2\right],
\end{eqnarray}
where $H$ is a $l\times l$ Hermitian random matrix with the measure
\begin{eqnarray}\label{3.4.2}
 d[H]=\prod\limits_{i=1}^ld H_{ii}\prod\limits_{1\leq i<j\leq l}2d\RE\,H_{ij}d\IM\,H_{ij}.
\end{eqnarray}
This random matrix integral can be drastically reduced to a small number of integration variables by the supersymmetry method \cite{Guh06,LSZ08,KGG09,Guh11}. A famous representation of the Hermite polynomials can be derived in this way,
\begin{eqnarray}\label{3.4.3}
 p_l^{(\pm)}(z)&=&\frac{l!}{2\pi}\int\limits_0^{2\pi}d\varphi\exp\left[-\frac{a^2}{2n}e^{\imath 2\varphi}+\left(z\mp\frac{a^2}{n}\mu_\lt\right)e^{\imath\varphi}\right]e^{-\imath l\varphi}.
\end{eqnarray}
The corresponding Rodrigues-formula for the Hermite polynomials is a simple lemma from this, i.e.
\begin{eqnarray}\label{3.4.4}
 p_l^{(\pm)}(z)&=&\left(-\frac{a^2}{n}\right)^l\exp\left[\frac{n}{2a^2}z^2\mp\mu_\lt z\right]\frac{\partial^l}{\partial z^l}\exp\left[-\frac{n}{2a^2}z^2\pm\mu_\lt z\right]\\
 &=&\left(\frac{a^2}{n}\right)^{l/2}H_l\left(\sqrt{\frac{n}{a^2}}z\mp\sqrt{\frac{a^2}{n}}\mu_l\right).\nonumber
\end{eqnarray}
The aim is to find the formulas analogous to Eqs.~\eref{3.4.1}, \eref{3.4.3} and \eref{3.4.4} for $q_{\nu+2l}^{(\pm)}$.

We compare the definition~\eref{3.1.4} with Eqs.~\eref{2.1.8} and \eref{2.2.7} for $k=k_\rt=k_\lt=0$, $n=l$ and $N_{\rm f}=1$. Since we were free of choosing the two-point weight $G_2^{(\pm)}$ and the polynomials $q_{\nu+l}^{(\pm)}$ at this step of the calculation, Eqs.~\eref{2.1.8} and \eref{2.2.7} are also valid when replacing $G_2^{(\pm)}$ by $g_2^{(\pm)}$ and $q_{\nu+l}^{(\pm)}$ by $p_{\nu+l}^{(\pm)}$. Moreover, $q_{\nu+2l}^{(\pm)}$ is also equal to
\begin{eqnarray}
  \fl q^{(\pm)}_{\nu+2l}(z)&=&\frac{(-1)^{\nu(\nu+1)/2}}{\Pf\left[\left(p_i^{(\pm)}|p_j^{(\pm)}\right)_{g^{(\pm)}_2}\right]_{\nu\leq i,j\leq\nu+2l-1}\prod\limits_{j=0}^{\nu-1}h_j}\label{3.4.5}\\
 \fl&\times&\hspace*{-0.1cm}\Pf\hspace*{-0.2cm}\left[\begin{array}{ccc} \hspace*{-0.2cm}\left\{\left(p_i^{(\pm)}|p_j^{(\pm)}\right)_{g^{(\pm)}_2}\right\}\underset{0\leq i,j\leq \nu+2l}{} & \hspace*{-0.35cm}\left\{\langle p_i^{(\pm)}|p_j^{(\pm)}\rangle_{g_1^{(\pm)}}\right\}\underset{0\leq j\leq \nu-1}{\underset{0\leq i\leq \nu+2l}{}} & \hspace*{-0.35cm}\left\{p_i^{(\pm)}(z)\right\}\underset{0\leq i\leq \nu+2l}{} \\ \hspace*{-0.2cm}\left\{-\langle p_i^{(\pm)}|p_j^{(\pm)}\rangle_{g_1^{(\pm)}}\right\}\underset{0\leq j\leq \nu+2l}{\underset{0\leq i\leq \nu-1}{}} & 0 & 0 \\ \hspace*{-0.2cm}\left\{-p_j(z)^{(\pm)}\right\}\underset{0\leq j\leq \nu+2l}{} & 0 & 0 \end{array}\right].\nonumber
\end{eqnarray}
Indeed this equation coincides with the ansatz~\eref{3.1.4} since the scalar products in the second row and column are either zero or equal to the normalization constants $h_j$. An expansion in these rows and columns yields Eq.~\eref{3.1.4}.

The polynomials $q_{\nu+2l}^{(\pm)}$ are the partition functions with one fermionic flavor,
\begin{eqnarray}\label{3.4.6}
 q_{\nu+2l}^{(\pm)}(z)&=&(-1)^\nu\frac{Z_1^{(l,\nu,\pm)}(-z)}{Z_0^{(l,\nu,\pm)}}.
\end{eqnarray}
By means of the supersymmetry technique~\cite{Guh06,LSZ08,KGG09,Guh11} we find the result
\begin{eqnarray}
 \fl q_{\nu+2l}^{(\pm)}(z)&=&\frac{(\pm1)^{l+\nu}l!(l+\nu)!}{(2\pi)^2}\int\limits_{[0,2\pi]^2}d\varphi_\rt d\varphi_\lt \exp\left[-\frac{a^2}{2n}(e^{\imath2\varphi_\rt}+e^{\imath2\varphi_\lt})+\frac{1}{n}e^{\imath(\varphi_\rt+\varphi_\lt)}\right]\label{3.4.11}\\
 \fl &\times&\exp\left[-\left(\frac{a^2\mu_\rt}{n}-z\right)e^{\imath\varphi_\rt}-\left(\frac{a^2\mu_\lt}{n}\mp z\right)e^{\imath\varphi_\lt}\right]e^{-\imath l\varphi_\rt}e^{-\imath (l+\nu)\varphi_\lt}\nonumber
\end{eqnarray}
in \ref{app6}. Notice the similarity with Eq.~\eref{3.4.3}.

Again we can ask for a Rodrigues formula and indeed it is a direct consequence of Eq.~\eref{3.4.11}. We find
\begin{eqnarray}
 \fl q_{\nu+2l}^{(\pm)}(z)&=&(\pm1)^{l+\nu}\left(\frac{a^2}{n}\right)^{(l+\nu)/2}\exp\left[\frac{a^2}{2n}\left(\mu_\rt-\frac{n}{a^2}z\right)^2\right]\label{3.4.12}\\
 \fl&\times&\left.\frac{\partial^l}{\partial\widetilde{x}^l}\right|_{\widetilde{x}=0}\exp\left[\frac{a^2}{2n}\left(\widetilde{x}+\mu_\rt-\frac{n}{a^2}z\right)^2\right]H_{l+\nu}\left(\pm\sqrt{\frac{n}{a^2}}z-\sqrt{\frac{a^2}{n}}\mu_\lt+\frac{\widetilde{x}}{\sqrt{na^2}}\right).\nonumber
\end{eqnarray}
Performing the derivatives we find an explicit expression in terms of Hermite polynomials for the skew-orthogonal polynomials $q_{\nu+2l}^{(\pm)}$,
\begin{eqnarray}
 \fl q_{\nu+2l}^{(\pm)}(z)&=&(\pm1)^{l+\nu}\left(\frac{1}{\sqrt{na^2}}\right)^l\left(\frac{a^2}{n}\right)^{(l+\nu)/2}\sum\limits_{j=0}^l\frac{l!(l+\nu)!}{j!(l-j)!(\nu+j)!}a^{2j}\label{3.4.13}\\
 \fl&\times&H_{\nu+j}\left(\pm\sqrt{\frac{n}{a^2}}z-\sqrt{\frac{a^2}{n}}\mu_\lt\right)H_{j}\left(\sqrt{\frac{n}{a^2}}z-\sqrt{\frac{a^2}{n}}\mu_\rt\right).\nonumber
\end{eqnarray}
The polynomials for $q_{\nu+2l+1}^{(\pm)}$ can be readily obtained with help of relation~\eref{3.2.0.1}. Remarkably the prefactors of the single summands are exactly the same as the ones of the modified Laguerre polynomials, $L_l^{(\nu)}$, when replacing the Hermite polynomials by monomials.

The limit $a\to0$ yields the generalized Laguerre polynomials,
\begin{eqnarray}\label{3.4.14}
 q_{\nu+2l}^{(\pm)}(z)&\overset{a\to0}{=}&\left(\pm\frac{1}{n}\right)^lz^\nu L_l^{(\nu)}(\mp nz^2),\\
 q_{\nu+2l+1}^{(\pm)}(z)&=&\left(\frac{a^2\widetilde{\epsilon}^{(\pm)}}{n}-\frac{a^2}{n}\widetilde{D}\right)q_{\nu+2l}^{(\pm)}(z)\overset{a\to0}{=}\left(\pm\frac{1}{n}\right)^lz^{\nu+1} L_l^{(\nu)}(\mp nz^2),\nonumber
\end{eqnarray}
which is in agreement with Ref.~\cite{Kieburg:2011ct}. The large $a$ limit with fixed variables $\sqrt{n/a^2}z$ and $\sqrt{a^2/n}\mu_{\lt/\rt}$ is a product of two Hermite polynomials
\begin{eqnarray}
 \fl q_{\nu+2l}^{(\pm)}(z)&\overset{a\gg1}{=}&\left(\frac{a^2}{n}\right)^{(2l+\nu)/2}H_{\nu+l}\left(\sqrt{\frac{n}{a^2}}z\mp\sqrt{\frac{a^2}{n}}\mu_\lt\right)H_{l}\left(\sqrt{\frac{n}{a^2}}z-\sqrt{\frac{a^2}{n}}\mu_\rt\right).\label{3.4.15}
\end{eqnarray}
Both limits can already be directly derived from the random matrix model, cf. Eqs.~(\ref{2.1}) and (\ref{2.2}). For $a=0$ we have a $\chi$GUE whose orthogonal polynomials are the Laguerre polynomials. Recently it was shown that the $\chi$GUE has also a non-trivial Pfaffian factorization whose skew-orthogonal polynomials of even order are the orthogonal polynomials itself. Hence the limit~\eref{3.4.14} agrees with the observation in Ref.~\cite{Kieburg:2011ct}.

In the large $a$ limit the off-diagonal blocks $W$ and $W^\dagger$, see \eref{2.1}, are suppressed. Therefore we end up with two decoupled GUE's. One is of dimension $l$ and the other one of dimension $l+\nu$. This indeed yields a product of two Hermite polynomials, cf. Eq.~\eref{3.4.15}.

A particular case of the polynomials can be obtained for the random matrix $D_5$. Let $\mu_\rt=-\mu_\lt=\mu$ and $a=1$. Then we have a $2n+\nu$ dimensional GUE, cf. Eq.~\eref{2.2}. Indeed we also get the corresponding Hermite polynomials. Equation~\eref{3.4.11} simplifies to
\begin{eqnarray}
 q_{\nu+2l}^{(-)}(z)&=&\frac{(\pm1)^{l+\nu}l!(l+\nu)!}{(2\pi)^2}\int\limits_{[0,2\pi]^2}d\varphi_\rt d\varphi_\lt \exp\left[-\frac{1}{2n}(e^{\imath\varphi_\rt}-e^{\imath\varphi_\lt})^2\right]\label{3.4.11b}\\
 &\times&\exp\left[\left(z-\frac{\mu}{n}\right)(e^{\imath\varphi_\rt}-e^{\imath\varphi_\lt})\right]e^{-\imath l\varphi_\rt}e^{-\imath (l+\nu)\varphi_\lt}\nonumber\\
 &=&n^{-(\nu+2l)/2}H_{\nu+2l}\left(\sqrt{n}z-\frac{\mu}{\sqrt{n}}\right)\nonumber
\end{eqnarray}
for the even polynomials and
\begin{eqnarray}
 q_{\nu+2l+1}^{(-)}(z)&=&\left(-\frac{\mu}{n}+z-\frac{1}{n}\frac{\partial}{\partial z}\right)q_{\nu+2l}^{(-)}(z)\label{3.4.11c}\\
 &=&n^{-(\nu+2l+1)/2}H_{\nu+2l+1}\left(\sqrt{n}z-\frac{\mu}{\sqrt{n}}\right)\nonumber
\end{eqnarray}
for the odd ones. Therefore all polynomials are given by Hermite polynomials corresponding to the same Gaussian distribution.

Another useful random matrix integral representation would be the one for the Christoffel Darboux-like formula~\eref{3.3.5}. For the orthogonal polynomials $p_l^{(\pm)}$ such a representation is well known,
\begin{eqnarray}
 \fl&&\frac{p_\nu^{(\pm)}(z_1)p_{\nu-1}^{(\pm)}(z_2)-p_\nu^{(\pm)}(z_2)p_{\nu-1}^{(\pm)}(z_1)}{z_1-z_2}\label{3.4.16}\\
 \fl&=&\left(\frac{2\pi a^2}{n}\right)^{(\nu-1)^2}\hspace*{-0.3cm}\int d[H]\det(z_1\eins_{\nu-1}-H)\det(z_2\eins_{\nu-1}-H)\exp\left[-\frac{n}{2a^2}\tr \left(H\mp\frac{a^2}{n}\mu_\lt\right)^2\right]\nonumber
\end{eqnarray}
with a $(\nu-1)\times(\nu-1)$ Hermitian matrix $H$. With the supersymmetry method \cite{Guh06,LSZ08,KGG09,Guh11} one can also find the representation
\begin{eqnarray}
 \fl&&\frac{p_\nu^{(\pm)}(z_1)p_{\nu-1}^{(\pm)}(z_2)-p_\nu^{(\pm)}(z_2)p_{\nu-1}^{(\pm)}(z_1)}{z_1-z_2}\label{3.4.17}\\
 \fl&=&\nu!(\nu-1)!\int\limits_{\U(2)}d\mu(U)\exp\left[-\frac{a^2}{2n}\tr U^2+\tr\left(\diag(z_1,z_2)\mp\frac{a^2\mu_\lt}{n}\eins_2\right)U\right]{\det}^{-\nu+1}U ,\nonumber
\end{eqnarray}
where $d\mu(U)$ is the normalized Haar measure of the unitary group $\U(2)$.

In \ref{app7} we show that the Christoffel Darboux-like formula~\eref{3.3.5} is essentially the partition function with two fermionic flavors, i.e.
\begin{eqnarray}\label{3.4.21}
 \Sigma_{n}^{(\pm)}(z_1,z_2)=(z_1-z_2)\frac{Z_2^{(n,\nu,\pm)}(-z_1,-z_2)}{o_n^{(\pm)}Z_0^{(n,\nu,\pm)}}.
\end{eqnarray}
Also the two-flavor partition function can be mapped to an integral over unitary groups by performing the same calculation as for the one-flavor partition function, see the discussion in \ref{app6}. Therefore $\Sigma_{n}^{(\pm)}$ is an integral over a compact set,
\begin{eqnarray}\label{3.4.22}
 \fl\Sigma_{n}^{(\pm)}(z_1,z_2)&=&-\frac{(n+1)!(n+\nu+1)!}{4}\sqrt{\frac{n(1\pm a^2)}{\pi}}\left(\frac{n}{1\pm a^2}\right)^{2n+\nu+1}\frac{1}{a}\\
 \fl&\times&\exp\left[-\frac{a^2}{4n}(\mu_\rt\pm\mu_\lt)^2\mp\frac{a^4}{4n(1\pm a^2)}(\mu_\rt\mp\mu_\lt)^2\right](z_1-z_2)\nonumber\\
 \fl&\times&\int\limits_{\U(2)\times\U(2)}d\mu(U_\rt)d\mu(U_\lt)\exp\left[-\frac{a^2}{2n}(\tr U_\rt^2+\tr U_\lt^2)+\frac{1}{n}\tr U_\rt U_\lt\right]\nonumber\\
 \fl&\times&\exp\left[-\tr\left(\frac{a^2\mu_\rt}{n}\eins_2-\diag(z_1,z_2)\right)U_\rt\right]{\det}^{-n}U_\rt\nonumber\\
 \fl&\times&\exp\left[-\tr\left(\frac{a^2\mu_\lt}{n}\eins_2\mp\diag(z_1,z_2)\right)U_\lt\right]{\det}^{-n-\nu}U_\lt.\nonumber
\end{eqnarray}
Equations~\eref{3.4.11} and \eref{3.4.22} are suitable for discussing the asymptotic behavior as it is done in subsection~\ref{sec3.5}.

\subsection{Asymptotics}\label{sec3.5}

The microscopic limit ($n\to\infty$, see discussion after Eq.~\eref{2.9d}) directly relates chiral random matrix theory with QCD in the $\epsilon$-regime. Hence we want to know the expressions of the polynomials $q_{\nu+2l}^{(\pm)}$ as well as the one of  the Christoffel Darboux-like formula $\Sigma_{n-1}^{(\pm)}$ in this limit.

For an arbitrary function $f$ which is $n$-independent and smooth on the group $\U(k)\times\U(k)$ the following asymptotic result exists
\begin{eqnarray}\label{3.5.1}
 \fl&&\int\limits_{\U(k)\times\U(k)}d\mu(U_\rt)d\mu(U_\lt) f(U_\rt,U_\lt)\exp\left[n\tr U_\rt U_\lt-n\tr {\rm ln}U_\rt U_\lt\right]\\
 \fl&\overset{n\gg1}{=}&(2\pi)^{-k/2}n^{-k^2/2}e^{nk}\prod\limits_{j=0}^{k-1}j!\int\limits_{\U(k)}d\mu(U) f(U,U^{-1}).\nonumber
\end{eqnarray}
This identity can be readily proven by a shift of the unitary matrix $U_\lt\rightarrow U_\rt^{-1}U_\lt$. Then the exponent only depends on $U_\lt$. The saddlepoint approximation yields an expansion of $U_\lt$ about the unit matrix yielding Eq.~\eref{3.5.1}.

Equations~\eref{3.4.11} and \eref{3.4.22} are particular cases of identity~\eref{3.5.1}. Hence we have
\begin{eqnarray}
 \fl q_{\nu+2n}^{(\pm)}\left(\frac{\widehat{z}}{2n}\right)&\overset{n\gg1}{=}&\frac{(-1)^\nu(\pm1)^{n+\nu}\sqrt{n}e^{-n}}{\sqrt{2\pi}}\int\limits_{[0,2\pi]}d\varphi \exp\left[-\widehat{a}^2(e^{\imath2\varphi}+e^{-\imath2\varphi})\right]\label{3.5.2}\\
 \fl &\times&\exp\left[\frac{1}{2}\left(\widehat{m}_6+\widehat{\lambda}_7-\widehat{z}\right)e^{\imath\varphi}+\frac{1}{2}\left(\widehat{m}_6-\widehat{\lambda}_7\mp \widehat{z}\right)e^{-\imath\varphi}\right]e^{\imath \nu\varphi}.\nonumber
\end{eqnarray}
for the polynomials which is the one-flavor partition function derived in Refs.~\cite{Damgaard:2010cz} and
\begin{eqnarray}\label{3.5.3}
 \fl\Sigma_{n}^{(\pm)}\left(\frac{\widehat{z}_1}{2n},\frac{\widehat{z}_2}{2n}\right)&\overset{n\gg1}{=}&-\frac{1}{ 8\sqrt{2\pi}}\frac{n^2}{\widehat{a}}\exp\left[-\frac{\widehat{a}^2}{2n^2}(\mu_\rt\pm\mu_\lt)^2\mp 4\widehat{a}^2\right](\widehat{z}_1-\widehat{z}_2)\\
 \fl&\times&\int\limits_{\U(2)}d\mu(U)\exp\left[\frac{1}{2}\tr\left((\widehat{m}_6-\widehat{\lambda}_7)\eins_2\mp \diag(\widehat{z}_1,\widehat{z}_2)\right)U^{-1}\right]{\det}^{\nu}U\nonumber\\
 \fl&\times&\exp\left[-\widehat{a}^2\tr (U^2+U^{-2})+\frac{1}{2}\tr\left((\widehat{m}_6+\widehat{\lambda}_7)\eins_2-\diag(\widehat{z}_1,\widehat{z}_2)\right)U\right]\nonumber
\end{eqnarray}
for the Christoffel Darboux-like formula, cf. Eqs.~\eref{2.1.2} and \eref{2.2.2}. In both equation we applied Stirling's formula to the factorials.

In the case of the polynomials $q_{\nu+2n}^{(\pm)}(z)$ we are able to integrate over the domain,
\begin{eqnarray}\label{3.5.4}
 \fl q_{\nu+2n}^{(\pm)}(z)&\overset{n\gg1}{=}&\frac{(-1)^\nu(\pm1)^{n+\nu}\sqrt{n}e^{-n-\nu}}{(2\pi)^{3/2}}\sum\limits_{j=-\infty}^\infty\int\limits_{[0,2\pi]^2}d\varphi_1d\varphi_2 \exp\left[-\widehat{a}^2(e^{\imath\varphi_1}+e^{-\imath\varphi_1})\right]\\
 \fl &\times&\exp\left[\frac{1}{2}\left(\widehat{m}_6+\widehat{\lambda}_7-\widehat{z}\right)e^{\imath\varphi_2}+\frac{1}{2}\left(\widehat{m}_6-\widehat{\lambda}_7\mp \widehat{z}\right)e^{-\imath\varphi_2}\right]e^{\imath \nu\varphi_2}e^{\imath j(\varphi_1-2\varphi_2)}\nonumber\\
 \fl&=&(-1)^\nu(\pm1)^{n+\nu}\sqrt{2\pi n}e^{-n-\nu}\sum\limits_{j=-\infty}^\infty \left(\frac{\widehat{m}_6-\widehat{\lambda}_7\mp\widehat{z}}{\widehat{m}_6+\widehat{\lambda}_7- \widehat{z}}\right)^{\nu/2+j}\nonumber\\
 \fl&\times&I_{\nu+2j}\left(\sqrt{(\widehat{m}_6+\widehat{\lambda}_7-\widehat{z})(\widehat{m}_6-\widehat{\lambda}_7\mp \widehat{z})}\right)I_j(-\widehat{a}^2).\nonumber
\end{eqnarray}
Due to the modified Bessel functions of the second kind $I_l(z)\propto(ze/2|l|)^{|l|}/\sqrt{2\pi |l|}\propto(z/2)^{|l|}/|l|!$, for $|l|\gg1$, the series rapidly converges and is numerically more stable than the integral~\eref{3.5.2} in simulations.

Unfortunately it is much harder to find such a sum for the Christoffel Darboux-like formula. However we can diagonalize the unitary matrix $U$ and find
\begin{eqnarray}\label{3.5.5}
 \fl\Sigma_{n}^{(+)}\left(\frac{\widehat{z}_1}{2n},\frac{\widehat{z}_2}{2n}\right)&\overset{n\gg1}{=}&-\frac{n^2}{ 4(2\pi)^{5/2}\widehat{a}}\int\limits_{[0,2\pi]^2}d\varphi_1d\varphi_2\sin^2\left[\frac{\varphi_1-\varphi_2}{2}\right]\\
 \fl&\times&\exp\left[\sum\limits_{j=1}^2\left(-\left(2\widehat{a}\cos\varphi_j+\frac{\widehat{m}_6}{4\widehat{a}}\right)^2+\imath\nu\varphi_j-\imath\widehat{\lambda}_7\sin\varphi_j\right)\right]\nonumber\\
 \fl&\times&\frac{\exp\left[\widehat{z}_1\cos\varphi_1+\widehat{z}_2\cos\varphi_2\right]-\exp\left[\widehat{z}_2 \cos\varphi_1+\widehat{z}_1\cos\varphi_2\right]}{\cos\varphi_1-\cos\varphi_2}\nonumber
\end{eqnarray}
for $D_\W$ and
\begin{eqnarray}\label{3.5.6}
 \fl\Sigma_{n}^{(-)}\left(\frac{\widehat{z}_1}{2n},\frac{\widehat{z}_2}{2n}\right)&\overset{n\gg1}{=}&\frac{\imath n^2}{ 4(2\pi)^{5/2}\widehat{a}}\int\limits_{[0,2\pi]^2}d\varphi_1d\varphi_2\sin^2\left[\frac{\varphi_1-\varphi_2}{2}\right]\\
 \fl&\times&\exp\left[\sum\limits_{j=1}^2\left(\left(2\widehat{a}\sin\varphi_j-\frac{\imath\widehat{\lambda}_7}{4\widehat{a}}\right)^2+\imath\nu\varphi_j-\imath\widehat{m}_6\cos\varphi_j\right)\right]\nonumber\\
 \fl&\times&\frac{\exp\left[\imath\widehat{z}_1\sin\varphi_1+\imath\widehat{z}_2\sin\varphi_2\right]-\exp \left[\imath\widehat{z}_2\sin\varphi_1+\imath\widehat{z}_1\sin\varphi_2\right]}{\sin\varphi_1-\sin\varphi_2}\nonumber
\end{eqnarray}
for $D_5$. These two formulas are quite suitable for the applications discussed in Sec.~\ref{sec4}. Both Christoffel-Darboux formulas are mostly two-flavor partition functions. In Ref.~\cite{Splittorff:2011bj} these functions are expressed as non-compact integrals over Bessel functions.

\section{Application to Wilson RMT}\label{sec4}

The results of the previous sections are helpful to simplify the $k$-point functions of $D_5$ as well as of $D_\W$. A Pfaffian factorization of the eigenvalue correlations of $D_5$ was already given in Ref.~\cite{Akemann:2011kj}. We obtain this structure in Sec.~\ref{sec4.1}, too. Moreover we express the kernels of the Pfaffian in terms of two-flavor partition functions which has proven fruitful in other random matrix ensembles, see Ref.~\cite{KieGuh09a} and the references therein. The unquenched $(k_\rt,k_\lt)$-point correlation function of $D_\W$ is shown in Sec.~\ref{sec4.2} which is a completely new result. Also this result displays a Pfaffian factorization whose entries are two-flavor partition functions.

\subsection{The Hermitian Wilson random matrix ensemble}\label{sec4.1}

In the $k$-point correlation function~\eref{2.1.8} we encounter an integral transform of the orthogonal and skew-orthogonal polynomials, cf. Eq.~\eref{2.1.9b}. Thus we define the integral transform of the skew-orthogonal polynomials,
\begin{eqnarray}\label{4.1.1}
 \widetilde{q}_{\nu+l}^{(-)}(x)&=&\int\limits_{\mathbb{R}}d\widetilde{x}q_{\nu+l}^{(-)}(\widetilde{x})G_2^{(-)}(\widetilde{x},x)\\
 &=&\int\limits_{\mathbb{R}}d\widetilde{x}q_{\nu+l}^{(-)}(\widetilde{x})g_2^{(-)}(\widetilde{x},x)-\sum\limits_{j=0}^{\nu-1}\frac{(q_{\nu+l}^{(-)}|p_j^{(-)})_{g_2^{(-)}}}{h_j}p_j^{(-)}(x)g_1^{(-)}(x)\nonumber\\
 &=&-\int\limits_{\mathbb{R}}d\widetilde{x}q_{\nu+l}^{(-)}(\widetilde{x})G_2^{(-)}(x,\widetilde{x}).\nonumber
\end{eqnarray}
The same integral transform for the orthogonal polynomials $p_l^{(-)}$, $0\leq l\leq\nu-1$, vanishes, i.e.
\begin{eqnarray}\label{4.1.2}
 \int\limits_{\mathbb{R}}d\widetilde{x}p_{l}^{(-)}(\widetilde{x})G_2^{(-)}(\widetilde{x},x)=0,
\end{eqnarray}
cf. Eq.~\eref{3.1.16}.

Using the identity
\begin{eqnarray}\label{3.4.19}
 \Pf\left[\begin{array}{cc} A & B \\ - B^T & C \end{array}\right]=\Pf A \Pf[ C+B^T A^{-1}B],
\end{eqnarray}
where $B$ and $C$ are arbitrary and $A$ is invertible, the $k$-point correlation function with an even number of fermionic flavors $N_{\rm f}=2n_{\rm f}$, see Eq.~\eref{2.1.8}, is
\begin{eqnarray}
 \fl &&R_{2n_{\rm f},k}^{(n,\nu,-)}(x^\prime)=\frac{(-1)^{k(k+1)/2}}{\Pf[K_3^{(-,n+n_{\rm f})}(-\lambda_i,-\lambda_j)]_{1\leq i,j\leq 2 n_{\rm f}}}\label{4.1.3}\\
 \fl&\times&\Pf\left[\begin{array}{c|c|c} \displaystyle K_1^{(-,n+n_{\rm f})}(x_i,x_j) & \displaystyle -K_2^{(-,n+n_{\rm f})}(x_j,x_i) & \displaystyle -K_2^{(-,n+n_{\rm f})}(-\lambda_j,x_i) \\ \hline \displaystyle K_2^{(-,n+n_{\rm f})}(x_i,x_j) & \displaystyle K_3^{(-)}(x_i,x_j) & \displaystyle K_3^{(-,n+n_{\rm f})}(x_i,-\lambda_j) \\ \hline \displaystyle K_2^{(-,n+n_{\rm f})}(-\lambda_i,x_j) & \displaystyle K_3^{(-)}(-\lambda_i,x_j) & \displaystyle K_3^{(-,n+n_{\rm f})}(-\lambda_i,-\lambda_j) \end{array}\right]\nonumber
\end{eqnarray}
which is the main result for the Hermitian Wilson random matrix $D_5$. The indices $i$ and $j$ of the Pfaffian in the denominator take the values $(1,\ldots,k,1,\ldots,k,1,\ldots,2n_{\rm f})$. The functions in the entries are
\begin{eqnarray}
 \fl K_1^{(-,n+n_{\rm f})}(x_1,x_2)&=&G_2^{(-)}(x_1,x_2)+\sum\limits_{l=0}^{n+n_{\rm f}-1}\frac{1}{o_l^{(-)}}\det\left[\begin{array}{cc} \widetilde{q}_{\nu+2l+1}^{(-)}(x_1) & \widetilde{q}_{\nu+2l}^{(-)}(x_1) \\ \widetilde{q}_{\nu+2l+1}^{(-)}(x_2) & \widetilde{q}_{\nu+2l}^{(-)}(x_2) \end{array}\right],\label{4.1.4}\\
 \fl K_2^{(-,n+n_{\rm f})}(x_1,x_2)&=&\sum\limits_{l=0}^{\nu-1}\frac{1}{h_l}p_l^{(-)}(x_1)p_l^{(-)}(x_2)g_1^{(-)}(x_2)\nonumber\\
  \fl&+&\sum\limits_{l=0}^{n+n_{\rm f}-1}\frac{1}{o_l^{(-)}}\det\left[\begin{array}{cc} q_{\nu+2l+1}^{(-)}(x_1) & q_{\nu+2l}^{(-)}(x_1) \\ \widetilde{q}_{\nu+2l+1}^{(-)}(x_2) & \widetilde{q}_{\nu+2l}^{(-)}(x_2) \end{array}\right],\nonumber\\
  \fl&&\label{4.1.5}\\
 \fl K_3^{(-,n+n_{\rm f})}(x_1,x_2)&=&\sum\limits_{l=0}^{n+n_{\rm f}-1}\frac{1}{o_l^{(-)}}\det\left[\begin{array}{cc} q_{\nu+2l+1}^{(-)}(x_1) & q_{\nu+2l}^{(-)}(x_1) \\ q_{\nu+2l+1}^{(-)}(x_2) & q_{\nu+2l}^{(-)}(x_2) \end{array}\right]\nonumber\\
 \fl&=&-\Sigma_{n+n_{\rm f}-1}^{(-)}(x_1,x_2).\label{4.1.6}
\end{eqnarray}
The $k$-point correlation function for an odd number of flavors can be derived by shifting one of the axial masses $\lambda$ to infinity. Then we get the skew-orthogonal polynomial $q_{\nu+2(n+n_{\rm f}-1)}^{(-)}$ and its integral transform $\widetilde{q}_{\nu+2(n+n_{\rm f}-1)}^{(-)}$ in one row and one column of the numerator and the denominator of Eq.~\eref{4.1.3}.

The case $k=0$ is the normalization. For $k=2n+\nu$ we have a compact representation of the joint probability density $p_5$ as a single Pfaffian determinant. 

The representation~(\ref{4.1.4}-\ref{4.1.6}) in terms of the Hermite polynomials $p_l^{(-)}$ and the skew-orthogonal polynomials $q_{\nu+l}^{(-)}$ can be easily interpreted. The $\nu$ former zero modes are broadened by a GUE of dimension $\nu$. The skew-orthogonal polynomials can be identified with the remaining modes and describe the spectral density thereof. Both spectra, the one of the GUE and the one of the remaining modes, are coupled by the sum in Eq.~\eref{4.1.1}. They manifest the repulsion of the former zero modes with the remaining modes which is given by the Vandermonde determinant in the joint probability density~\eref{2.6}.

Not only the kernel $K_3^{(-,n+n_{\rm f})}$ can be expressed in terms of two-flavor partition functions, note that  the Christoffel-Darboux-like formula, $\Sigma_{n+n_{\rm f}-1}^{(-)}$, is mostly such a partition function. Also the kernels $K_1^{(-,n+n_{\rm f})}$ and $K_2^{(-,n+n_{\rm f})}$ can be traced back to partition functions. In \ref{app3} we derive the following results
\begin{eqnarray}
\fl K_1^{(-,n)}(x_1,x_2)&=&\frac{o_n^{(-)}}{\pi^2}(x_1-x_2)\underset{\varepsilon_2\to0}{\underset{\varepsilon_1\to0}{\IM}}\label{4.1.7}\\
 \fl&&\hspace*{-1.3cm}\times\left\langle\frac{1}{\det(D_5-(x_1+\imath\varepsilon_1)\eins_{2n+\nu+2})\det(D_5-(x_2+\imath\varepsilon_2)\eins_{2n+\nu+2})}\right\rangle_{n+1,\nu},\nonumber\\
\fl K_2^{(-,n)}(x_1,x_2)&=&\frac{1}{\pi}\frac{1}{x_2-x_1}\underset{\varepsilon\to0}{\IM}\left\langle\frac{\det(D_5-x_1\eins_{2n+\nu})}{\det(D_5-(x_2+\imath\varepsilon)\eins_{2n+\nu})}\right\rangle_{n,\nu}.\label{4.1.8}
\end{eqnarray}
We employ the notations
\begin{eqnarray}
 \underset{\varepsilon\to0}{\IM}\int dx\frac{f(x)}{x-\imath\varepsilon}=\underset{\varepsilon\to0}{\lim}\int dx\frac{\varepsilon f(x)}{x^2+\varepsilon^2}=\pi f(0)\label{4.1.9}
\end{eqnarray}
and
\begin{eqnarray}
 \left\langle F(D_\W)\right\rangle_{N,\nu}= \left\langle F(\gamma_5 D_5)\right\rangle_{N,\nu}=\int d[D_\W] F(D_\W)P(D_\W)\label{4.1.10}
\end{eqnarray}
for two arbitrary sufficiently integrable functions $f$ and $F$ and the definition of the probability density $P$ in Eq.~\eref{2.2}. The random matrix on the right hand side of Eq.~\eref{4.1.10} has the dimension $(2N+\nu)\times(2N+\nu)$ with index $\nu$. Hence we have to take the averages~\eref{4.1.7} and \eref{4.1.8} over a Wilson random matrix with $N=n+n_{\rm f}$.

Considering Eqs.~\eref{3.4.21}, \eref{4.1.6}, \eref{4.1.7} and \eref{4.1.8} we traced the unquenched $k$-point correlation functions of $D_5$ back to partition functions with two fermionic, two bosonic and one fermionic and one bosonic determinant. Hence the structure of the eigenvalue correlations of $D_5$ is in the same class of matrix ensembles as the $\beta=1$ and $\beta=4$ standard ensembles, e.g. GOE, GSE, the real and quaternion Ginibre ensemble, the chiral GOE and the chiral GSE, see Ref.~\cite{KieGuh09a} and the references therein. When taking the continuum limit, $a\to0$, the Pfaffian determinant will persist though we have then chiral GUE. This observation agrees with the result found in Ref.~\cite{Kieburg:2011ct}. Therein a non-trivial Pfaffian was derived for all random matrix ensembles corresponding to orthogonal polynomials. Exactly this structure carries over to the finite lattice spacing result~\eref{4.1.3}.

The kernel $K_2^{(-,n)}(x,x)$ is equal to the quenched one point function of $D_5$, denoted by $\rho_5(x)$ in Refs.~\cite{Damgaard:2010cz,Akemann:2010zp}. Due to the prefactor $1/(x_1-x_2)$, see Eq.~\eref{4.1.8}, we have to apply l'Hospital's rule which exactly agrees with the common definition of $\rho_5$.

The Pfaffian factorization~\eref{4.1.3} was already discovered in Ref.~\cite{Akemann:2011kj} but we made the connection to two-flavor partition functions. Furthermore the structure as well as the expression in two-flavor partition functions carry over to the microscopic limit. In this limit Wilson random matrix theory is directly related to the $\epsilon$-regime of Wilson fermions in lattice QCD \cite{Sharpe:1998xm,RS02,Bar:2003mh,Sharpe06,Necco:2011vx}. Hence we found a neat representation which drastically simplifies the numerical realization of the $k$-point correlation functions.

The microscopic limit of the kernel $K_3^{(-,n)}$ is shown in Sec.~\ref{sec3.5}, see Eq.~\eref{3.5.6}. A derivation of the other kernels as well as a qualitative discussion of the results will be done elsewhere \cite{KZV12}.

In the notation of Refs.~\cite{Damgaard:2010cz,Akemann:2010zp,Splittorff:2011bj} the kernels are proportional to the two-flavor partition functions of the chiral Lagrangian,
\begin{eqnarray}
  \fl K_1^{(-,\infty)}(x_1,x_2)&\propto&(\widehat{x}_1-\widehat{x}_2)\underset{\varepsilon_2\to0}{\underset{\varepsilon_1\to0}{\IM}}Z_{0/2}^{\nu}(\widehat{m}_6,\widehat{m}_6;\widehat{\lambda}_7-\widehat{x}_1,\widehat{\lambda}_7-\widehat{x}_2;\widehat{a}_8,\widehat{a}_{6/7}=0),\label{4.1.11}\\
  \fl K_2^{(-,\infty)}(x_1,x_2)&\propto&\underset{\varepsilon\to0}{\IM}\frac{Z_{1/1}^{\nu}(\widehat{m}_6,\widehat{m}_6;\widehat{\lambda}_7-\widehat{x}_1,\widehat{\lambda}_7-\widehat{x}_2;\widehat{a}_8,\widehat{a}_{6/7}=0)}{\widehat{x}_2-\widehat{x}_1},\label{4.1.12}\\
 \fl K_3^{(-,\infty)}(x_1,x_2)&\propto&(\widehat{x}_1-\widehat{x}_2)Z_{2/0}^{\nu}(\widehat{m}_6,\widehat{m}_6;\widehat{\lambda}_7-\widehat{x}_1,\widehat{\lambda}_7-\widehat{x}_2;\widehat{a}_8,\widehat{a}_{6/7}=0),\label{4.1.13}
\end{eqnarray}
 in the microscopic limit.  Please recall that $\widehat{x}=2nx$ is fixed. The constants $\widehat{a}_i$ are essentially the product of the lattice spacing $a$ times the square roots of the low energy constants, $\sqrt{W_i}$, \cite{Damgaard:2010cz,Akemann:2010zp,Splittorff:2011bj}. We get the case $\widehat{a}_{6/7}\neq0$ when we multiply the expression~\eref{4.1.3} with the partition function of $N_{\rm f}$ fermionic flavors cancelling the Pfaffian in the denominator. Then we have to integrate over Gaussian distributions of $\widehat{m}_6$ and $\widehat{\lambda}_7$. Finally we divide the result by the partition function of $N_{\rm f}$ fermionic flavors with $\widehat{a}_{6/7}\neq0$ which is also the two Gaussian integrals over  $\widehat{m}_6$ and $\widehat{\lambda}_7$ of the partition function with $\widehat{a}_{6/7}=0$, cf. Ref.~\cite{KSV12}. Please notice that we will lose the Pfaffian factorization when going from $\widehat{a}_{6/7}=0$ to $\widehat{a}_{6/7}\neq0$.

\subsection{The non-Hermitian Wilson random matrix ensemble}\label{sec4.2}

As in the Hermitian version we define the integral transform of the skew-orthogonal polynomials $q_{\nu+l}^{(+)}$. However we have to distinguish between left and right transformation because $G_2^{(+)}$ is not anti-symmetric anymore,
\begin{eqnarray}
 \widetilde{q}_{\nu+l}^{(\lt,+)}(z)&=&\int\limits_{\mathbb{C}}d[\widetilde{z}]q_{\nu+l}^{(+)}(\widetilde{z})G_2^{(+)}(\widetilde{z},z)\label{4.2.1}\\
 &=&\int\limits_{\mathbb{C}}d[\widetilde{z}]q_{\nu+l}^{(+)}(\widetilde{z})g_2^{(+)}(\widetilde{z},z)-\sum\limits_{j=0}^{\nu-1}\frac{(q_{\nu+l}^{(+)}|p_j^{(+)})_{g_2^{(+)}}}{h_j}p_j^{(+)}(z)g_1(z),\nonumber\\
 \widetilde{q}_{\nu+l}^{(\rt,+)}(z)&=&\int\limits_{\mathbb{C}}d[\widetilde{z}]q_{\nu+l}^{(+)}(\widetilde{z})G_2^{(+)}(z,\widetilde{z})\label{4.2.2}\\
  &=&\int\limits_{\mathbb{C}}d[\widetilde{z}]q_{\nu+l}^{(+)}(\widetilde{z})g_2^{(+)}(z,\widetilde{z}).\nonumber
\end{eqnarray}
Another difference to the Hermitian case is a non-vanishing integral transform of the orthogonal polynomials
\begin{eqnarray}\label{4.2.3}
 \widetilde{p}_{l}(z)=\int\limits_{\mathbb{C}}d[\widetilde{z}]p_l^{(+)}(\widetilde{z})G_2^{(+)}(\widetilde{z},z)=\int\limits_{\mathbb{C}}d[\widetilde{z}]p_l^{(+)}(\widetilde{z})G_2^{(+)}(z,\widetilde{z})
\end{eqnarray}
for $0\leq l\leq\nu-1$ due to Eq.~\eref{3.1.17}.

Again we consider an even number of fermionic flavors. Then we arrive at our main result for the non-Hermitian Wilson random matrix which is  the $(k_\rt,k_\lt)$-point correlation function~\eref{2.2.7},
\begin{eqnarray}
 \fl &&R_{2n_{\rm f},k_\rt,k_\lt}^{(n,\nu,+)}(Z^\prime,-m)=\frac{1}{\Pf[K_6^{(+,n+n_{\rm f})}(m_i,m_j)]_{1\leq i,j\leq 2 n_{\rm f}}}\label{4.2.4}\\
 \fl&\times&\Pf\left[\begin{array}{c|c|c} \displaystyle \underset{}{\widehat{K}_1^{(+,n+n_{\rm f})}(z^{(\rt)}_i,z^{(\rt)}_j)}  & \displaystyle \widehat{K}_3^{(+,n+n_{\rm f})}(z^{(\lt)}_j,z^{(\rt)}_i) & \displaystyle \widehat{K}_4^{(+,n+n_{\rm f})}(m_j,z^{(\rt)}_i) \\ \hline \displaystyle -\overset{}{\underset{}{\widehat{K}_3^{(+,n+n_{\rm f})T}(z^{(\lt)}_i,z^{(\rt)}_j)}} & \displaystyle \widehat{K}_2^{(+,n+n_{\rm f})}(z^{(\lt)}_i,z^{(\lt)}_j) & \displaystyle \widehat{K}_5^{(+,n+n_{\rm f})}(m_j,z^{(\lt)}_i) \\ \hline \displaystyle -\overset{}{\widehat{K}_4^{(+,n+n_{\rm f})T}(m_i,z^{(\rt)}_j)}  & \displaystyle -\widehat{K}_5^{(+,n+n_{\rm f})T}(m_i,z^{(\lt)}_j)  & \displaystyle K_6^{(+,n+n_{\rm f})}(m_i,m_j)  \end{array}\right],\nonumber
\end{eqnarray}
with
\begin{eqnarray}
\fl\widehat{K}_1^{(+,n+n_{\rm f})}(z^{(\rt)}_i,z^{(\rt)}_j) &=&\left[\begin{array}{cc} K_1^{(+,n+n_{\rm f})}(z^{(\rt)}_i,z^{(\rt)}_j) & -K_3^{(+,n+n_{\rm f})}(z^{(\rt)}_j,z^{(\rt)}_i) \\  K_3^{(+,n+n_{\rm f})}(z^{(\rt)}_i,z^{(\rt)}_j) & K_6^{(+,n+n_{\rm f})}(z^{(\rt)}_i,z^{(\rt)}_j) \end{array}\right],\label{4.2.4a}\\
\fl\widehat{K}_2^{(+,n+n_{\rm f})}(z^{(\lt)}_i,z^{(\lt)}_j) &=&\left[\begin{array}{cc} K_4^{(+,n+n_{\rm f})}(z^{(\lt)}_i,z^{(\lt)}_j) & K_5^{(+,n+n_{\rm f})}(z^{(\lt)}_j,z^{(\lt)}_i) \\  -K_5^{(+,n+n_{\rm f})}(z^{(\lt)}_i,z^{(\lt)}_j) & K_6^{(+,n+n_{\rm f})}(z^{(\lt)}_i,z^{(\lt)}_j) \end{array}\right],\label{4.2.4b}\\
\fl\widehat{K}_3^{(+,n+n_{\rm f})}(z^{(\lt)}_j,z^{(\rt)}_i) &=&\left[\begin{array}{cc} K_2^{(+,n+n_{\rm f})}(z^{(\lt)}_j,z^{(\rt)}_i) & -K_3^{(+,n+n_{\rm f})}(z^{(\lt)}_j,z^{(\rt)}_i) \\  -K_5^{(+,n+n_{\rm f})}(z^{(\rt)}_i,z^{(\lt)}_j) & K_6^{(+,n+n_{\rm f})}(z^{(\rt)}_i,z^{(\lt)}_j) \end{array}\right],\label{4.2.4c}\\
\fl\widehat{K}_4^{(+,n+n_{\rm f})}(m_j,z^{(\rt)}_i)  &=&\left[\begin{array}{c}  -K_3^{(+,n+n_{\rm f})}(m_j,z^{(\rt)}_i)  \\  K_6^{(+,n+n_{\rm f})}(z^{(\rt)}_i,m_j)  \end{array}\right],\label{4.2.4d}\\
\fl\widehat{K}_5^{(+,n+n_{\rm f})}(m_j,z^{(\lt)}_i) &=&\left[\begin{array}{c}  K_5^{(+,n+n_{\rm f})}(m_j,z^{(\lt)}_i) \\  K_6^{(+,n+n_{\rm f})}(z^{(\lt)}_i,m_j)  \end{array}\right],\label{4.2.4e}
\end{eqnarray}
where the indices $i$ and $j$ take the values $(1,\ldots,k_\rt,1,\ldots,k_\lt,1,\ldots,2n_{\rm f})$ from left to right and top to bottom. The functions are given by
\begin{eqnarray}
 \fl K_1^{(+,n+n_{\rm f})}(z_1,z_2)&=&\sum\limits_{l=0}^{n+n_{\rm f}-1}\frac{1}{o_l^{(+)}}\det\left[\begin{array}{cc} \widetilde{q}_{\nu+2l+1}^{(\rt,+)}(z_1) & \widetilde{q}_{\nu+2l}^{(\rt,+)}(z_1) \\ \widetilde{q}_{\nu+2l+1}^{(\rt,+)}(z_2) & \widetilde{q}_{\nu+2l}^{(\rt,+)}(z_2) \end{array}\right]\nonumber\\
 \fl&=&-\int\limits_{\mathbb{C}^2}d[\widetilde{z}_1]d[\widetilde{z}_2]\Sigma_{n+n_{\rm f}-1}^{(+)}(\widetilde{z}_1,\widetilde{z}_2)g_2^{(+)}(z_1,\widetilde{z}_1)g_2^{(+)}(z_2,\widetilde{z}_2),\label{4.2.5}\\
 \fl K_2^{(+,n+n_{\rm f})}(z_1,z_2)&=&G_2^{(+)}(z_2,z_1)+\sum\limits_{l=0}^{n+n_{\rm f}-1}\frac{1}{o_l^{(+)}}\det\left[\begin{array}{cc} \widetilde{q}_{\nu+2l+1}^{(\lt,+)}(z_1) & \widetilde{q}_{\nu+2l}^{(\lt,+)}(z_1) \\ \widetilde{q}_{\nu+2l+1}^{(\rt,+)}(z_2) & \widetilde{q}_{\nu+2l}^{(\rt,+)}(z_2) \end{array}\right]\nonumber\\
 \fl&-&\sum\limits_{l=0}^{\nu-1}\frac{1}{h_l}p_l^{(+)}(x_1)\widetilde{p}_l(z_2)g_1^{(+)}(x_1)\delta(y_1),\label{4.2.6}\\
 \fl K_3^{(+,n+n_{\rm f})}(z_1,z_2)&=&\sum\limits_{l=0}^{n+n_{\rm f}-1}\frac{1}{o_l^{(+)}}\det\left[\begin{array}{cc} q_{\nu+2l+1}^{(+)}(z_1) & q_{\nu+2l}^{(+)}(z_1) \\ \widetilde{q}_{\nu+2l+1}^{(\rt,+)}(z_2) & \widetilde{q}_{\nu+2l}^{(\rt,+)}(z_2) \end{array}\right]\nonumber\\
 \fl&=&-\int\limits_{\mathbb{C}}d[\widetilde{z}]\Sigma_{n+n_{\rm f}-1}^{(+)}(z_1,\widetilde{z})g_2^{(+)}(z_2,\widetilde{z}),\label{4.2.7}\\
 \fl K_4^{(+,n+n_{\rm f})}(z_1,z_2)&=&\sum\limits_{l=0}^{n+n_{\rm f}-1}\frac{1}{o_l^{(+)}}\det\left[\begin{array}{cc} \widetilde{q}_{\nu+2l+1}^{(\lt,+)}(z_1) & \widetilde{q}_{\nu+2l}^{(\lt,+)}(z_1) \\ \widetilde{q}_{\nu+2l+1}^{(\lt,+)}(z_2) & \widetilde{q}_{\nu+2l}^{(\lt,+)}(z_2) \end{array}\right]\nonumber\\
 \fl&+&\sum\limits_{l=0}^{\nu-1}\frac{1}{h_l}\det\left[\begin{array}{cc} \widetilde{p}_l(z_1) & p_l^{(+)}(x_1)g_1^{(+)}(x_1) \delta(y_1)\\ \widetilde{p}_l(z_2) & p_l^{(+)}(x_2)g_1^{(+)}(x_2) \delta(y_2)\end{array}\right],\label{4.2.8}\\
 \fl K_5^{(+,n+n_{\rm f})}(z_1,z_2)&=&\sum\limits_{l=0}^{\nu-1}\frac{1}{h_l}p_l^{(+)}(z_1)p_l^{(+)}(x_2)g_1^{(+)}(x_2)\delta(y_2)\nonumber\\
 \fl&+&\sum\limits_{l=0}^{n+n_{\rm f}-1}\frac{1}{o_l^{(+)}}\det\left[\begin{array}{cc} q_{\nu+2l+1}^{(+)}(z_1) & q_{\nu+2l}^{(+)}(z_1) \\ \widetilde{q}_{\nu+2l+1}^{(\lt,+)}(z_2) & \widetilde{q}_{\nu+2l}^{(\lt,+)}(z_2) \end{array}\right],\label{4.2.9}\\
 \fl K_6^{(+,n+n_{\rm f})}(z_1,z_2)&=&\sum\limits_{l=0}^{n+n_{\rm f}-1}\frac{1}{o_l^{(+)}}\det\left[\begin{array}{cc} q_{\nu+2l+1}^{(+)}(z_1) & q_{\nu+2l}^{(+)}(z_1) \\ q_{\nu+2l+1}^{(+)}(z_2) & q_{\nu+2l}^{(+)}(z_2) \end{array}\right]\nonumber\\
 \fl&=&-\Sigma_{n+n_{\rm f}-1}^{(+)}(z_1,z_2).\label{4.2.10}
\end{eqnarray}
Note that although some of the sums seem to look identical they slightly differ by the integral transforms which have to be taken.

The result for an odd number of flavors can again be obtained by taking the limit of one mass to infinity. Then one row and one column only depend on the skew-orthogonal polynomial $q_{\nu+2(n+n_{\rm f}-1)}^{(+)}$ and its two integral transforms $\widetilde{q}_{\nu+2(n+n_{\rm f}-1)}^{(\lt,+)}$ and $\widetilde{q}_{\nu+2(n+n_{\rm f}-1)}^{(\rt,+)}$.

For $k_\rt=k_\lt=0$ we find the normalization and in the case $(k_\rt,k_\lt)=(n,n+\nu)$ we have a representation of joint probability density $p_\W$ as a Pfaffian similar to the one of $p_5$. Additionally, we can consider the particular cases $(k_\rt,k_\lt)=(n,0)$ and $(k_\rt,k_\lt)=(0,n+\nu)$ which are the joint probability densities for the eigenvalues $z^{(\rt)}$ and $z^{(\lt)}$ separately. These two joint probability densities are the ones for the right handed and the half of the complex modes, namely $z^{(\rt)}$, and for the left handed and the other half of the complex modes, which is $z^{(\lt)}$.

Again we recognize a natural splitting of the spectral properties. There are those terms, the sums with Hermite polynomials $p_l^{(+)}$, which describe the broadening of the former zero modes. They are again manifested by the same GUE which we found when discussing $D_5$ and are located on the real axis only, notice the Dirac delta functions. Moreover we have the terms for the remaining modes given by the skew-orthogonal polynomials, $q_{\nu+l}^{(+)}$. The corresponding eigenvalues to these modes do not necessarily  lie on the real axis. On the contrary most eigenvalues are distributed in the complex plane, see Ref.~\cite{Kieburg:2011yg,Kieburg:2011uf,KSV12}.

There is an interaction between these two kinds of spectra in the integral transform, cf. Eqs.~\eref{4.2.1} and \eref{4.2.3}. This interaction directly follows from the Vandermonde determinant in the joint probability density~\eref{2.10}. The repulsion obtained by this coupling effects the spectrum located on the real axis as well as the complex one.

As for $D_5$ all kernels of the result~\eref{4.2.4} can be traced back to two-flavor partition functions. For the kernels $K_1^{(+,n+n_{\rm f})}$, Eq.~\eref{4.2.5}, $K_3^{(+,n+n_{\rm f})}$, Eq.~\eref{4.2.7}, and $K_6^{(+,n+n_{\rm f})}$, Eq.~\eref{4.2.10}, we know already appropriate expressions. In \ref{app3} we derive the results for the other kernels,
\begin{eqnarray}
\fl \Delta K_2^{(+,n)}(z_1,z_2)&=&K_2^{(+,n)}(z_1,z_2)+K_1^{(+,n)}(z_1,z_2)\label{4.2.11}\\
\fl&=&g_2^{(+)}(z_2,z_1)+\frac{1}{\pi}\int\limits_{\mathbb{C}}d[\widetilde{z}]\frac{g_2^{(+)}(z_2,\widetilde{z})}{x_1-\widetilde{z}}\nonumber\\
\fl&\times&\underset{\varepsilon\to0}{\IM}\left\langle\frac{\det(D_\W-\widetilde{z}\eins_{2n+\nu})}{\det(D_\W-x_1\eins_{2n+\nu}-\imath\varepsilon\gamma_5)}\right\rangle_{n,\nu}\delta(y_1),\nonumber\\
\fl \Delta K_4^{(+,n)}(z_1,z_2)&=&K_4^{(+,n)}(z_1,z_2)-\Delta K_2^{(+,n)}(z_1,z_2)+\Delta K_2^{(+,n)}(z_2,z_1)-K_1^{(+,n)}(z_1,z_2)\nonumber\\
\fl&=&\frac{o_n^{(+)}}{\pi^2}(x_1-x_2)\delta(y_1)\delta(y_2)\underset{\varepsilon_2\to0}{\underset{\varepsilon_1\to0}{\IM}}\label{4.2.12}\\
\fl&&\hspace*{-1.1cm}\times\left\langle\frac{1}{\det(D_\W-x_1\eins_{2n+\nu+2}-\imath\varepsilon_1\gamma_5)\det(D_\W-x_2\eins_{2n+\nu+2}-\imath\varepsilon_2\gamma_5)}\right\rangle_{n+1,\nu},\nonumber\\
\fl \Delta K_5^{(+,n)}(z_1,z_2)&=&K_5^{(+,n)}(z_1,z_2)+K_3^{(+,n)}(z_1,z_2)\label{4.2.13}\\
\fl&=&\frac{1}{\pi}\frac{1}{z_1-x_2}\underset{\varepsilon\to0}{\IM}\left\langle\frac{\det(D_\W-z_1\eins_{2n+\nu})}{\det(D_\W-x_2\eins_{2n+\nu}-\imath\varepsilon\gamma_5)}\right\rangle_{n,\nu}\delta(y_2).\nonumber
\end{eqnarray}
The kernel $\Delta K_4^{(+,n+n_{\rm f})}$ describes the correlation of the chiral distribution over the real eigenvalues with itself. This can be seen by the $\gamma_5$ weight of the imaginary increments in the denominators and the Dirac delta functions, cf. Refs.~\cite{Splittorff:2011bj}. The other two kernels~\eref{4.2.11} and \eref{4.2.13} represent the interaction of the chiral distribution over the real eigenvalues with the remaining spectrum describing the additional real modes and the complex ones.

The quenched one point functions presented in Refs.~\cite{Splittorff:2011bj,Kieburg:2011yg,Kieburg:2011uf} are given by the kernels $K_3^{(+,n)}(z,z)$ and  $K_5^{(+,n)}(z,z)$. The kernel $K_3^{(+,n)}(z,z)$ was denoted by $\rho_r(x)\delta(y)+\rho_c(z)/2$ in Refs.~\cite{Kieburg:2011yg,Kieburg:2011uf} which is the sum of the distribution of the right handed modes and the half of the distribution of the complex eigenvalues. Then the kernel $\Delta K_5^{(+,n+n_{\rm f})}(x,x)$ is equal to the chirality distribution over the real eigenvalues $\rho_\chi(x)$, see Refs.~\cite{Splittorff:2011bj,Kieburg:2011yg,Kieburg:2011uf}.

Again the Pfaffian determinant as well as  the identification with two-flavor partition functions of $D_\W$ carry over to the microscopic limit and, thus, to eigenvalue correlations of the Wilson-Dirac operator in the $\epsilon$-regime. The microscopic limit of the three kernels $K_1^{(+,n)}$, $K_3^{(+,n)}$ and $K_6^{(+,n)}$ are trivial corollaries of Eq.~\eref{3.5.5}. The derivation of this limit for the other three kernels will be made elsewhere \cite{KZV12}. Also the discussion of the results will not be done here.

Again we can look what our results mean in the notation of Refs.~\cite{Damgaard:2010cz,Akemann:2010zp,Splittorff:2011bj}. In the microscopic limit the following kernels are proportional to the two-flavor partition functions of the chiral Lagrangian
\begin{eqnarray}
  \fl \Delta K_4^{(+,\infty)}(z_1,z_2)&\propto&(\widehat{x}_1-\widehat{x}_2)\underset{\varepsilon_2\to0}{\underset{\varepsilon_1\to0}{\IM}}Z_{0/2}^{\nu}(\widehat{m}_6-\widehat{x}_1,\widehat{m}_6-\widehat{x}_2;\widehat{\lambda}_7,\widehat{\lambda}_7;\widehat{a}_8,\widehat{a}_{6/7}=0)\delta(\widehat{y}_1)\delta(\widehat{y}_2),\nonumber\\
  \fl&&\label{4.2.14}\\
  \fl \Delta K_5^{(+,\infty)}(z_1,z_2)&\propto&\underset{\varepsilon\to0}{\IM}\frac{Z_{1/1}^{\nu}(\widehat{m}_6-\widehat{z}_1,\widehat{m}_6-\widehat{x}_2;\widehat{\lambda}_7,\widehat{\lambda}_7;\widehat{a}_8,\widehat{a}_{6/7}=0)}{\widehat{x}_2-\widehat{z}_1}\delta(\widehat{y}_2),\label{4.2.15}\\
 \fl K_6^{(+,\infty)}(z_1,z_2)&\propto&(\widehat{z}_1-\widehat{z}_2)Z_{2/0}^{\nu}(\widehat{m}_6-\widehat{z}_1,\widehat{m}_6-\widehat{z}_2;\widehat{\lambda}_7,\widehat{\lambda}_7;\widehat{a}_8,\widehat{a}_{6/7}=0).\label{4.2.16}
\end{eqnarray}
The other kernels are only integral transforms of these three partition functions. As for $D_5$ we can create the case  $\widehat{a}_{6/7}\neq0$ by multiplying the expression~\eref{4.2.4} with the partition function of $N_{\rm f}$ fermionic flavors and integrating over Gaussian distributions of $\widehat{m}_6$ and $\widehat{\lambda}_7$. At the end we divide the resulting expression by the partition function with $N_{\rm f}$ fermionic flavor and with $\widehat{a}_{6/7}\neq0$.

\section{Conclusions}\label{sec5}

We derived the orthogonal and skew-orthogonal polynomials corresponding to the Hermitian as well as the non-Hermitian Wilson random matrix ensemble. The orthogonal polynomials are the Hermite polynomials from order $0$ to $\nu-1$ in both cases. They result from the $\nu$-dimensional GUE describing the broadening of the $\nu$ generic real modes which are at zero lattice spacing the zero modes. Such a GUE was already discovered in the chirality distribution over the real eigenvalues \cite{Splittorff:2011bj,Kieburg:2011yg,Kieburg:2011uf} as well as in the level density of the Hermitian Wilson random matrix ensemble and, thus, the Wilson Dirac operator \cite{Damgaard:2010cz,Akemann:2010zp,Damgaard:2011eg}. Surprisingly this GUE is already the universal result and is a dominant part in the eigenvalue correlations at small lattice spacing since it forms the Dirac delta functions at zero with weight $\nu$ in the continuum limit, see Refs.~\cite{Akemann:2010zp,Akemann:2011kj}.

The remaining spectrum is described by skew-orthogonal polynomials starting from order $\nu$. They describe the remaining spectrum apart from the $\nu$ generic real modes. In a unifying way we constructed these polynomials and derived recursion relations which enable us to obtain the odd polynomials by simply acting with a derivative operator on the even ones, cf. Eq.~\eref{3.2.0.1}. This derivative operator can be identified by a creation operator of a harmonic oscillator. Moreover we derived a Christoffel Darboux-like formula~\eref{3.3.5} which is equivalent to the partition function of two fermionic flavors, see Eq.~\eref{3.4.21}. The even skew-orthogonal polynomials are equal to one-flavor partition functions, see Eq.~\eref{3.4.6}. With help of this knowledge we were able to derive the Rodrigues formula~\eref{3.4.12} interpolating between the one of the Laguerre polynomials and the one of the Hermite polynomials.

As an application we considered the unquenched $k$-point correlation functions of the Hermitian and non-Hermitian Wilson random matrix ensemble. We derived a Pfaffian factorization in both cases. The one of the Hermitian matrix was already known before~\cite{Akemann:2011kj} but we traced the entries back to the two-flavor partition functions, see subsection~\ref{sec4.1}, which is a better expression for numerical evaluations. The Pfaffian of the non-Hermitian random matrix is a completely new result. We identified its kernels as two-flavor partition functions, too, see subsection~\ref{sec4.2}. These partition functions can be readily interpreted as correlations of the complex conjugated pairs, the real eigenvalues corresponding to the right handed modes and the average chirality over the real eigenvalues.

Although the random matrix $D_\W$ is non-Hermitian we did not need a Hermitization as it was introduced in Ref.~\cite{Feinberg:1997dk}. We circumvented this approach by splitting the kernels with bosonic flavors into two kinds of terms. One kind corresponds to the chirality over the real eigenvalues which exhibits no singularities in the bosonic determinants. The other term are integral transforms of partition functions with fermionic flavors instead of bosonic ones. Hence there are no problems of integrability anymore. Especially we have not to double the number of the bosonic dimensions in the superspace when applying the supersymmetry method.

The Pfaffian factorization as well as the  identification with two-flavor partition functions carry over to the microscopic limit and, thus, to the spectral properties of the Wilson-Dirac operator in the $\epsilon$-regime \cite{Sharpe:1998xm,RS02,Bar:2003mh,Sharpe06}. Hence the results shown in Sec.~\ref{sec4} are a good starting point for an analytical study of the Hermitian and non-Hermitian Wilson-Dirac operator. In particular the calculation of the individual eigenvalue distributions will benefit of the structure since a representation as Fredholm Pfaffians are  possible, see Ref.~\cite{Akemann:2012pn}. Fredholm determinants and Pfaffians are compact expressions simplifying the perturbative expansion of the gap probability in the $k$-point correlations function to obtain the individual eigenvalue distributions.

Moreover, the skew-orthogonal polynomials and the Christoffel Darboux-like formula also appearing as kernels of the Pfaffian determinants reduce to a quickly converging sum, see Eq.~\eref{3.5.4}, and two-fold integrals over phases, see Eqs.~\eref{3.5.5} and \eref{3.5.6}, respectively.

The Pfaffian of the $k$-point correlation function will persist in the continuum limit. It is in agreement with Ref.~\cite{Kieburg:2011ct} where a non-trivial Pfaffian determinant was derived for $\beta=2$ random matrix ensembles. A similar but not completely equivalent structure was derived in Refs.~\cite{Sin,ForSin} for $\beta=2$ ensembles, too. Hence Pfaffians seem to be more universal than the determinantal structures in the eigenvalues statistics of RMT.

The Pfaffian determinants we found reflect the breaking of the generic pairing of eigenvalues in the continuum limit to no reflection symmetry at all in the Hermitian case and the reflection symmetry at the real axis in the non-Hermitian one. For example at $a=0$ and fixed matrix $D_5$ we can say if $\lambda$ is an eigenvalue of $D_5$ then $-\lambda$ is also one. This is not anymore true at finite $a$.

\section*{Acknowledgements}

I thank Gernot Akemann, Kim Splittorff, Jacobus J.M. Verbaarschot and Savvas Zafeiropoulos for fruitful discussions and helpful comments. Furthermore I acknowledge financial support by the Alexander-von-Humboldt Foundation.

\appendix

\section{De Bruijn-like integration theorems}\label{app1}

We generalize the de Bruijn-like integration theorem \cite{Bru55} to an integrand which is a product of one determinant and one Pfaffian, see \ref{app1.1}, and of two determinants, see \ref{app1.2}.

\subsection{With a Pfaffian integrand}\label{app1.1}

Let $N_1$, $N_2$ and $N_3$ be three positive integers fulfilling the condition $2N_3, N_2\geq N_1>0$. We consider the following integral
\begin{eqnarray}
 I_1&=&\int \prod\limits_{j=1}^{N_1}d[z_j]\det\left[\begin{array}{c} \displaystyle\{A_c(z_b)\}\underset{1\leq c\leq N_2}{\underset{1\leq b\leq N_1}{\ }} \\ \displaystyle\{B_{bc}\}\underset{1\leq c\leq N_2}{\underset{1\leq b\leq N_2-N_1}{\ }} \end{array}\right]\nonumber\\
  &\times&\Pf\left[\begin{array}{cc} \displaystyle\{C(z_b,z_c)\}\underset{1\leq b,c\leq N_1}{\ } & \displaystyle\{D_c(z_b)\}\underset{1\leq c\leq 2N_3-N_1}{\underset{1\leq b\leq N_1}{\ }} \\ \displaystyle\{-D_b(z_c)\}\underset{1\leq c\leq N_1}{\underset{1\leq b\leq 2N_3-N_1}{\ }} & \displaystyle\{E_{bc}\}\underset{1\leq b,c\leq 2N_3-N_1}{\ } \end{array}\right].\nonumber\\
 \fl\label{a1.1.1}
\end{eqnarray}
The matrix $B$ is an arbitrary constant matrix whereas $E$ is an anti-symmetric constant matrix. The matrix valued functions $A$, $C$ and $D$ are sufficiently integrable and $C$ is anti-symmetric in its entries.

After an expansion of the first determinant in Eq.~\eref{a1.1.1} in the entries $A_c(z_b)$ we can integrate over the variables $z$ \cite{KieGuh09a}, i.e.
\begin{eqnarray}\label{a1.1.2}
 \fl &&I_1=\frac{1}{(N_2-N_1)!}\sum\limits_{\omega\in\widetilde{S}(N_2)}\sign\omega\det[B_{b\omega(c)}]\underset{N_1+1\leq c\leq N_2}{\underset{1\leq b\leq N_2-N_1}{\ }}\\
 \fl&\times&\hspace*{-0.1cm}\Pf\hspace*{-0.2cm}\left[\begin{array}{cc} \hspace*{-0.2cm}\displaystyle\left\{\int d[z] A_{\omega(b)}(z_1)A_{\omega(c)}(z_2)C(z_1,z_2)\right\}\underset{1\leq b,c\leq N_1}{\ } & \hspace*{-0.4cm}\displaystyle\left\{\int d[z]A_{\omega(b)}(z)D_c(z)\right\}\underset{1\leq c\leq 2N_3-N_1}{\underset{1\leq b\leq N_1}{\ }}\hspace*{-0.25cm} \\ \displaystyle\left\{-\int d[z]A_{\omega(c)}(z)D_b(z)\right\}\underset{1\leq c\leq N_1}{\underset{1\leq b\leq 2N_3-N_1}{\ }} & \displaystyle\{E_{bc}\}\underset{1\leq b,c\leq 2N_3-N_1}{\ }\hspace*{-0.25cm} \end{array}\right]\hspace*{-0.1cm}.\nonumber
\end{eqnarray}
The remaining determinant can be combined with the Pfaffian by the sum. Thereby we use the identity
\begin{eqnarray}\label{a1.1.2a}
 \fl\det[B_{b\omega(c)}]\underset{N_1+1\leq c\leq N_2}{\underset{1\leq b\leq N_2-N_1}{\ }}&=&(-1)^{(N_2-N_1)(N_2-N_1-1)/2}\\
 \fl&\times&\Pf\left[\begin{array}{cccc} 0 & \left\{B_{c\omega(b)}\right\}\underset{1\leq c\leq N_2-N_1}{\underset{N_1+1\leq b\leq N_2}{\ }} \\ \left\{-B_{b\omega(c)}\right\}\underset{N_1+1\leq c\leq N_2}{\underset{1\leq b\leq N_2-N_1}{\ }} & 0 \end{array} \right].\nonumber
\end{eqnarray}
This yields the result
\begin{eqnarray}
 \fl I_1&=&(-1)^{(N_2-N_1)(N_1+N_2-1)/2}N_1!\nonumber\\
 \fl&\times&\Pf\left[\begin{array}{c|c|c} \displaystyle\int d[z]A_{b}(z_1)A_{c}(z_2)C(z_1,z_2) & \displaystyle\int d[z]A_{b}(z)D_c(z) & B_{cb} \\ \hline \displaystyle-\int d[z]A_{c}(z)D_b(z) & \displaystyle E_{bc} & 0 \\ \hline -B_{bc} & 0 & 0 \end{array}\right].\label{a1.1.3}
\end{eqnarray}
The number of the first set of columns and rows is $N_2$, the one of the second set is $2N_3-N_1$ and the one of the third part $N_2-N_1$. Hence, we take the Pfaffian of a $2(N_2+N_3-N_1)\times2(N_2+N_3-N_1)$ anti-symmetric matrix.

\subsection{With a determinantal integrand}\label{app1.2}

Now we study the integral with a determinant instead of a Pfaffian, cf. Eq.~\eref{a1.1.1}, i.e.
\begin{eqnarray}
 I_2&=&\int\prod\limits_{j=1}^{N_R}d[z_{jR}]\prod\limits_{j=1}^{N_L}d[z_{jL}]\det\left[\begin{array}{c}\displaystyle\{A_c(z_{bR})\}\underset{1\leq c\leq N_1}{\underset{1\leq b\leq N_R}{\ }} \\ \displaystyle\{B_c(z_{bL})\}\underset{1\leq c\leq N_1}{\underset{1\leq b\leq N_L}{\ }} \\ \displaystyle\{C_{bc}\}\underset{1\leq c\leq N_1}{\underset{1\leq b\leq N_1-N_R-N_L}{\ }} \end{array}\right]\nonumber\\
 &\times&\det\left[\begin{array}{cc} \displaystyle\{D(z_{bR},z_{cL})\}\underset{1\leq c\leq N_L}{\underset{1\leq b\leq N_R}{\ }} & \displaystyle\{E_c(z_{bR})\}\underset{1\leq c\leq N_2-N_L}{\underset{1\leq b\leq N_R}{\ }}\\ \displaystyle\{F_b(z_{cL})\}\underset{1\leq c\leq N_L}{\underset{1\leq b\leq N_2-N_R}{\ }} & \displaystyle\{H_{bc}\}\underset{1\leq c\leq N_2-N_L}{\underset{1\leq b\leq N_2-N_R}{\ }} \end{array}\right].\label{a1.2.1}
\end{eqnarray}
The matrices $C$ and $H$ are arbitrary constant matrices and the matrix valued functions $A$, $B$, $D$, $E$ and $F$ are chosen such that the integrals exist. The positive integers $N_1$, $N_2$, $N_R$ and $N_L$ have the relations $N_1\geq N_R+N_L$ and $N_2\geq N_R,N_L$. Without loss of generality we can assume $N_L\geq N_R\geq 0$.

In the first step we split both matrices $F$ and $H$ into two blocks, i.e.
\begin{eqnarray}
 [F_b(z_{cL})]\underset{1\leq c\leq N_L}{\underset{1\leq b\leq N_2-N_R}{\ }}&=&\left[\begin{array}{c} \{F_{b1}(z_{cL})\}\underset{1\leq c\leq N_L}{\underset{1\leq b\leq N_L-N_R}{\ }} \\ \{F_{b2}(z_{cL})\}\underset{1\leq c\leq N_L}{\underset{1\leq b\leq N_2-N_L}{\ }} \end{array}\right]\,,\label{a1.2.2}\\
 \left[H_{bc}\right]\underset{1\leq c\leq N_2-N_L}{\underset{1\leq b\leq N_2-N_R}{\ }}&=&\left[\begin{array}{c} \{H_{bc1}\}\underset{1\leq c\leq N_2-N_L}{\underset{1\leq b\leq N_L-N_R}{\ }} \\ \{H_{bc2}\}\underset{1\leq b,c\leq N_2-N_L}{\ } \end{array}\right]\,,\label{a1.2.3}
\end{eqnarray}
where we assume that $H_2$ is invertible. Later on we will relax this restriction since $I_2$ is a polynomial in the constant matrices $C$ and $H$. We pull $H_2$ out of the second determinant and have
\begin{eqnarray}
 \fl I_2&=&\det H_2\int\prod\limits_{j=1}^{N_R}d[z_{jR}]\prod\limits_{j=1}^{N_L}d[z_{jL}]\det\left[\begin{array}{c} \displaystyle\{A_c(z_{bR})\}\underset{1\leq c\leq N_1}{\underset{1\leq b\leq N_R}{\ }} \\ \displaystyle\{B_c(z_{bL})\}\underset{1\leq c\leq N_1}{\underset{1\leq b\leq N_L}{\ }} \\ \displaystyle\{C_{bc}\}\underset{1\leq c\leq N_1}{\underset{1\leq b\leq N_1-N_R-N_L}{\ }} \end{array}\right]\label{a1.2.4}\\
 \fl&\times&\det\left[\begin{array}{cc} \displaystyle\left\{D(z_{bR},z_{cL})-\sum\limits_{1\leq i,j\leq N_2-N_L}E_i(z_{bR})\left(H_2^{-1}\right)_{ij}F_{j2}(z_{cL})\right\}\underset{1\leq c\leq N_L}{\underset{1\leq b\leq N_R}{\ }} \\ \displaystyle\left\{F_{b1}(z_{cL})-\sum\limits_{1\leq i,j\leq N_2-N_L}H_{bi1}\left(H_2^{-1}\right)_{ij}F_{j2}(z_{cL})\right\}\underset{1\leq c\leq N_L}{\underset{1\leq b\leq N_L-N_R}{\ }} \end{array}\right]\,.\nonumber
\end{eqnarray}
After an expansion in both determinants we obtain
\begin{eqnarray}\label{a1.2.5}
 \fl I_2&=&N_L!\det H_2\sum\limits_{\omega\in\widetilde{S}(N_1)}\sign\omega\prod\limits_{b=1}^{N_R}\int d[z_R]d[z_L]A_{\omega(b)}(z_{R})B_{\omega(b+N_R)}(z_{L})\\
 \fl&\times&\left[D(z_{R},z_{L})-\sum\limits_{1\leq i,j\leq N_2-N_L}E_i(z_{R})\left(H_2^{-1}\right)_{ij}F_{j2}(z_{L})\right]\nonumber\\
 \fl&\times&\prod\limits_{b=1}^{N_L-N_R}\int d[z_L] B_{\omega(b+2N_R)}(z_{L})\left[F_{b1}(z_{L})-\sum\limits_{1\leq i,j\leq N_2-N_L}H_{bi1}\left(H_2^{-1}\right)_{ij}F_{j2}(z_L)\right]\nonumber\\
 \fl&\times&\prod\limits_{b=1}^{N_1-N_R-N_L}C_{b\omega(b+N_R+N_L)}\,.\nonumber
\end{eqnarray}
Notice that the sum over the permutation of the second determinant can be absorbed into the first one which gives $N_L!$\,.

To shorten the notation we define the following matrices which are integrals over one or two variables
\begin{eqnarray}
 \fl O_{bc}&=&\int d[z_R]d[z_L]\left(A_{b}(z_{R})B_{c}(z_{L})-A_{c}(z_{R})B_{b}(z_{L})\right)D(z_{R},z_{L})\,,\label{a1.2.6}\\
 \fl P_{bc}&=&\int d[z_R]A_{b}(z_{R})E_c(z_{R})\,,\label{a1.2.7}\\
 \fl Q_{bc\alpha}&=&\int d[z_L]F_{b\alpha}(z_{L})B_{c}(z_{L})\quad,\ \alpha\in\{1,2\}\,.\label{a1.2.8}
\end{eqnarray}
Then the integral~\eref{a1.2.5} reads
\begin{eqnarray}\label{a1.2.9}
 \fl I_2&=&N_L!2^{-N_R}\det H_2\sum\limits_{\omega\in\widetilde{S}(N_1)}\sign\omega\\
 \fl&\times&\prod\limits_{b=1}^{N_R}\left[O_{\omega(b)\omega(b+N_R)}-\sum\limits_{1\leq i,j\leq N_2-N_L}\left(P_{\omega(b)i}Q_{j\omega(b+N_R)2}-P_{\omega(b+N_R)i}Q_{j\omega(b)2}\right)\left(H_2^{-1}\right)_{ij}\right]\nonumber\\
 \fl&\times&\prod\limits_{b=1}^{N_L-N_R}\left[Q_{b\omega(b+2N_R)1}-\hspace*{-0.3cm}\sum\limits_{1\leq i,j\leq N_2-N_L}H_{bi1}\left(H_2^{-1}\right)_{ij}Q_{j\omega(b+2N_R)2}\right]\prod\limits_{b=1}^{N_1-N_R-N_L}C_{b\omega(b+N_R+N_L)}\,.\nonumber
\end{eqnarray}
This sum can be represented as a Pfaffian, i.e.
\begin{eqnarray}\label{a1.2.10}
 \fl I_2&=&(-1)^{N_1(N_1-1)/2+N_R(N_R+1)/2}N_L!N_R!\det H_2\ \Pf\left[\begin{array}{c|c} R_{bc} & \begin{array}{c} C_{cb} \\ 0 \end{array} \\ \hline \begin{array}{cc} -C_{bc} & 0 \end{array} & 0 \end{array}\right]
\end{eqnarray}
with
\begin{eqnarray}\label{a1.2.11}
 \fl[R_{bc}]&=&\left(\begin{array}{cc} O_{bc} & Q_{cb1} \\ -Q_{bc1} & 0 \end{array}\right)\\
 \fl&+& \sum\limits_{1\leq i,j\leq N_2-N_L}\left(\begin{array}{cc} Q_{ib2} & P_{bi} \\ 0 & -H_{bi1} \end{array}\right)\left(\begin{array}{cc} 0 & \left(H_2^{-1}\right)_{ji} \\ -\left(H_2^{-1}\right)_{ij} & 0 \end{array}\right)\left(\begin{array}{cc} Q_{jc2} & 0 \\ P_{cj} & -H_{cj1} \end{array}\right)\,.\nonumber
\end{eqnarray}
Pushing the determinant of $H_2$ into the Pfaffian we have the final result
\begin{eqnarray}\label{a1.2.12}
 \fl I_2&=&(-1)^{N_1(N_1-1)/2+N_R(N_R+1)/2+(N_2-N_L)(N_2-N_L+1)/2}N_L!N_R!\\
 \fl&\times&\Pf\left[\begin{array}{c|c|c|c|c} O_{bc} & Q_{cb1} & C_{cb} & Q_{cb2} & P_{bc} \\ \hline -Q_{bc1} & 0 & 0 & 0 & -H_{bc1} \\ \hline -C_{bc} & 0 & 0 & 0 & 0 \\ \hline - Q_{bc2} & 0 & 0 & 0 & -H_{bc2} \\ \hline -P_{cb} & H_{cb1} & 0 & H_{cb2} & 0 \end{array}\right]\nonumber\\
 \fl&=&(-1)^{N_1(N_1-1)/2+N_R(N_R+1)/2+(N_2-N_L)(N_2-N_L+1)/2}N_L!N_R!\nonumber\\
 \fl&\times&\Pf\left[\begin{array}{c|c|c|c} O_{bc} & Q_{cb}  & P_{bc} & C_{cb} \\ \hline -Q_{bc} & 0 & -H_{bc} & 0 \\ \hline -P_{cb} & H_{cb} & 0 & 0  \\ \hline -C_{bc} & 0 & 0 & 0 \end{array}\right]\nonumber
\end{eqnarray}
The dimensions of rows and columns are from top to bottom and left to right $(N_1,N_2-N_R,N_2-N_L,N_1-N_R-N_L)$. In Eq.~\eref{a1.2.12} we drop the invertibility of the matrix $H_2$ because $I_2$ is a polynomial of this matrix.

\section{Derivation of the coefficients in the recursion relation}\label{app2}

In \ref{app2.1} we show that the recursion relations of the polynomials $q_{\nu+l}^{(\pm)}$ take the form~\eref{3.2.0.1} and \eref{3.2.0.2}. The coefficients $\epsilon_l^{(\pm)}$ and $\widetilde{\epsilon}_l^{(\pm)}$ are derived in \ref{app2.2}.

\subsection{The general form}\label{app2.1}

In the first step we take the scalar product~\eref{2.1.6} of Eq.~\eref{3.2.12} with $p_k^{(\pm)}$, $0\leq k\leq\nu-1$. We find
\begin{eqnarray}\label{3.2.13}
 \fl\beta^{(\pm)}_{lk}&=&\frac{1}{h_k}\langle\widetilde{D}^{(\pm)}q^{(\pm)}_{\nu+l}|p_k^{(\pm)}\rangle_{g_1^{(\pm)}}=-\frac{1}{h_k}\langle q^{(\pm)}_{\nu+l}| p_k^{(\pm)\,\prime}\rangle_{g_1^{(\pm)}}+\frac{\mu_\rt\mp\mu_\lt}{2h_k}\langle q^{(\pm)}_{\nu+l}| p_k^{(\pm)}\rangle_{g_1^{(\pm)}}=0,
\end{eqnarray}
Thereby we used the fact that $q^{(\pm)}_{\nu+l}$ and $p_k^{(\pm)}$, $0\leq k\leq\nu-1$, are orthogonal to each other.

The monic normalization of the polynomials  enforces the condition
\begin{eqnarray}\label{3.2.14}
 \alpha^{(\pm)}_{l,l+1}&=&-\frac{n}{a^2}.
\end{eqnarray}
For the other conditions we take the anti-symmetric product of Eq.~\eref{3.2.12} with $q^{(\pm)}_{\nu+l^\prime}$. Let $l=2i$ and $l^\prime=2i^\prime$ we have
\begin{eqnarray}
 \left( \widetilde{D}^{(\pm)} q^{(\pm)}_{\nu+2i}|q^{(\pm)}_{\nu+2i^\prime}\right)_{g^{(\pm)}_2}&=&-\left( q^{(\pm)}_{\nu+2i}|\widetilde{D}^{(\pm)} q^{(\pm)}_{\nu+2i^\prime}\right)_{g^{(\pm)}_2}\nonumber\\
 -o^{(\pm)}_{i^\prime}\alpha^{(\pm)}_{2i,2i^\prime+1}\Theta_{i-i^\prime}&=&-o^{(\pm)}_{i}\alpha^{(\pm)}_{2i^\prime,2i+1}\Theta_{i^\prime-i},\label{3.2.15}
\end{eqnarray}
where the integrated Kronecker delta is
\begin{eqnarray}\label{3.2.16}
 \Theta_l=\left\{\begin{array}{cl} 1, & l\in\mathbb{N}_0, \\ 0, & {\rm else}.\end{array}\right.
\end{eqnarray}
Combining Eqs.~\eref{3.2.14} and \eref{3.2.15} we find
\begin{eqnarray}\label{3.2.17}
 \alpha^{(\pm)}_{2i,2i^\prime+1}&=&\alpha^{(\pm)}_{2i^\prime,2i+1}=-\frac{n}{a^2}\delta_{ii^\prime}.
\end{eqnarray}
With $l=2i+1$ and $l^\prime=2i^\prime+1$ we get another relation
\begin{eqnarray}
 \left( \widetilde{D}^{(\pm)} q^{(\pm)}_{\nu+2i+1}|q^{(\pm)}_{\nu+2i^\prime+1}\right)_{g^{(\pm)}_2}&=&-\left( q^{(\pm)}_{\nu+2i+1}|\widetilde{D}^{(\pm)} q^{(\pm)}_{\nu+2i^\prime+1}\right)_{g^{(\pm)}_2}\nonumber\\
 o^{(\pm)}_{i^\prime}\alpha^{(\pm)}_{2i+1,2i^\prime}\Theta_{i-i^\prime+1}&=&o^{(\pm)}_{i}\alpha^{(\pm)}_{2i^\prime+1,2i}\Theta_{i^\prime-i+1},\label{3.2.18}
\end{eqnarray}
With this we conclude
\begin{eqnarray}\label{3.2.19}
 \alpha^{(\pm)}_{2i+1,2i^\prime}=\left\{\begin{array}{cl} \displaystyle\underset{}{-\frac{n}{a^2}\frac{o^{(\pm)}_{i}}{o^{(\pm)}_{i-1}}}, & i^\prime=i-1, \\ \epsilon^{(\pm)}_i, & i^\prime=i, \\ \displaystyle-\frac{n}{a^2}, & i^\prime=i+1, \\ 0, & {\rm else}. \end{array}\right.
\end{eqnarray}
The constants $\epsilon^{(\pm)}_i$ cannot be specified by Eq.~\eref{3.2.18}.

The last relation which we get by the skew-orthogonality of the polynomials is the one for the choice $l=2i$ and $l^\prime=2i^\prime+1$, i.e.
\begin{eqnarray}
 \left( \widetilde{D}^{(\pm)} q^{(\pm)}_{\nu+2i}|q^{(\pm)}_{\nu+2i^\prime+1}\right)_{g^{(\pm)}_2}&=&-\left( q^{(\pm)}_{\nu+2i}|\widetilde{D}^{(\pm)} q^{(\pm)}_{\nu+2i^\prime+1}\right)_{g^{(\pm)}_2}\nonumber\\
 o^{(\pm)}_{i^\prime}\alpha^{(\pm)}_{2i,2i^\prime}\Theta_{i-i^\prime}&=&-o^{(\pm)}_{i}\alpha^{(\pm)}_{2i^\prime+1,2i+1}\Theta_{i^\prime-i},\label{3.2.20}
\end{eqnarray}
The identity yields
\begin{eqnarray}\label{3.2.21}
 \alpha^{(\pm)}_{2i,2i^\prime}=-\alpha^{(\pm)}_{2i+1,2i^\prime+1}=\widetilde{\epsilon}^{(\pm)}_i\delta_{ii^\prime}.
\end{eqnarray}
Again the constants $\widetilde{\epsilon}^{(\pm)}_i$ have to be determined.

Collecting the intermediate results~\eref{3.2.13}, \eref{3.2.17}, \eref{3.2.19} and \eref{3.2.21} the expansion~\eref{3.2.12} reduces to the results~\eref{3.2.0.1} and \eref{3.2.0.2}.
The derivation of the constants $\epsilon^{(\pm)}_l$ and $\widetilde{\epsilon}^{(\pm)}_l$ remains.

\subsection{The coefficients $\epsilon_l^{(\pm)}$ and $\widetilde{\epsilon}_l^{(\pm)}$}\label{app2.2}

Considering the anti-symmetric product of Eq.~\eref{3.2.0.1} with $p_{\nu+2l+1}^{(\pm)}$ we find
\begin{eqnarray}
 \fl\left( \widetilde{D} ^{(\pm)}q^{(\pm)}_{\nu+2l}|p_{\nu+2l+1}^{(\pm)}\right)_{g^{(\pm)}_2}&=&\left(\left.-\frac{n}{a^2}q^{(\pm)}_{\nu+2l+1}+\widetilde{\epsilon}^{(\pm)}_lq^{(\pm)}_{\nu+2l}\right|p_{\nu+2l+1}^{(\pm)}\right)_{g^{(\pm)}_2}\nonumber\\
 \fl-\left( q^{(\pm)}_{\nu+2l}| \widetilde{D}^{(\pm)}p_{\nu+2l+1}^{(\pm)}\right)_{g^{(\pm)}_2}&=&o^{(\pm)}_{l}\widetilde{\epsilon}^{(\pm)}_l\nonumber\\
 \fl\left( q^{(\pm)}_{\nu+2l}\left| \frac{n}{a^2}p_{\nu+2l+2}^{(\pm)}-\frac{\mu_\rt\mp\mu_\lt}{2} p_{\nu+2l+1}^{(\pm)}\right.\right)_{g^{(\pm)}_2}&=&o^{(\pm)}_{l}\widetilde{\epsilon}^{(\pm)}_l\nonumber\\
 \fl\frac{n}{a^2}\left( q^{(\pm)}_{\nu+2l}|p_{\nu+2l+2}^{(\pm)}\right)_{g^{(\pm)}_2}-o^{(\pm)}_{l}\frac{\mu_\rt\mp\mu_\lt}{2}&=&o^{(\pm)}_{l}\widetilde{\epsilon}^{(\pm)}_l.\label{3.2.24}
\end{eqnarray}
A similar calculation can be done with the scalar product of Eq.~\eref{3.2.0.1} with $p_{\nu+2l}^{(\pm)}$
\begin{eqnarray}
 \fl\langle \widetilde{D}^{(\pm)} q^{(\pm)}_{\nu+2l}|p_{\nu+2l}^{(\pm)}\rangle_{g_1^{(\pm)}}&=&\langle \left.-\frac{n}{a^2}q^{(\pm)}_{\nu+2l+1}+\widetilde{\epsilon}^{(\pm)}_lq^{(\pm)}_{\nu+2l}\right|p_{\nu+2l}^{(\pm)}\rangle_{g_1^{(\pm)}}\nonumber\\
 \fl-\langle q^{(\pm)}_{\nu+2l}| p^{(\pm)\,\prime}_{\nu+2l}\rangle_{g_1^{(\pm)}}+\frac{\mu_\rt\mp\mu_\lt}{2}\langle q^{(\pm)}_{\nu+2l}| p_{\nu+2l}^{(\pm)}\rangle_{g_1^{(\pm)}}&=&h_{\nu+2l}\widetilde{\epsilon}^{(\pm)}_l\nonumber\\
 \fl-(\nu+2l)\langle q^{(\pm)}_{\nu+2l}| p_{\nu+2l-1}^{(\pm)}\rangle_{g_1^{(\pm)}}+h_{\nu+2l}\frac{\mu_\rt\mp\mu_\lt}{2}&=&h_{\nu+2l}\widetilde{\epsilon}^{(\pm)}_l\nonumber\\
 \fl\frac{(\nu+2l)h_{\nu+2l-1}}{o^{(\pm)}_{l-1}}\left( q^{(\pm)}_{\nu+2l-2}| p_{\nu+2l}^{(\pm)}\right)_{g^{(\pm)}_2}+h_{\nu+2l}\frac{\mu_\rt\mp\mu_\lt}{2}&=&h_{\nu+2l}\widetilde{\epsilon}^{(\pm)}_l.\label{3.2.25}
\end{eqnarray}
In both calculations we employed the definition~\eref{3.1.4}, the orthogonality of the polynomials $p_l^{(\pm)}$ and Eqs.~\eref{3.2.1}, \eref{3.2.8}, \eref{3.2.10} and \eref{3.2.11}. The combination of both results yields the recursion relation
\begin{eqnarray}\label{3.2.26}
 \frac{\left( q^{(\pm)}_{\nu+2l}| p_{\nu+2l+2}^{(\pm)}\right)_{g^{(\pm)}_2}}{o^{(\pm)}_{l}}=\frac{a^2}{n}(\mu_\rt\mp\mu_\lt)+\frac{\left( q^{(\pm)}_{\nu+2l-2}| p_{\nu+2l}^{(\pm)}\right)_{g^{(\pm)}_2}}{o^{(\pm)}_{l-1}}.
\end{eqnarray}
The starting point of this recursion is $l=0$. Due to the definition~\eref{3.1.4} we know that $q^{(\pm)}_\nu=p_\nu^{(\pm)}$ and $q^{(\pm)}_{\nu+1}=p_{\nu+1}^{(\pm)}$. We conclude
\begin{eqnarray}
 \left( q^{(\pm)}_{\nu}| p_{\nu+2}^{(\pm)}\right)_{g^{(\pm)}_2}&=&\left( p_{\nu}^{(\pm)}| p_{\nu+2}^{(\pm)}\right)_{g^{(\pm)}_2}\nonumber\\
 &=&\left( p_{\nu}^{(\pm)}\left| \left(-\frac{a^2}{n}\widetilde{D}^{(\pm)}+\frac{a^2(\mu_\rt\mp\mu_\lt)}{2n}\right)p_{\nu+1}^{(\pm)}\right.\right)_{g^{(\pm)}_2}\nonumber\\
 &=&\left(\left.\left(\frac{a^2}{n}\widetilde{D}^{(\pm)}+\frac{a^2(\mu_\rt\mp\mu_\lt)}{2n}\right) p_{\nu}^{(\pm)}\right|p_{\nu+1}^{(\pm)} \right)_{g^{(\pm)}_2}\nonumber\\
 &=&\left(-p_{\nu+1}^{(\pm)}+\frac{a^2}{n}(\mu_\rt\mp\mu_\lt) p_{\nu}^{(\pm)}|p_{\nu+1}^{(\pm)} \right)_{g^{(\pm)}_2}\nonumber\\
 &=&\frac{a^2 o^{(\pm)}_{0}}{n}(\mu_\rt\mp\mu_\lt).\label{3.2.27}
\end{eqnarray}
Hence, we can solve the recursion and find
\begin{eqnarray}
 \left( q^{(\pm)}_{\nu+2l}| p_{\nu+2l+2}^{(\pm)}\right)_{g^{(\pm)}_2}&=&\frac{(l+1)a^2 o^{(\pm)}_{l}}{n}(\mu_\rt\mp\mu_\lt),\label{3.2.28}\\
 \widetilde{\epsilon}^{(\pm)}_l&=&(2l+1)\frac{\mu_\rt\mp\mu_\lt}{2}.\label{3.2.29}
\end{eqnarray}
In a similar way we derive an identity for the constants $\epsilon^{(\pm)}_l$. We take the scalar product of Eq.~\eref{3.2.0.2} with $p_{\nu+2l}^{(\pm)}$ and obtain
\begin{eqnarray}
 \fl&&\epsilon^{(\pm)}_l=\frac{1}{h_{\nu+2l}}\left(\langle \widetilde{D}^{(\pm)}q_{\nu+2l+1}^{(\pm)}|p_{\nu+2l}^{(\pm)}\rangle_{g_1^{(\pm)}}+\frac{n}{a^2}\langle q_{\nu+2l+2}^{(\pm)}|p_{\nu+2l}^{(\pm)}\rangle_{g_1^{(\pm)}}\right)\label{3.2.30}\\
 \fl&=&\frac{1}{h_{\nu+2l}}\left(-\langle q_{\nu+2l+1}^{(\pm)}|p_{\nu+2l}^{(\pm)\,\prime}\rangle_{g_1^{(\pm)}}+\frac{\mu_\rt\mp\mu_\lt}{2}\langle q_{\nu+2l+1}^{(\pm)}|p_{\nu+2l}^{(\pm)}\rangle_{g_1^{(\pm)}}\right.\nonumber\\
  \fl&+&\left.\frac{n}{(\nu+2l+1)a^2}\langle q_{\nu+2l+2}^{(\pm)}|p_{\nu+2l}^{(\pm)\,\prime}\rangle_{g_1^{(\pm)}}\right)\nonumber\\
 \fl&=&\frac{1}{h_{\nu+2l}}\left(-(\nu+2l)\langle q_{\nu+2l+1}^{(\pm)}|p_{\nu+2l-1}^{(\pm)}\rangle_{g_1^{(\pm)}}-\frac{n}{(\nu+2l+1)a^2}\langle \widetilde{D}^{(\pm)}q_{\nu+2l+2}^{(\pm)}|p_{\nu+2l}^{(\pm)}\rangle_{g_1^{(\pm)}}\right.\nonumber\\
 \fl&-&\left.\frac{n}{(\nu+2l+1)a^2}\frac{\mu_\rt\mp\mu_\lt}{2}\langle q_{\nu+2l+2}^{(\pm)}|p_{\nu+2l+1}^{(\pm)}\rangle_{g_1^{(\pm)}}\right)\nonumber\\
 \fl&=&\frac{n}{a^2}\left(\frac{\langle q_{\nu+2l+3}^{(\pm)}|p_{\nu+2l+1}^{(\pm)}\rangle_{g_1^{(\pm)}}}{h_{\nu+2l+1}}-\frac{\langle q_{\nu+2l+1}^{(\pm)}|p_{\nu+2l-1}^{(\pm)}\rangle_{g_1^{(\pm)}}}{h_{\nu+2l-1}}\right)\nonumber\\
 \fl&-&\frac{(l+1)(\mu_\rt\mp\mu_\lt)}{h_{\nu+2l+1}}\langle q_{\nu+2l+2}^{(\pm)}|p_{\nu+2l+1}^{(\pm)}\rangle_{g_1^{(\pm)}}\nonumber\\
 \fl&=&\frac{n}{a^2}\left(\frac{\langle q_{\nu+2l+3}^{(\pm)}|p_{\nu+2l+1}^{(\pm)}\rangle_{g_1^{(\pm)}}}{h_{\nu+2l+1}}-\frac{\langle q_{\nu+2l+1}^{(\pm)}|p_{\nu+2l-1}^{(\pm)}\rangle_{g_1^{(\pm)}}}{h_{\nu+2l-1}}\right)\nonumber\\
 \fl&-&\frac{(l+1)(\mu_\rt\mp\mu_\lt)}{o_{l}^{(\pm)}}\left( q_{\nu+2l}^{(\pm)}|p_{\nu+2l+2}^{(\pm)}\right)_{g_2^{(\pm)}}\nonumber\\
 \fl&=&\frac{n}{a^2}\left(\frac{\langle q_{\nu+2l+3}^{(\pm)}|p_{\nu+2l+1}^{(\pm)}\rangle_{g_1^{(\pm)}}}{h_{\nu+2l+1}}-\frac{\langle q_{\nu+2l+1}^{(\pm)}|p_{\nu+2l-1}^{(\pm)}\rangle_{g_1^{(\pm)}}}{h_{\nu+2l-1}}\right)-\frac{(l+1)^2(\mu_\rt\mp\mu_\lt)^2a^2}{n}\nonumber
\end{eqnarray}
For a further simplification we need more information, i.e. we have to perform the integral $\langle q_{\nu+2l+1}^{(\pm)}|p_{\nu+2l-1}^{(\pm)}\rangle_{g_1^{(\pm)}}$ for all $l\in\mathbb{N}_0$.

\section{Derivation of the Christoffel Darboux-like formula}\label{app5}

Let $z_1$ and $z_2$  be restricted to the real axis, i.e. $z_{1/2}=x_{1/2}$. The action of the sum of the two differential operators $\widetilde{D}^{(\pm)}$ with respect to $x_1$ and $x_2$ on $\Sigma_{n-1}^{(\pm)}$ is
\begin{eqnarray}
 \fl&&\left(\frac{\partial}{\partial x_1}+\frac{\partial}{\partial x_2}-\frac{n}{a^2}(x_1+x_2)+(\mu_\rt\pm\mu_\lt)\right)\Sigma_{n-1}^{(\pm)}(x_1,x_2)\nonumber\\
 \fl&=&\sum_{l=0}^{n-1}\frac{1}{o_l^{(\pm)}}\left(\widetilde{\epsilon}^{(\pm)}_l\det\left[\begin{array}{cc} q_{\nu+2l}^{(\pm)}(x_1) & q_{\nu+2l+1}^{(\pm)}(x_1) \\ q_{\nu+2l}^{(\pm)}(x_2) & q_{\nu+2l+1}^{(\pm)}(x_2) \end{array}\right]+\frac{n}{a^2}\det\left[\begin{array}{cc} q_{\nu+2l+2}^{(\pm)}(x_1) & q_{\nu+2l}^{(\pm)}(x_1) \\ q_{\nu+2l+2}^{(\pm)}(x_2) & q_{\nu+2l}^{(\pm)}(x_2) \end{array}\right]\right.\nonumber\\
 \fl&+&\left.\widetilde{\epsilon}^{(\pm)}_l\det\left[\begin{array}{cc} q_{\nu+2l+1}^{(\pm)}(x_1) & q_{\nu+2l}^{(\pm)}(x_1) \\ q_{\nu+2l+1}^{(\pm)}(x_2) & q_{\nu+2l}^{(\pm)}(x_2) \end{array}\right]+\frac{no^{(\pm)}_l}{a^2o^{(\pm)}_{l-1}}\det\left[\begin{array}{cc} q_{\nu+2l-2}^{(\pm)}(x_1) & q_{\nu+2l}^{(\pm)}(x_1) \\ q_{\nu+2l-2}^{(\pm)}(x_2) & q_{\nu+2l}^{(\pm)}(x_2) \end{array}\right]\right)\nonumber\\
 \fl&=&-\frac{n}{a^2o_{n-1}^{(\pm)}}\det\left[\begin{array}{cc} q_{\nu+2n-2}^{(\pm)}(x_1) & q_{\nu+2n}^{(\pm)}(x_1) \\ q_{\nu+2n-2}^{(\pm)}(x_2) & q_{\nu+2n}^{(\pm)}(x_2) \end{array}\right].\label{3.3.3}
\end{eqnarray}
Let $X=(x_1+x_2)/2$ and $\Delta x=(x_1-x_2)/2$. Then we rewrite the differential equation to
\begin{eqnarray}\label{3.3.4}
 \fl&&\frac{\partial}{\partial X}\exp\left[-\frac{n}{a^2}X^2+(\mu_\rt\pm\mu_\lt)X\right]\Sigma_{n-1}^{(\pm)}(X+\Delta x,X-\Delta x)\\
 \fl&=&-\frac{n}{a^2o_{n-1}^{(\pm)}}\exp\left[-\frac{n}{a^2}X^2+(\mu_\rt\pm\mu_\lt)X\right]\det\left[\begin{array}{cc} q_{\nu+2n-2}^{(\pm)}(X+\Delta x) & q_{\nu+2n}^{(\pm)}(X+\Delta x) \\ q_{\nu+2n-2}^{(\pm)}(X-\Delta x) & q_{\nu+2n}^{(\pm)}(X-\Delta x) \end{array}\right].\nonumber
\end{eqnarray}
In the next step we integrate this equation from $X$ to $\infty$ and take into account that the upper boundary vanishes due to the Gaussian. This yields Eq.~\eref{3.3.5} for real entries.  The restriction to real $z_1$ and $z_2$ can be relaxed since the integrand is absolutely integrable.

\section{Derivation of Eq.~\eref{3.4.11}}\label{app6}

We consider Eq.~\eref{3.4.6}. The characteristic polynomial in $Z_1^{(l,\nu,\pm)}$ can be raised into the exponent by a Gaussian integral over a complex vector of Grassmann (anti-commuting) variables,
\begin{eqnarray}\label{3.4.7}
 \fl\xi=\left[\begin{array}{c} \xi_{\rt} \\ \xi_{\lt} \end{array}\right]=\left[\begin{array}{c} \xi_{1\rt} \\ \vdots \\ \xi_{l\rt} \\ \xi_{1\lt} \\ \vdots \\ \xi_{l+\nu,\lt} \end{array}\right],\ \xi^\dagger=\left[\begin{array}{cc} \xi_{\rt}^\dagger, & \xi_{\lt}^\dagger \end{array}\right]=\left[\begin{array}{cccccc} \hspace*{-0.1cm}\xi_{1\rt}^*, & \hspace*{-0.1cm}\cdots &\hspace*{-0.1cm},\,\xi_{l\rt}^*, &\hspace*{-0.2cm} \xi_{1\lt}^*, & \hspace*{-0.1cm}\cdots &\hspace*{-0.1cm},\,\xi_{l+\nu,\lt}^* \hspace*{-0.1cm}\end{array}\right].
\end{eqnarray}
The integration is defined by
\begin{eqnarray}\label{3.4.8}
 \int\xi_id\xi_j=\int\xi_i^*d\xi_j^*=\frac{1}{\sqrt{2\pi}}\ {\rm and}\ \int d\xi_j=\int d\xi_j^*=0.
\end{eqnarray}
Moreover we employ the conjugation of the second kind, i.e.
\begin{eqnarray}\label{3.4.9}
 (\xi_i^*)^*=-\xi_i\ {\rm and}\ (\xi_i\xi_j)^*=\xi_i^*\xi_j^*.
\end{eqnarray}
Good introductions in the standard techniques of supersymmetry can be found in Refs.~\cite{VWZ85,Ber87}.

We find
\begin{eqnarray}
 \fl q_{\nu+2l}^{(\pm)}(z)&\propto&\int d[H]d[\xi]\exp\left[-\frac{n}{2a^2}(\tr A^2+\tr B^2)-n\tr WW^\dagger+\tr A(\mu_\rt+\xi_\rt\xi_\rt^\dagger)\right]\label{3.4.10}\\
 \fl &\times&\exp\left[\tr B(\mu_\lt+\xi_\lt\xi_\lt^\dagger)+\tr W\xi_\lt\xi_\rt^\dagger-\tr W^\dagger\xi_\rt\xi_\lt^\dagger+z(\xi_\rt^\dagger\xi_\rt\pm\xi_\lt^\dagger\xi_\lt)\right]\nonumber\\
 \fl&\propto&\int d[\xi]\exp\left[\frac{a^2}{2n}(\tr (\mu_\rt+\xi_\rt\xi_\rt^\dagger)^2+\tr (\mu_\lt+\xi_\lt\xi_\lt^\dagger)^2)-\frac{1}{n}\tr \xi_\rt\xi_\lt^\dagger\xi_\lt\xi_\rt^\dagger\right]\nonumber\\
 \fl &\times&\exp\left[z(\xi_\rt^\dagger\xi_\rt\pm\xi_\lt^\dagger\xi_\lt)\right]\nonumber\\
 \fl&\propto&\int d[\xi]\exp\left[-\frac{a^2}{2n}((\xi_\rt^\dagger\xi_\rt)^2+(\xi_\lt^\dagger\xi_\lt)^2)+\frac{1}{n}\xi_\rt^\dagger\xi_\rt\xi_\lt^\dagger\xi_\lt\right]\nonumber\\
 \fl &\times&\exp\left[-\left(\frac{a^2\mu_\rt}{n}-z\right)\xi_\rt^\dagger\xi_\rt-\left(\frac{a^2\mu_\lt}{n}\mp z\right)\xi_\lt^\dagger\xi_\lt\right].\nonumber
\end{eqnarray}
With help of the superbosonization formula \cite{Som07,LSZ08} we express the integral over $\xi$ by a two-fold integral over two phases~\eref{3.4.11}.

\section{Derivation of Eq.~\eref{3.4.21}}\label{app7}

Equations~\eref{2.1.8} and \eref{2.2.7} with the parameters $k=k_\rt=k_\lt=0$ and $N_{\rm f}=2$ read
\begin{eqnarray}\label{3.4.18}
 \fl&& Z_2^{(n,\nu,\pm)}(-z_1,-z_2)\propto\frac{1}{z_1-z_2}\\
 \fl&\times&\Pf\left[\begin{array}{c|c|c|c|c} 0 & 0 & h_i\delta_{ij} & p_i^{(\pm)}(z_1) & p_i^{(\pm)}(z_2) \\ \hline 0 & \begin{array}{cc} 0 & o_i^{(\pm)}\delta_{ij} \\ -o_i^{(\pm)}\delta_{ij} & 0 \end{array} & 0 & \begin{array}{c} q_{\nu+2i}^{(\pm)}(z_1) \\ q_{\nu+2i+1}^{(\pm)}(z_1) \end{array} & \begin{array}{c} q_{\nu+2i}^{(\pm)}(z_2) \\ q_{\nu+2i+1}^{(\pm)}(z_2) \end{array} \\ \hline -h_i\delta_{ij} & 0 & 0 & 0 & 0 \\ \hline -p_j^{(\pm)}(z_1) & \begin{array}{cc} -q_{\nu+2j}^{(\pm)}(z_1), & -q_{\nu+2j+1}^{(\pm)}(z_1) \end{array} & 0 & 0 & 0 \\ \hline -p_j^{(\pm)}(z_2) & \begin{array}{cc} -q_{\nu+2j}^{(\pm)}(z_2), & -q_{\nu+2j+1}^{(\pm)}(z_2) \end{array} & 0 & 0 & 0 \end{array}\right]\nonumber
\end{eqnarray}
combined with the derived knowledge in subsection~\ref{sec3.1}. This Pfaffian can be expanded in the normalization constants $h_j$. Then the polynomials $p_l^{(\pm)}$, $0\leq l<\nu-1$, drop out. Furthermore, we make use of identity~\eref{3.4.19} and get
\begin{eqnarray}
 \fl&& Z_2^{(n,\nu,\pm)}(-z_1,-z_2)\propto\frac{1}{z_1-z_2}\nonumber\\
 \fl&\times&\Pf\left[\sum\limits_{j=0}^{n}\frac{1}{o_j^{(\pm)}}\left(\begin{array}{cc} q_{\nu+2j}^{(\pm)}(z_1) & q_{\nu+2j+1}^{(\pm)}(z_1) \\ q_{\nu+2j}^{(\pm)}(z_2) & q_{\nu+2j+1}^{(\pm)}(z_2) \end{array}\right)\left(\begin{array}{cc} 0 & -1 \\ 1 & 0 \end{array}\right)\left(\begin{array}{cc} q_{\nu+2j}^{(\pm)}(z_1) &q_{\nu+2j}^{(\pm)}(z_2) \\ q_{\nu+2j+1}^{(\pm)}(z_1) &q_{\nu+2j+1}^{(\pm)}(z_2) \end{array}\right)\right]\nonumber\\
 \fl&\propto&\frac{1}{z_1-z_2}\sum\limits_{j=0}^{n}\frac{1}{o_j^{(\pm)}}\left(q_{\nu+2j}^{(\pm)}(z_1)q_{\nu+2j+1}^{(\pm)}(z_2)-q_{\nu+2j+1}^{(\pm)}(z_1)q_{\nu+2j}^{(\pm)}(z_2)\right).\label{3.4.20}
\end{eqnarray}
This result is proportional to the sum $\Sigma_{n}^{(\pm)}$, cf.~\eref{3.3.1}.

\section{Simplification of the kernels}\label{app3}

In \ref{app3.1} and \ref{app3.2} we simplify the kernels of $D_5$. Derivations of the kernels of $D_\W$ are given in \ref{app3.3}, \ref{app3.4} and \ref{app3.5}.

\subsection{The kernel $K_1^{(-,n)}$}\label{app3.1}

With help of Eqs.~(\ref{2.1.8}-\ref{2.1.9c}) it can be readily shown that
\begin{eqnarray}\label{a3.1.1}
 \fl K_1^{(-,n)}(x_1,x_2)&=&\frac{1}{(2n+\nu)!}\prod\limits_{j=0}^{\nu-1}\frac{1}{h_j}\prod\limits_{j=0}^{n-1}-\frac{1}{o_j^{(-)}}\int\limits_{\mathbb{R}^{2n+\nu}}d[\widetilde{x}]\Delta_{2n+\nu}(\widetilde{x})\\
  \fl&\times&\Pf\left[\begin{array}{c|cc|c} g_2^{(-)}(\widetilde{x}_i,\widetilde{x}_j) & g_2^{(-)}(\widetilde{x}_i,x_1) & g_2^{(-)}(\widetilde{x}_i,x_2) & \widetilde{x}_i^{j-1}g_1^{(-)}(\widetilde{x}_i) \\ \hline g_2^{(-)}(x_1,\widetilde{x}_j) & 0 & g_2^{(-)}(x_1,x_2) & x_1^{j-1}g_1^{(-)}(x_1) \\ g_2^{(-)}(x_2,\widetilde{x}_j) & g_2^{(-)}(x_2,x_1) & 0 & x_2^{j-1}g_1^{(-)}(x_2) \\ \hline -\widetilde{x}_j^{i-1}g_1^{(-)}(\widetilde{x}_j) & -x_1^{i-1}g_1^{(-)}(x_1) & -x_2^{i-1}g_1^{(-)}(x_2) & 0 \end{array}\right].\nonumber
\end{eqnarray}
The indices $i$ and $j$ run from $1$ to $2n+\nu$ in the first row and column and $1$ to $\nu$ in the last ones.

In the next step we extend the Vandermonde determinant by two Dirac delta functions such that the integration is over $2n+\nu+2$ variables,
\begin{eqnarray}\label{a3.1.2}
 \fl K_1^{(-,n)}(x_1,x_2)&=&\frac{(-1)^{(2n+\nu)(2n+\nu-1)/2}}{(2n+\nu+2)!}\prod\limits_{j=0}^{\nu-1}\frac{1}{h_j}\prod\limits_{j=0}^{n-1}-\frac{1}{o_j^{(-)}}\int\limits_{\mathbb{R}^{2n+\nu+2}}d[\widetilde{x}]\\
  \fl&\times&\det\left[\begin{array}{c} \widetilde{x}_j^{i-1} \\ \delta(\widetilde{x}_j-x_1) \\  \delta(\widetilde{x}_j-x_2) \end{array}\right]\Pf\left[\begin{array}{c|c} g_2^{(-)}(\widetilde{x}_i,\widetilde{x}_j) & \widetilde{x}_i^{j-1}g_1^{(-)}(\widetilde{x}_i) \\ \hline -\widetilde{x}_j^{i-1}g_1^{(-)}(\widetilde{x}_j) & 0 \end{array}\right].\nonumber
\end{eqnarray}
The two Dirac delta functions can be expressed by the imaginary parts of the Cauchy transforms in two variables. Using the identity~\cite{KieGuh09a}
\begin{eqnarray}\label{a3.1.3}
 \fl\det\left[\begin{array}{c} \widetilde{x}_j^{i-1} \\ \displaystyle\frac{1}{\widetilde{x}_j-x_1-\imath\varepsilon_1} \\  \displaystyle\frac{1}{\widetilde{x}_j-x_2-\imath\varepsilon_2} \end{array}\right]=(-1)^{(2n+\nu+2)(2n+\nu+1)/2}\frac{(x_1+\imath\varepsilon_1-x_2-\imath\varepsilon_2)\Delta_{2n+\nu+2}(\widetilde{x})}{\prod\limits_{j=1}^{2n+\nu+2}(\widetilde{x}_j-x_1-\imath\varepsilon_1)(\widetilde{x}_j-x_2-\imath\varepsilon_2)}
\end{eqnarray}
we find the expression
\begin{eqnarray}\label{a3.1.4}
 \fl K_1^{(-,n)}(x_1,x_2)&=&\frac{1}{(2n+\nu+2)!}\prod\limits_{j=0}^{\nu-1}\frac{1}{h_j}\prod\limits_{j=0}^{n-1}-\frac{1}{o_j^{(-)}}\frac{x_2-x_1}{\pi^2}\underset{\varepsilon_2\to0}{\underset{\varepsilon_1\to0}{\IM}}\ \int\limits_{\mathbb{R}^{2n+\nu+2}}d[\widetilde{x}]\\
  \fl&&\hspace*{-1cm}\times\frac{\Delta_{2n+\nu+2}(\widetilde{x})}{\prod\limits_{j=1}^{2n+\nu+2}(\widetilde{x}_j-x_1-\imath\varepsilon_1)(\widetilde{x}_j-x_2-\imath\varepsilon_2)}\Pf\left[\begin{array}{c|c} g_2^{(-)}(\widetilde{x}_i,\widetilde{x}_j) & \widetilde{x}_i^{j-1}g_1^{(-)}(\widetilde{x}_i) \\ \hline -\widetilde{x}_j^{i-1}g_1^{(-)}(\widetilde{x}_j) & 0 \end{array}\right].\nonumber
\end{eqnarray}
This result is the partition function of $D_5$ with two bosonic flavors, see Eq.~\eref{4.1.7}.

\subsection{The kernel $K_2^{(-,n)}$}\label{app3.2}

Again we start from an identity between the kernel and an integral weighted by the joint probability density $p_5$, see Eq.~\eref{2.6}, i.e.
\begin{eqnarray}\label{a3.2.1}
 \fl K_2^{(-,n)}(x_1,x_2)&=&\frac{1}{(2n+\nu-1)!}\prod\limits_{j=0}^{\nu-1}\frac{1}{h_j}\prod\limits_{j=0}^{n-1}-\frac{1}{o_j^{(-)}}\int\limits_{\mathbb{R}^{2n+\nu-1}}d[\widetilde{x}]\Delta_{2n+\nu}(\widetilde{x},x_1)\\
  \fl&\times&\Pf\left[\begin{array}{c|c|c} g_2^{(-)}(\widetilde{x}_i,\widetilde{x}_j) & g_2^{(-)}(\widetilde{x}_i,x_2) & \widetilde{x}_i^{j-1}g_1^{(-)}(\widetilde{x}_i) \\ \hline g_2^{(-)}(x_2,\widetilde{x}_j) & 0 & x_2^{j-1}g_1^{(-)}(x_2) \\ \hline -\widetilde{x}_j^{i-1}g_1^{(-)}(\widetilde{x}_j) & -x_2^{i-1}g_1^{(-)}(x_2) & 0 \end{array}\right].\nonumber
\end{eqnarray}
Notice that we integrate this time over $2n+\nu-1$ variables. Hence the range of the indices $i$ and $j$ is from $1$ to $2n+\nu-1$ in the first row and column and from $1$ to $\nu$ in the last ones.

The integral is extended to $2n+\nu$ variables by introducing a Dirac delta function,
\begin{eqnarray}\label{a3.2.2}
 \fl K_1^{(-,n)}(x_1,x_2)&=&\frac{(-1)^{(2n+\nu-1)(2n+\nu-2)/2}}{(2n+\nu+2)!}\prod\limits_{j=0}^{\nu-1}\frac{1}{h_j}\prod\limits_{j=0}^{n-1}-\frac{1}{o_j^{(-)}}\int\limits_{\mathbb{R}^{2n+\nu}}d[\widetilde{x}]\frac{\prod\limits_{j=1}^{2n+\nu}(\widetilde{x}_j-x_1)}{x_2-x_1}\\
  \fl&\times&\det\left[\begin{array}{c} \widetilde{x}_j^{i-1} \\  \delta(\widetilde{x}_j-x_2) \end{array}\right]\Pf\left[\begin{array}{c|c} g_2^{(-)}(\widetilde{x}_i,\widetilde{x}_j) & \widetilde{x}_i^{j-1}g_1^{(-)}(\widetilde{x}_i) \\ \hline -\widetilde{x}_j^{i-1}g_1^{(-)}(\widetilde{x}_j) & 0 \end{array}\right].\nonumber
\end{eqnarray}
We employ again the Cauchy integral as a representation of the Dirac delta function and an equation similar to Eq.~\eref{a3.1.3}. This yields the result~\eref{4.1.8} which is the partition function of $D_5$ with one fermionic flavor and one bosonic one.

\subsection{The kernel $K_2^{(+,n)}$}\label{app3.3}

Also for $D_\W$ the kernels have a representation as an integral over the eigenvalues weighted by the joint probability density $p_\W$~\eref{2.10},
\begin{eqnarray}\label{a3.3.1}
 \fl K_2^{(+,n)}(z_1,z_2)&=&\frac{(-1)^{n(n+1)/2+\nu(\nu+1)/2}}{n!(n+\nu)!}\prod\limits_{j=0}^{\nu-1}\frac{1}{h_j}\prod\limits_{j=0}^{n-1}\frac{1}{o_j^{(+)}}\int\limits_{\mathbb{C}^{2n+\nu}}d[\widetilde{z}]\Delta_{2n+\nu}(\widetilde{z})\\
  \fl&\times&\det\left[\begin{array}{c|c} g_2^{(+)}(\widetilde{z}_{i}^{(\rt)},\widetilde{z}_{j}^{(\lt)}) & g_2^{(+)}(\widetilde{z}_{i}^{(\rt)},z_1) \\ \hline g_2^{(+)}(z_2,\widetilde{z}_{j}^{(\lt)}) & g_2^{(+)}(z_2,z_1) \\ \hline -(\widetilde{x}_{j}^{(\lt)})^{i-1}g_1^{(+)}(\widetilde{x}_{j}^{(\lt)})\delta(\widetilde{y}_{j}^{(\lt)}) & -x_1^{i-1}g_1^{(+)}(x_1)\delta(y_1) \end{array}\right].\nonumber
\end{eqnarray}
The index $j$ in the first column takes the values $1$ to $n+\nu$ while $i$ goes from $1$ to $n$ in the first row and from $1$ to $\nu$ in the last one.

We expand the determinant in the row with the variable $z_2$ and have
\begin{eqnarray}\label{a3.3.2}
 \fl K_2^{(+,n)}(z_1,z_2)&=&g_2^{(+)}(z_2,z_1)\\
 \fl&+&\frac{(-1)^{n(n+1)/2+(\nu+2)(\nu+1)/2}}{n!(n+\nu-1)!}\prod\limits_{j=0}^{\nu-1}\frac{1}{h_j}\prod\limits_{j=0}^{n-1}\frac{1}{o_j^{(+)}}\hspace*{-0.1cm}\int\limits_{\mathbb{C}^{2n+\nu}}d[\widetilde{z}]d[\widehat{z}]g_2^{(+)}(z_2,\widehat{z})\nonumber\\
  \fl&\times&\Delta_{2n+\nu}(\widetilde{z},\widehat{z})\det\left[\begin{array}{c|c} g_2^{(+)}(\widetilde{z}_{i}^{(\rt)},\widetilde{z}_{j}^{(\lt)}) & g_2^{(+)}(\widetilde{z}_{i}^{(\rt)},z_1) \\ \hline -(\widetilde{x}_{j}^{(\lt)})^{i-1}g_1^{(+)}(\widetilde{x}_{j}^{(\lt)})\delta(\widetilde{y}_{j}^{(\lt)}) & -x_1^{i-1}g_1^{(+)}(x_1)\delta(y_1) \end{array}\right].\nonumber
\end{eqnarray}
The integration of the second term is extended by a Dirac delta function. However this distribution can only be symmetrized with respect to the $\widetilde{z}^{(\lt)}$ integration in contrast to the calculation in \ref{app3.1} and \ref{app3.2}. We add and subtract a Dirac delta function for the integration over $\widetilde{z}^{(\rt)}$. Collecting these steps we find
\begin{eqnarray}\label{a3.3.3}
 \fl K_2^{(+,n)}(z_1,z_2)&=&g_2^{(+)}(z_2,z_1)\\
 \fl&+&\frac{(-1)^{n(n+1)/2+\nu(\nu+1)/2}}{(n-1)!(n+\nu)!}\prod\limits_{j=0}^{\nu-1}\frac{1}{h_j}\prod\limits_{j=0}^{n-1}\frac{1}{o_j^{(+)}}\int\limits_{\mathbb{C}^{2n+\nu}}d[\widetilde{z}]d[\widehat{z}]g_2^{(+)}(z_2,\widehat{z})\nonumber\\
  \fl&\times&\Delta_{2n+\nu}(\widetilde{z},\widehat{z})\det\left[\begin{array}{c} g_2^{(+)}(z_1,\widetilde{z}_{j}^{(\lt)}) \\  g_2^{(+)}(\widetilde{z}_{i}^{(\rt)},\widetilde{z}_{j}^{(\lt)}) \\ -(\widetilde{x}_{j}^{(\lt)})^{i-1}g_1^{(+)}(\widetilde{x}_{j}^{(\lt)})\delta(\widetilde{y}_{j}^{(\lt)}) \end{array}\right]\nonumber\\
 \fl&+&\frac{(-1)^{n(n-1)/2+\nu}}{n!(n+\nu)!}\prod\limits_{j=0}^{\nu-1}\frac{1}{h_j}\prod\limits_{j=0}^{n-1}\frac{1}{o_j^{(+)}}\int\limits_{\mathbb{C}^{2n+\nu+1}}\hspace*{-0.2cm}d[\widetilde{z}]d[\widehat{z}]\frac{g_2^{(+)}(z_2,\widehat{z})}{z_1-\widehat{z}}\det(\widetilde{z}-\widehat{z}\eins_{2n+\nu})\nonumber\\
  \fl&\times&\det\left[\begin{array}{c|c} (\widetilde{z}_{i}^{(\rt)})^{j-1} & -\delta^{(2)}(\widetilde{z}_{i}^{(\rt)}-z_1) \\ \hline (\widetilde{z}_{i}^{(\lt)})^{j-1} &  \delta^{(2)}(\widetilde{z}_{i}^{(\lt)}-z_1) \end{array}\right]\det\left[\begin{array}{c} g_2^{(+)}(\widetilde{z}_{i}^{(\rt)},\widetilde{z}_{j}^{(\lt)}) \\ -(\widetilde{x}_{j}^{(\lt)})^{i-1}g_1^{(+)}(\widetilde{x}_{j}^{(\lt)})\delta(\widetilde{y}_{j}^{(\lt)}) \end{array}\right]\nonumber
\end{eqnarray}
with $\delta^{(2)}(z)=\delta(x)\delta(y)$.

The minus sign in front of $\delta^{(2)}(\widetilde{z}_{i}^{(\rt)}-z_1)$ is needful to construct the chirality distribution over the real eigenvalues. Thereby we need the following relation \cite{Splittorff:2011bj} between the real eigenvalues of $D_\W+m\eins_{2n+\nu}$, $\lambda_i^{(\W)}+m$, and the eigenvalues of $D_5+m\gamma_5$, $\lambda_i^{(5)}(m)$,
\begin{eqnarray}\label{a3.3.4}
 \left.\frac{\partial\lambda_i^{(5)}}{\partial m}\right|_{\lambda_i^{(5)}=\lambda_i^{(\W)}+m=0}&=&\langle\psi_i|\gamma_5|\psi_i\rangle,\\
 \lambda_i^{(5)}(m)&=&\langle\psi_i|\gamma_5|\psi_i\rangle(\lambda_i^{(\W)}+m)+{\it o}(\lambda_i^{(\W)}+m),\label{a3.3.4a}
\end{eqnarray}
where $\psi_i$ is the eigenvector to the eigenvalue $\lambda_i^{(\W)}$ of $D_\W$. The right hand side of Eq.~\eref{a3.3.4} is  the chirality of the corresponding eigenvector. Since the eigenvectors of the complex eigenvalues have vanishing chirality, Eq.~\eref{a3.3.4} is only applicable to the real modes of $D_\W$. The following short calculation will show the connection between the chiral distribution over the real eigenvalues and the third term in Eq.~\eref{a3.3.3}, 
\begin{eqnarray}\label{a3.3.5}
 \fl\frac{1}{\pi}\underset{\varepsilon\to0}{\IM}\frac{\delta(y)}{\det(D_\W-x\eins_{2n+\nu}-\imath\varepsilon\gamma_5)}&=&\frac{1}{\pi}\underset{\varepsilon\to0}{\IM}\frac{\delta(y)}{\prod\limits_{j=1}^{2n+\nu}(\lambda_j^{(5)}(-x)-\imath\varepsilon)}\\
 \fl&=&\frac{\delta(y)}{\prod\limits_{j=1}^{2n+\nu}\lambda_j^{(5)}(-x)}\sum_{j=1}^{2n+\nu}\lambda_j^{(5)}(-x)\delta(\lambda_j^{(5)}(-x))\nonumber\\
 \fl&=&\hspace*{-0.5cm}\sum_{\lambda_j^{(\W)}\ {\rm is\ real}}\frac{\lambda_j^{(\W)}-x}{\prod\limits_{i=1}^{2n+\nu}(\lambda_i^{(\W)}-x)}\delta(\lambda_j^{(\W)}-x)\delta(y){\rm sign}\langle\psi_j|\gamma_5|\psi_j\rangle\nonumber\\
 \fl&=&\sum_{\widetilde{z}_j^{(\rt)}\ {\rm is\ real}}\frac{\widetilde{z}_j^{(\rt)}-z}{\det(D_\W-z\eins_{2n+\nu})}\delta^{(2)}(\widetilde{z}_j^{(\rt)}-z)\nonumber\\
 \fl&-&\sum_{\widetilde{z}_j^{(\lt)}\ {\rm is\ real}}\frac{\widetilde{z}_j^{(\lt)}-z}{\det(D_\W-z\eins_{2n+\nu})}\delta^{(2)}(\widetilde{z}_j^{(\lt)}-z)\nonumber
\end{eqnarray}
Hereby we have to understand the whole calculation, in particular the limit of the imaginary increment $\imath\varepsilon$, in a weak sense. The complex conjugated pairs of the integration variables in the third term of Eq.~\eref{a3.3.3}  do not contribute. We recognize this by expanding the second determinant in the two-point weights and the first determinant in the Dirac delta distribution such that we consider the integral
\begin{eqnarray}\label{a3.3.6}
 \fl I&=&\int\limits_{\mathbb{C}^{2n+\nu}}d[\widetilde{z}]\det(\widetilde{z}-\widehat{z}\eins_{2n+\nu})\prod\limits_{j=1}^n g_2^{(+)}(\widetilde{z}_{j}^{(\rt)},\widetilde{z}_{j}^{(\lt)})\prod\limits_{j=n+1}^{n+\nu} g_1^{(+)}(\widetilde{x}_{j}^{(\lt)})\delta(\widetilde{y}_{j}^{(\lt)})\Delta_\nu(\widetilde{z}^{(\lt)}_{>n})\\
  \fl&\times&\det\left[\begin{array}{c|c} (\widetilde{z}_{1}^{(\rt)})^{j-1} & -\delta^{(2)}(\widetilde{z}_{1}^{(\rt)}-z_1) \\ \hline (\widetilde{z}_{i}^{(\rt)})^{j-1} & 0 \\ \hline (\widetilde{z}_{1}^{(\lt)})^{j-1} &  \delta^{(2)}(\widetilde{z}_{1}^{(\lt)}-z_1) \\ \hline (\widetilde{z}_{i}^{(\lt)})^{j-1} &  0 \end{array}\right]\nonumber
\end{eqnarray}
If $\widetilde{z}_1^{(\lt)}=\widetilde{z}_1^{(\rt)\,*}=z$ the integrand is anti-symmetric under the complex conjugation of $z$, i.e. $z\leftrightarrow z^*$. The determinant is symmetric under $z\leftrightarrow z^*$ while $g_2^{(+)}(z,z^*)\sim g_{\rm c}(z)=-g_{\rm c}(z^*)$, see Eq.~\eref{2.13}. Thus the integral over the imaginary part of $z$ vanishes. The same discussion can be made for all complex conjugated pairs.

Expanding the determinant in the Dirac delta function and using the calculation~\eref{a3.3.5} we find
\begin{eqnarray}\label{a3.3.7}
 \fl K_2^{(+,n)}(z_1,z_2)&=&g_2^{(+)}(z_2,z_1)-\frac{(-1)^{n(n-1)/2+\nu(\nu-1)/2}}{(n-1)!(n+\nu-1)!}\prod\limits_{j=0}^{\nu-1}\frac{1}{h_j}\prod\limits_{j=0}^{n-1}\frac{1}{o_j^{(+)}}\\
 \fl&\times&\int\limits_{\mathbb{C}^{2}}d[\widehat{z}]g_2^{(+)}(z_1,\widehat{z}_1)g_2^{(+)}(z_2,\widehat{z}_2)(\widehat{z}_1-\widehat{z}_2)\int\limits_{\mathbb{C}^{2n+\nu-2}}d[\widetilde{z}]\Delta_{2n+\nu-2}(\widetilde{z})\nonumber\\
  \fl&\times&\det(\widetilde{z}-\widehat{z}_1\eins_{2n+\nu-2})\det(\widetilde{z}-\widehat{z}_2\eins_{2n+\nu-2})\det\left[\begin{array}{c} g_2^{(+)}(\widetilde{z}_{i}^{(\rt)},\widetilde{z}_{j}^{(\lt)}) \\ \hspace*{-0.2cm}-(\widetilde{x}_{j}^{(\lt)})^{i-1}g_1^{(+)}(\widetilde{x}_{j}^{(\lt)})\delta(\widetilde{y}_{j}^{(\lt)})\hspace*{-0.2cm} \end{array}\right]\nonumber\\
 \fl&+&\frac{(-1)^{n(n+1)/2+\nu(\nu-1)/2}}{n!(n+\nu)!}\prod\limits_{j=0}^{\nu-1}\frac{1}{h_j}\prod\limits_{j=0}^{n-1}\frac{1}{o_j^{(+)}}\frac{\delta(y_1)}{\pi}\int\limits_{\mathbb{C}}d[\widehat{z}]\frac{g_2^{(+)}(z_2,\widehat{z})}{x_1-\widehat{z}}\underset{\varepsilon\to0}{\IM}\int\limits_{\mathbb{C}^{2n+\nu}}d[\widetilde{z}]\nonumber\\
  \fl&\times&\Delta_{2n+\nu}(\widetilde{z})\frac{\det(D_\W-\widehat{z}\eins_{2n+\nu})}{\det(D_\W-x_1\eins_{2n+\nu}-\imath\varepsilon\gamma_5)}\det\left[\begin{array}{c} g_2^{(+)}(\widetilde{z}_{i}^{(\rt)},\widetilde{z}_{j}^{(\lt)}) \\ -(\widetilde{x}_{j}^{(\lt)})^{i-1}g_1^{(+)}(\widetilde{x}_{j}^{(\lt)})\delta(\widetilde{y}_{j}^{(\lt)}) \end{array}\right]\nonumber
\end{eqnarray}
The second term is an integral transform of the partition of $D_\W$ with two fermionic flavors and the last term is an integral over the partition function with one bosonic and one fermionic flavor. Notice that the integral over $\widehat{z}$ does not commute with the limit $\varepsilon\to0$ because of the singularity at $x_1$. This singularity cancels with a term after we take the limit. Hence, the expression~\eref{a3.3.7} is equal to the result~\eref{4.2.11}.

\subsection{The kernel $K_4^{(+,n)}$}\label{app3.4}

The starting point for this kernel is the identity
\begin{eqnarray}\label{a3.4.1}
 \fl K_4^{(+,n)}(z_1,z_2)&=&\frac{(-1)^{n(n-1)/2+\nu(\nu-1)/2}}{(n+1)!(n+\nu-1)!}\prod\limits_{j=0}^{\nu-1}\frac{1}{h_j}\prod\limits_{j=0}^{n-1}\frac{1}{o_j^{(+)}}\int\limits_{\mathbb{C}^{2n+\nu}}d[\widetilde{z}]\Delta_{2n+\nu}(\widetilde{z})\\
  \fl&&\hspace*{-1.5cm}\times\det\left[\begin{array}{c|cc} g_2^{(+)}(\widetilde{z}_{i}^{(\rt)},\widetilde{z}_{j}^{(\lt)}) & g_2^{(+)}(\widetilde{z}_{i}^{(\rt)},z_1)  & g_2^{(+)}(\widetilde{z}_{i}^{(\rt)},z_2) \\ \hline -(\widetilde{x}_{j}^{(\lt)})^{i-1}g_1^{(+)}(\widetilde{x}_{j}^{(\lt)})\delta(\widetilde{y}_{j}^{(\lt)}) & -x_1^{i-1}g_1^{(+)}(x_1)\delta(y_1)  & -x_2^{i-1}g_1^{(+)}(x_2)\delta(y_2) \end{array}\right].\nonumber
\end{eqnarray}
The index $j$ runs from $1$ to $n+\nu-1$ and the index $i$ takes the values $1$ to $n+1$ in the upper row and from $1$ to $\nu$ in the lower one. We introduce two Dirac delta functions and, thus, extend the integral by two additional $\widetilde{z}_{j}^{(\lt)}$ variables,
\begin{eqnarray}\label{a3.4.2}
 \fl &&K_4^{(+,n)}(z_1,z_2)=\frac{(-1)^{n(n+1)/2}}{(n+1)!(n+\nu+1)!}\prod\limits_{j=0}^{\nu-1}\frac{1}{h_j}\prod\limits_{j=0}^{n-1}\frac{1}{o_j^{(+)}}\int\limits_{\mathbb{C}^{2n+\nu+2}}d[\widetilde{z}]\\
 \fl&\times&\det\left[\begin{array}{c} g_2^{(+)}(\widetilde{z}_{i}^{(\rt)},\widetilde{z}_{j}^{(\lt)}) \\ -(\widetilde{x}_{j}^{(\lt)})^{i-1}g_1^{(+)}(\widetilde{x}_{j}^{(\lt)})\delta(\widetilde{y}_{j}^{(\lt)}) \end{array}\right]\det\left[\begin{array}{c|cc} (\widetilde{z}_{i}^{(\rt)})^{j-1} & 0 & 0 \\ \hline (\widetilde{z}_{i}^{(\lt)})^{j-1} & \delta^{(2)}(\widetilde{z}_{i}^{(\lt)}-z_1) & \delta^{(2)}(\widetilde{z}_{i}^{(\lt)}-z_2) \end{array}\right].\nonumber
\end{eqnarray}
We extend the determinant by Dirac delta functions of $\widetilde{z}_{i}^{(\rt)}$ similar to the calculation in \ref{app3.3},
\begin{eqnarray}\label{a3.4.3}
 \fl&& K_4^{(+,n)}(z_1,z_2)=\frac{(-1)^{n(n+1)/2}}{(n+1)!(n+\nu+1)!}\prod\limits_{j=0}^{\nu-1}\frac{1}{h_j}\prod\limits_{j=0}^{n-1}\frac{1}{o_j^{(+)}}\int\limits_{\mathbb{C}^{2n+\nu+2}}d[\widetilde{z}]\\
 \fl&\times&\det\left[\begin{array}{c|cc} (\widetilde{z}_{i}^{(\rt)})^{j-1} & -\delta^{(2)}(\widetilde{z}_{i}^{(\rt)}-z_1) & -\delta^{(2)}(\widetilde{z}_{i}^{(\rt)}-z_2) \\ \hline (\widetilde{z}_{i}^{(\lt)})^{j-1} & \delta^{(2)}(\widetilde{z}_{i}^{(\lt)}-z_1) & \delta^{(2)}(\widetilde{z}_{i}^{(\lt)}-z_2) \end{array}\right]\nonumber\\
  \fl&\times&\det\left[\begin{array}{c} g_2^{(+)}(\widetilde{z}_{i}^{(\rt)},\widetilde{z}_{j}^{(\lt)}) \\ -(\widetilde{x}_{j}^{(\lt)})^{i-1}g_1^{(+)}(\widetilde{x}_{j}^{(\lt)})\delta(\widetilde{y}_{j}^{(\lt)}) \end{array}\right]-\frac{(-1)^{n(n+1)/2+\nu}}{n!(n+\nu+1)!}\prod\limits_{j=0}^{\nu-1}\frac{1}{h_j}\prod\limits_{j=0}^{n-1}\frac{1}{o_j^{(+)}}\nonumber\\
 \fl&\times&\int\limits_{\mathbb{C}^{2n+\nu+1}}d[\widetilde{z}]\det\left[\begin{array}{c} g_2^{(+)}(z_2,\widetilde{z}_{j}^{(\lt)}) \\ g_2^{(+)}(\widetilde{z}_{i}^{(\rt)},\widetilde{z}_{j}^{(\lt)}) \\ -(\widetilde{x}_{j}^{(\lt)})^{i-1}g_1^{(+)}(\widetilde{x}_{j}^{(\lt)})\delta(\widetilde{y}_{j}^{(\lt)}) \end{array}\right]\det\left[\begin{array}{c|c} (\widetilde{z}_{i}^{(\rt)})^{j-1} & -\delta^{(2)}(\widetilde{z}_{i}^{(\rt)}-z_1) \\ \hline (\widetilde{z}_{i}^{(\lt)})^{j-1} & \delta^{(2)}(\widetilde{z}_{i}^{(\lt)}-z_1) \end{array}\right]\nonumber\\
 \fl&+&\frac{(-1)^{n(n+1)/2+\nu}}{n!(n+\nu+1)!}\prod\limits_{j=0}^{\nu-1}\frac{1}{h_j}\prod\limits_{j=0}^{n-1}\frac{1}{o_j^{(+)}}\nonumber\\
 \fl&\times&\int\limits_{\mathbb{C}^{2n+\nu+1}}d[\widetilde{z}]\det\left[\begin{array}{c} g_2^{(+)}(z_1,\widetilde{z}_{j}^{(\lt)}) \\ g_2^{(+)}(\widetilde{z}_{i}^{(\rt)},\widetilde{z}_{j}^{(\lt)}) \\ -(\widetilde{x}_{j}^{(\lt)})^{i-1}g_1^{(+)}(\widetilde{x}_{j}^{(\lt)})\delta(\widetilde{y}_{j}^{(\lt)}) \end{array}\right]\det\left[\begin{array}{c|c} (\widetilde{z}_{i}^{(\rt)})^{j-1} & -\delta^{(2)}(\widetilde{z}_{i}^{(\rt)}-z_2) \\ \hline (\widetilde{z}_{i}^{(\lt)})^{j-1} & \delta^{(2)}(\widetilde{z}_{i}^{(\lt)}-z_2) \end{array}\right]\nonumber\\
 \fl&+&\frac{(-1)^{n(n-1)/2+\nu(\nu-1)/2}}{(n-1)!(n+\nu+1)!}\prod\limits_{j=0}^{\nu-1}\frac{1}{h_j}\prod\limits_{j=0}^{n-1}\frac{1}{o_j^{(+)}}\hspace*{-0.2cm}\int\limits_{\mathbb{C}^{2n+\nu}}\hspace*{-0.3cm}d[\widetilde{z}]\Delta_{2n+\nu}(\widetilde{z})\det\left[\begin{array}{c} g_2^{(+)}(z_1,\widetilde{z}_{j}^{(\lt)}) \\ g_2^{(+)}(z_2,\widetilde{z}_{j}^{(\lt)}) \\ g_2^{(+)}(\widetilde{z}_{i}^{(\rt)},\widetilde{z}_{j}^{(\lt)}) \\ \hspace*{-0.2cm}-(\widetilde{x}_{j}^{(\lt)})^{i-1}g_1^{(+)}(\widetilde{x}_{j}^{(\lt)})\delta(\widetilde{y}_{j}^{(\lt)})\hspace*{-0.2cm} \end{array}\right]\hspace*{-0.1cm}.\nonumber
\end{eqnarray}
In the final step we expand the last three terms in $g_2^{(+)}(z_j,\widetilde{z}_{j}^{(\lt)})$. The Dirac delta functions can be rewritten as limits of Cauchy transforms, see Eq.~\eref{a3.3.5},
\begin{eqnarray}\label{a3.4.4}
 \fl&& K_4^{(+,n)}(z_1,z_2)=\frac{(-1)^{n(n-1)/2+
 \nu(\nu-1)/2}}{(n+1)!(n+\nu+1)!}\prod\limits_{j=0}^{\nu-1}\frac{1}{h_j}\prod\limits_{j=0}^{n-1}\frac{1}{o_j^{(+)}}\frac{x_2-x_1}{\pi^2}\delta(y_1)\delta(y_2)\underset{\varepsilon_2\to0}{\underset{\varepsilon_1\to0}{\IM}}\\
 \fl&\times&\int\limits_{\mathbb{C}^{2n+\nu+2}}d[\widetilde{z}]\Delta_{2n+\nu+2}(\widetilde{z})\frac{\det\left[\begin{array}{c} g_2^{(+)}(\widetilde{z}_{i}^{(\rt)},\widetilde{z}_{j}^{(\lt)}) \\ -(\widetilde{x}_{j}^{(\lt)})^{i-1}g_1^{(+)}(\widetilde{x}_{j}^{(\lt)})\delta(\widetilde{y}_{j}^{(\lt)}) \end{array}\right]}{\det(D_\W-x_1\eins_{2n+\nu+2}-\imath\varepsilon_1\gamma_5)\det(D_\W-x_2\eins_{2n+\nu+2}-\imath\varepsilon_2\gamma_5)}\nonumber\\
  \fl&-&g_2(z_2,z_1)-\frac{(-1)^{n(n+1)/2+\nu(\nu-1)/2}}{n!(n+\nu)!}\prod\limits_{j=0}^{\nu-1}\frac{1}{h_j}\prod\limits_{j=0}^{n-1}\frac{1}{o_j^{(+)}}\frac{\delta(y_1)}{\pi}\int\limits_{\mathbb{C}}d[\widehat{z}]\frac{g_2^{(+)}(z_2,\widehat{z}) }{x_1-\widehat{z}}\underset{\varepsilon\to0}{\IM}\nonumber\\
 \fl&\times&\int\limits_{\mathbb{C}^{2n+\nu}}d[\widetilde{z}]\Delta_{2n+\nu}(\widetilde{z})\frac{\det(D_\W-\widehat{z}\eins_{2n+\nu})}{\det(D_\W-x_1\eins_{2n+\nu}-\imath\varepsilon\gamma_5)}\det\left[\begin{array}{c} g_2^{(+)}(\widetilde{z}_{i}^{(\rt)},\widetilde{z}_{j}^{(\lt)}) \\ -(\widetilde{x}_{j}^{(\lt)})^{i-1}g_1^{(+)}(\widetilde{x}_{j}^{(\lt)})\delta(\widetilde{y}_{j}^{(\lt)}) \end{array}\right]\nonumber\\
  \fl&+&g_2(z_1,z_2)+\frac{(-1)^{n(n+1)/2+\nu(\nu-1)/2}}{n!(n+\nu)!}\prod\limits_{j=0}^{\nu-1}\frac{1}{h_j}\prod\limits_{j=0}^{n-1}\frac{1}{o_j^{(+)}}\frac{\delta(y_2)}{\pi}\int\limits_{\mathbb{C}}d[\widehat{z}]\frac{g_2^{(+)}(z_1,\widehat{z}) }{x_2-\widehat{z}}\underset{\varepsilon\to0}{\IM}\nonumber\\
 \fl&\times&\int\limits_{\mathbb{C}^{2n+\nu}}d[\widetilde{z}]\Delta_{2n+\nu}(\widetilde{z})\frac{\det(D_\W-\widehat{z}\eins_{2n+\nu})}{\det(D_\W-x_1\eins_{2n+\nu}-\imath\varepsilon\gamma_5)}\det\left[\begin{array}{c} g_2^{(+)}(\widetilde{z}_{i}^{(\rt)},\widetilde{z}_{j}^{(\lt)}) \\ -(\widetilde{x}_{j}^{(\lt)})^{i-1}g_1^{(+)}(\widetilde{x}_{j}^{(\lt)})\delta(\widetilde{y}_{j}^{(\lt)}) \end{array}\right]\nonumber\\
 \fl&+&\int\limits_{\mathbb{C}^2}d[\widehat{z}]g_2^{(+)}(z_1,\widehat{z}_1)g_2^{(+)}(z_2,\widehat{z}_2)(\widehat{z}_1-\widehat{z}_2)\Sigma_{n-1}^{(+)}(\widehat{z}_1,\widehat{z}_2).\nonumber
\end{eqnarray}
Only the first term is new in comparison to the kernel $K_2^{(+,n)}$. It is the partition function of $D_\W$ with two bosonic flavors which agrees with the result~\eref{4.2.12}.

\subsection{The kernel $K_5^{(+,n)}$}\label{app3.5}

Also for this kernel we start with
\begin{eqnarray}\label{a3.5.1}
 \fl K_5^{(+,n)}(z_1,z_2)&=&\frac{(-1)^{n(n+1)/2+\nu(\nu-1)/2}}{n!(n+\nu-1)!}\prod\limits_{j=0}^{\nu-1}\frac{1}{h_j}\prod\limits_{j=0}^{n-1}\frac{1}{o_j^{(+)}}\int\limits_{\mathbb{C}^{2n+\nu-1}}d[\widetilde{z}]\Delta_{2n+\nu}(\widetilde{z},z_1)\\
  \fl&\times&\det\left[\begin{array}{c|c} g_2^{(+)}(\widetilde{z}_{i}^{(\rt)},\widetilde{z}_{j}^{(\lt)})  & g_2^{(+)}(\widetilde{z}_{i}^{(\rt)},z_2) \\ \hline -(\widetilde{x}_{j}^{(\lt)})^{i-1}g_1^{(+)}(\widetilde{x}_{j}^{(\lt)})\delta(\widetilde{y}_{j}^{(\lt)}) & -x_2^{i-1}g_1^{(+)}(x_2)\delta(y_2) \end{array}\right],\nonumber
\end{eqnarray}
where the indices of the determinant are $j\in\{1,\ldots,n+\nu-1\}$, $i\in\{1,\ldots,n\}$ in the first row and $i\in\{1,\ldots,\nu\}$ in the last one. The extension with a Dirac delta function yields
\begin{eqnarray}\label{a3.5.2}
 \fl K_5^{(+,n)}(z_1,z_2)&=&\frac{(-1)^{n(n-1)/2+\nu}}{n!(n+\nu)!}\prod\limits_{j=0}^{\nu-1}\frac{1}{h_j}\prod\limits_{j=0}^{n-1}\frac{1}{o_j^{(+)}}\int\limits_{\mathbb{C}^{2n+\nu}}d[\widetilde{z}]\frac{\det(\widetilde{z}-z_1\eins_{2n+\nu})}{z_1-z_2}\\
  \fl&\times&\det\left[\begin{array}{c|c} (\widetilde{z}_{i}^{(\rt)})^{j-1} & 0 \\ \hline (\widetilde{z}_{i}^{(\lt)})^{j-1} & \delta^{(2)}(\widetilde{z}_{i}^{(\lt)}-z_2) \end{array}\right]\det\left[\begin{array}{c} g_2^{(+)}(\widetilde{z}_{i}^{(\rt)},\widetilde{z}_{j}^{(\lt)})  \\ -(\widetilde{x}_{j}^{(\lt)})^{i-1}g_1^{(+)}(\widetilde{x}_{j}^{(\lt)})\delta(\widetilde{y}_{j}^{(\lt)}) \end{array}\right].\nonumber
\end{eqnarray}
We proceed in the same way as in \ref{app3.3} by extending the first determinant by $-\delta^{(2)}(\widetilde{z}_{i}^{(\lt)}-z_2)$ and expanding the resulting correction in $g_2^{(+)}(z_2,\widetilde{z}_{i}^{(\lt)})$. Then we find
\begin{eqnarray}\label{a3.5.3}
 \fl K_5^{(+,n)}(z_1,z_2)&=&\frac{(-1)^{n(n+1)/2+\nu(\nu-1)/2}}{n!(n+\nu)!}\prod\limits_{j=0}^{\nu-1}\frac{1}{h_j}\prod\limits_{j=0}^{n-1}\frac{1}{o_j^{(+)}}\frac{1}{\pi}\frac{\delta(y_2)}{z_1-x_2}\underset{\varepsilon\to0}{\IM}\int\limits_{\mathbb{C}^{2n+\nu}}d[\widetilde{z}]\\
  \fl&\times&\Delta_{2n+\nu}(\widetilde{z})\frac{\det(D_\W-z_1\eins_{2n+\nu})}{\det(D_\W-x_2\eins_{2n+\nu}-\imath\varepsilon\gamma_5)}\det\left[\begin{array}{c} g_2^{(+)}(\widetilde{z}_{i}^{(\rt)},\widetilde{z}_{j}^{(\lt)})  \\ -(\widetilde{x}_{j}^{(\lt)})^{i-1}g_1^{(+)}(\widetilde{x}_{j}^{(\lt)})\delta(\widetilde{y}_{j}^{(\lt)}) \end{array}\right]\nonumber\\
  \fl&+&\int\limits_{\mathbb{C}}d[\widehat{z}]g_2^{(+)}(z_2,\widehat{z})(\widehat{z}-z_1)\Sigma_{n-1}^{(+)}(\widehat{z},z_1).\nonumber
\end{eqnarray}
The first term is the partition function of $D_\W$ with one bosonic and one fermionic flavor and the second term is the kernel $K_3^{(+,n)}$. Therefore Eq.~\eref{a3.5.3} is the result~\eref{4.2.13}.

\end{document}